\documentclass[
 pra,
 amsmath,amssymb,
 aps,
 twocolumn,
 notitlepage,
 superscriptaddress
]{revtex4-1}

\usepackage[toc,page]{appendix}
\newcommand{\nocontentsline}[3]{}
\newcommand{\tocless}[2]{\bgroup\let\addcontentsline=\nocontentsline#1{#2}\egroup}

\usepackage{styles}
\usepackage{fancyhdr}
\usepackage{txfonts}
\usepackage{xurl} 
\usepackage{color}

\newcommand{\benchmark}[1]{\texttt{#1}}

\begin{document}

\title{Application-Oriented Performance Benchmarks for Quantum Computing}

\author{Thomas Lubinski}
\affiliation{Quantum Circuits Inc, 25 Science Park, New Haven, CT 06511}
\affiliation{QED-C Technical Advisory Committee on Standards and Performance Benchmarks Chairman}

\author{Sonika Johri}
\affiliation{IonQ Inc, 4505 Campus Dr, College Park, MD 20740, USA}

\author{Paul Varosy }
\affiliation{Department of Physics, Colorado School of Mines, Golden, CO 80401, USA}

\author{Jeremiah Coleman}
\affiliation{Department of Electrical and Computer Engineering, Princeton University, Princeton, NJ, 08544, USA}

\author{Luning Zhao}
\affiliation{IonQ Inc, 4505 Campus Dr, College Park, MD 20740, USA}

\author{Jason Necaise}
\affiliation{D-Wave Systems, Burnaby, British Columbia, Canada, V5G 4M9, Canada}

\author{Charles H. Baldwin}
\affiliation{Quantinuum, 303 S. Technology Ct, Broomfield, CO 80021, USA}

\author{Karl Mayer}
\affiliation{Quantinuum, 303 S. Technology Ct, Broomfield, CO 80021, USA}

\author{Timothy Proctor}
\affiliation{Quantum Performance Laboratory, Sandia National Laboratories, Livermore, CA 94550, USA}

\collaboration{Quantum Economic Development Consortium (QED-C) collaboration} 

\thanks{This work was sponsored by the Quantum Economic Development Consortium (QED-C) and was performed under the auspices of the QED-C Technical Advisory Committee on Standards and Performance Benchmarks. The authors acknowledge many committee members for their input to and feedback on the project and this manuscript.}

\date{\rule[15pt]{0pt}{0pt}\today}
             
\begin{abstract}

\vspace{0.0cm}
In this work we introduce an open source suite of quantum application-oriented performance benchmarks that is designed to measure the effectiveness of quantum computing hardware at executing quantum applications. These benchmarks probe a quantum computer's performance on various algorithms and small applications as the problem size is varied, by mapping out the fidelity of the results as a function of circuit width and depth using the framework of volumetric benchmarking. In addition to estimating the fidelity of results generated by quantum execution, the suite is designed to benchmark certain aspects of the execution pipeline in order to provide end-users with a practical measure of both the quality of and the time to solution. Our methodology is constructed to anticipate advances in quantum computing hardware that are likely to emerge in the next five years. This benchmarking suite is designed to be readily accessible to a broad audience of users and provides benchmarks that correspond to many well-known quantum computing algorithms. 
\end{abstract}

\keywords{Quantum Computing \and Benchmarks \and Benchmarking \and Algorithms \and Application Benchmarks}

\maketitle

\tableofcontents


\pagestyle{fancy}

\renewcommand{\headrulewidth}{0.0pt}
\lhead{}
\rhead{\thepage}

\renewcommand{\footrulewidth}{0.4pt}
\cfoot{}
\lfoot{Application-Oriented Performance Benchmarks for Quantum Computing}
\rfoot{\today}
\vspace{2cm}
\begin{figure*}[t!]
\includegraphics[width=18cm]{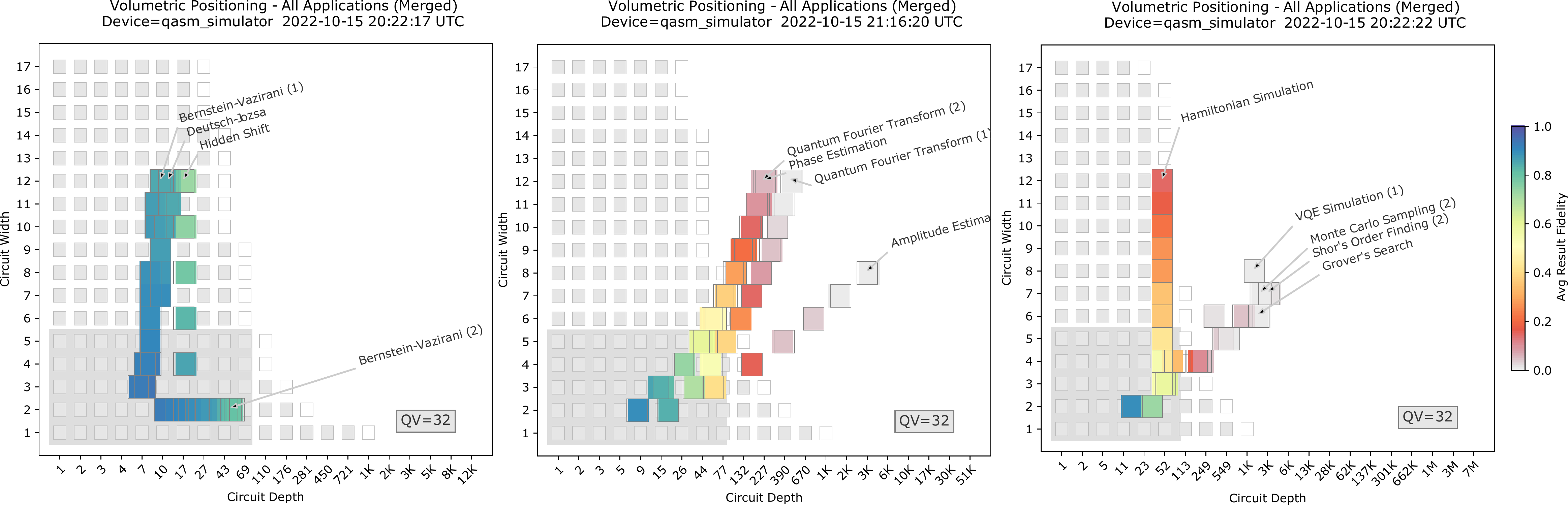}
\caption{\textbf{ Quantum Application-Oriented Performance Benchmarks.} The results of executing our quantum application-oriented performance benchmarking suite on a simulator of a noisy quantum computer, with results split into benchmarks based on three loose categories of algorithm: tutorial, subroutine, and functional. For each benchmark, circuits are run for a variety of problem sizes. This typically correspond to the circuit's width, i.e., the number of qubits it acts on, which here range from 2 to 12 qubits. The result fidelity, a measure of the result quality, is computed for each circuit execution, and is shown as a colored square positioned at the corresponding circuit's width and normalized depth. Results for circuits of equal width and similar depth are averaged together. The results of the application-oriented benchmarks are shown on top of a `volumetric background' (grey-scale squares). Here and throughout this paper, except where stated, this volumetric background is a heuristic extrapolation of a device's quantum volume (here, 32) to predict the region in which a circuit's result fidelity will be above $\nicefrac{1}{2}$ (the grey squares). Note that this extrapolation is not expected to always be accurate (see discussion in main text) but it is nevertheless useful. This is because any deviations between the performance of the algorithmic benchmarks and the prediction of the volumetric background signify that, for the processor in question, the performance of these algorithms is difficult to predict from the processor's quantum volume alone.
}
\label{fig:benchmarks_all_vp_sim_1}
\end{figure*}

\section{Introduction}
\label{sec:introduction}
Over the past decade, quantum computers have evolved from exploratory physics experiments into cloud-accessible hardware, heralding their transformation into early-stage commercial products. A broad audience of users has now begun to explore and interact with quantum computers, inspiring research into practical quantum algorithms \cite{preskill2018quantum} and garnering attention from the academic, government, and business communities. With recent demonstrations of quantum advantage \cite{Arute2019-mk, Zhong2020-rk}, it seems increasingly likely that quantum computers will someday outperform their classical counterparts for practically relevant tasks. Unlike contemporary classical computers, though, the information processing capabilities of current quantum computers are limited primarily by errors, not by their size or speed.
 
Quantum computers can experience a wide range of complex errors. Many of these errors result from noise in or miscalibrations of the lowest-level components in the system---quantum gate operations and the qubits they act on. Component-level benchmarks and characterization protocols, such as randomized benchmarking \cite{PhysRevA.77.012307,PhysRevLett.106.180504} or gate set tomography \cite{Blume-Kohout2017-no}, can provide insight into the type and magnitude of these errors. These tools are critical for experimental efforts to improve hardware. But many users seeking to run an application on a quantum computer are not concerned with low-level details, but rather how likely their application is to execute successfully. Extrapolating low-level performance metrics to predict the performance of a specific application is challenging, and these extrapolations are often inaccurate \cite{murphy2019controlling, proctor2020measuring}.
 
The limited predictive power of low-level performance metrics has contributed to an increasing focus on benchmarks that concisely and directly summarize the holistic performance of a quantum computer \cite{Boixo_2018, Cross_2019, proctor2020measuring}. A notable example is the quantum volume benchmark \cite{Cross_2019}, which is designed to probe the effective useful size of a specific quantum computer and summarize it in one number: the quantum volume. The quantum volume has been widely adopted and reported by the community. However, due to the complexity of errors in quantum hardware, neither a device's quantum volume nor any other single metric is likely to accurately predict its performance on all applications \cite{murphy2019controlling, proctor2020measuring}. There is thus a pressing need for a diverse set of application-centric metrics and benchmarks that test the performance of quantum computers on practically relevant tasks. Such benchmarks will enable hardware developers to quantify their progress in commercially relevant ways, and will make it possible for end users to more accurately predict how the available hardware will perform on their application. 

In this paper, we introduce an extensible suite of quantum performance benchmarks that are \emph{application-oriented}. These benchmarks are designed to test quantum computers on a diverse range of tasks that are based on quantum computing applications and algorithms. Our suite complements other recent benchmarking methods \cite{mccaskey2019quantum, MICHIELSEN201744, Wright_2019, Koch_2020, Mills_2021, Amoretti2021effectiveframework, cornelissen2021scalable} that use, e.g., small chemistry problems or basic quantum circuits as benchmarks. Demonstrated in Figure~\ref{fig:benchmarks_all_vp_sim_1}, each benchmark in our suite is derived from an algorithm or application and specifies a scalable family of quantum circuits. The benchmark suite is intended to evaluate the capability of quantum hardware to successfully execute meaningful computational tasks, and to reflect likely use cases for quantum computers. The benchmarking suite is available as a public, open-source repository with extensive documentation \cite{qc-proto-benchmarks}. Validated implementations of our benchmarks are supplied in multiple common quantum programming languages, including Qiskit \cite{qiskit_org}, Cirq \cite{google_cirq}, Braket \cite{amazon_braket} and {Q\#} \cite{microsoft_qsharp}, and can be rapidly deployed on nearly all major cloud quantum computers.

Our benchmarking suite consists of a set of benchmarks that are each based on an algorithm or application, and each benchmark is designed to measure proxies for the quality of and the time to solution for its corresponding application.
In analyzing the results of these benchmarks, we primarily adopt the framework of volumetric benchmarking \cite{BlumeKohout2020volumetricframework}, as shown in Figure~\ref{fig:benchmarks_all_vp_sim_1}. This approach, which generalizes the quantum volume, displays the result quality for each application as a function of the circuit width (often this is equal to the problem size) and the computation's length (circuit depth). To enable algorithmic volumetric benchmarks, we define a normalized measure of result quality that makes results from different applications comparable, and we define a normalized, device-independent measure of circuit depth that removes many of the complications associated with diverse native gate sets and dissimilar device connectivities. This analysis makes it possible to easily compare the performance of difference quantum computing devices on the same tasks.

Our benchmarking suite is intended to be an evolving code base, accepting contributions from the quantum computing research community. Throughout this paper we highlight the limitations of the first version of our benchmarking suite, reported here, and suggest paths towards enhancing the suite. Indeed, there is evidence from classical computing that a suite of successful benchmarks will evolve significantly over time as weaknesses are uncovered and as hardware evolves.
A compelling example is the evolution of the SPEC (Standard Performance Evaluation Corporation) benchmark suite \cite{spec_org, hennessy_patterson_2019_all}, one of the most successful suites for benchmarking classical computers. Starting in the late 1980s with 10 benchmarks, this suite went through at least 5 major revisions over the next two decades, stabilizing in 2006 with 25 benchmarks, only 6 of which have some resemblance to those of the early suite.
 
The remainder of this paper is structured as follows. Section \ref{sec:background} provides background on the fundamentals guiding this work. Section \ref{sec:application_oriented_benchmarks} introduces our application-oriented benchmarks and our methodology. In Section \ref{sec:benchmark_results} we present results from executing subsets of the benchmark suite on quantum simulators and quantum hardware. We discuss the challenges of measuring execution time in Section \ref{sec:execution_time}, and we conclude in Section \ref{sec:summary-and-conclusions}. Appendix \ref{apdx:algorithms_and_applications} contains a detailed description of each of our application-oriented benchmarks. The limitations of our current benchmarking suite are highlighted throughout the paper, and Appendices~\ref{apdx:issues_apis} and~\ref{apdx:fidelity} address some of them in detail. In Appendix \ref{apdx:advances} we analyze the adaptations to our suite that will be required to address recent technological enhancements such as mid-circuit measurements, parameterized gate operations, and hybrid classical-quantum computations. 

\section{Background}
\label{sec:background}
In this section we review the methods on which our work is based.
To place our work in context, in Section~\ref{sec:benchmarking_quantum_computers} we begin with a brief discussion of the component-level performance metrics that are typically available to users of the current generation of hardware. Throughout this paper we will compare a quantum computer's performance on our benchmarks with its quantum volume, so in Section~\ref{sec:quantum_volume} we review the quantum volume metric. To present results from our benchmarks we will primarily use the framework of volumetric benchmarking. In Section~\ref{sec:volumetric_benchmarks} we review this methodology and explain how it provides the underpinning for the analysis of our application benchmarks.

\subsection{Component-Level Specifications}
\label{sec:benchmarking_quantum_computers}
\begin{figure}[t!]
\includegraphics[width=0.43\textwidth]{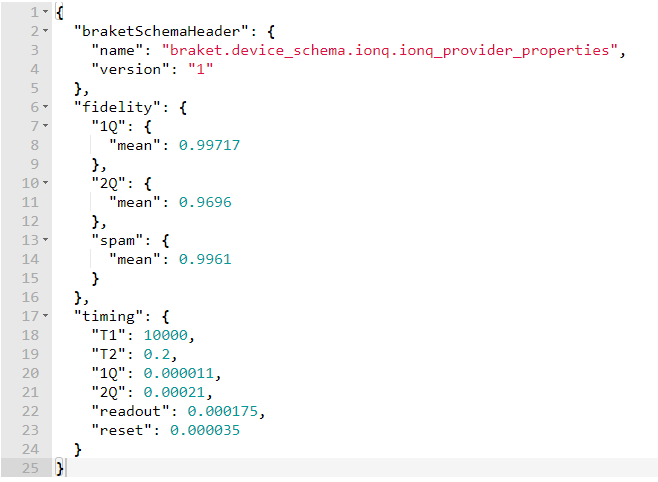}
\caption{\textbf{An example of component-level performance metrics.} The component-level performance metrics provided by Amazon Braket for the IonQ Quantum Processing Unit.}
\label{fig:amazon_braket_ionq_specs}
\end{figure}

\begin{figure}[t!]
\includegraphics[width=0.36\textwidth]{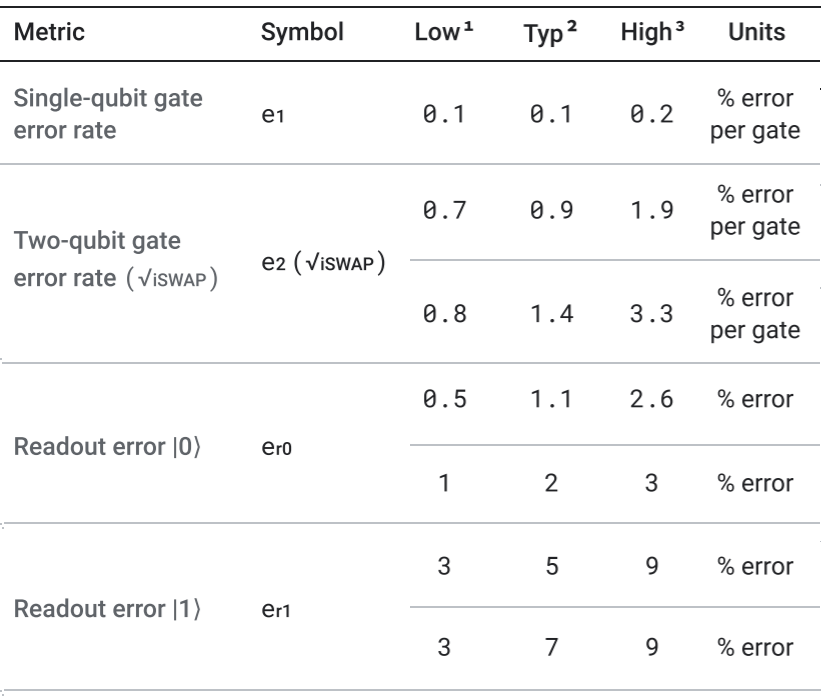}
\caption{\textbf{An example of component-level performance metrics.} The component-level performance metrics provided by Google for its Weber device.}
\label{fig:google_weber_specs}
\end{figure}

There are many techniques for estimating component-level performance of quantum computing devices, e.g., randomized benchmarking \cite{PhysRevA.77.012307,PhysRevLett.106.180504} and gate set tomography \cite{Blume-Kohout2017-no}. Hardware providers use these techniques to calibrate and test their systems, and they often publish performance metrics extracted from the results of these experiments. The measures of performance that providers report to users of their systems typically include one- and two-qubit gate fidelity, state preparation and measurement (SPAM) fidelity, and measures of how long each qubit can retain quantum information (T1 and T2 coherence times). For example, Figure \ref{fig:amazon_braket_ionq_specs} shows the contents of the text file that the Amazon Braket cloud service \cite{amazon_braket_ionq} presents to users of the IonQ Quantum Processing Unit (QPU), and Figure \ref{fig:google_weber_specs} shows similar information that Google \cite{google_weber_specs_1} provide for its Weber device (Sycamore class).
Other quantum computing cloud services, such as IBM Quantum Services \cite{ibmq2021}, provide similar types of information about their machines.

Component-level performance metrics provide a lot of information about a machine, but they have at least two important limitations. Firstly, component-level performance metrics are typically not sufficient to accurately predict a device's performance on an algorithm \cite{murphy2019controlling, proctor2020measuring}. This is because summary metrics cannot capture the full complexity of errors in quantum computing components (e.g., individual gates and readouts), and not all sources of error are even apparent when testing a single component (e.g., crosstalk \cite{sarovar2019detecting, gambetta2012characterization}). Second, component-level metrics are challenging for non-specialists to interpret. It is often difficult for a prospective quantum computer user to extrapolate from component-level metrics to predict how their application will perform.

\subsection{Quantum Volume}
\label{sec:quantum_volume}
\begin{figure}[t!]
\includegraphics[width=\columnwidth]{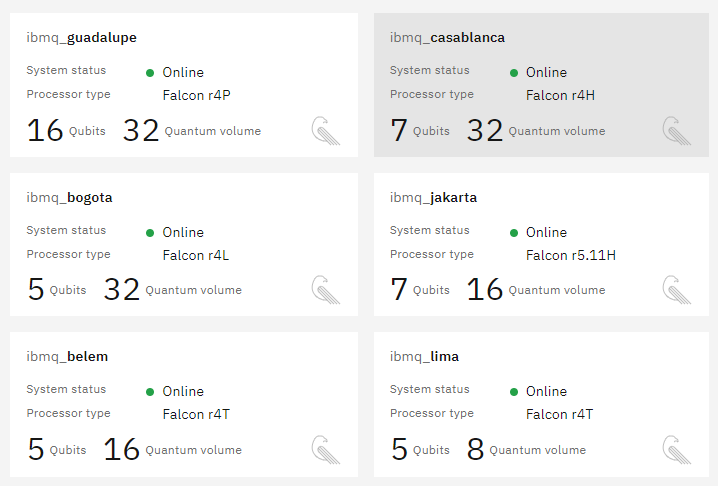}
\caption{\textbf{Quantum volume examples.} The number of qubits and the quantum volume for a selection of IBM Q machines, accessed via IBM Quantum Services.}
\label{fig:ibm_some_systems_qv}
\end{figure}

In this work we compare results from our benchmarks to predictions extrapolated from a device's quantum volume, and so we now review the quantum volume. IBM introduced the quantum volume (QV) metric \cite{Cross_2019} in recognition of the need for a simple way to quantify and compare the capabilities of quantum computing devices.
The quantum volume is a single metric that is influenced by multiple factors contributing to the overall performance of a quantum computer, including its number of qubits, systemic errors, device connectivity, and compiler efficiency.
Several vendors and users have welcomed this as a step towards quantifying what a quantum computer can do \cite{honeywell_qv_quote}. IBM publishes the quantum volume for each machine that it makes available to external users through IBM Quantum Services \cite{ibmq2021}, as shown in 
Figure \ref{fig:ibm_some_systems_qv}. Note that a device with more qubits does not always have a larger quantum volume.

A quantum computer's quantum volume is defined in terms of its average performance on a particular set of random circuits, which we refer to as the quantum volume circuits. An $n$-qubit quantum volume circuit is a square circuit consisting of $n$ layers of Haar-random $\textrm{SU}(4)$ unitaries between  $\lfloor \nicefrac{n}{2} \rfloor$ randomly selected qubit pairs (where $\lfloor \nicefrac{n}{2} \rfloor$ rounds $\nicefrac{n}{2}$ down to the nearest integer). A device's quantum volume is defined to be $ 2^n $ with $n$ being both the width (the number of qubits) and the depth (the number of layers) of the largest quantum volume circuits that can be executed `successfully' on a specified quantum device.
So a quantum computer that can successfully execute $5$ layer quantum volume circuits on 5 qubits, but not $6$ layer quantum volume circuits on 6 qubits, would have a quantum volume of $ 2^5 $ or 32. A device is considered to implement $n$-qubit quantum volume circuits successfully if, on average, it produces more `heavy outputs' than a set threshold of $\nicefrac2{3}$ with two-sigma confidence. Quantum volume is defined as the exponential of the circuit depth as this is intended to represent the challenge of simulating the same circuit on a classical computer. Note, however, that no single number can summarize the many dimensions that make up a quantum computer's performance characteristics \cite{murphy2019controlling, aaronson_2020}. For example, the quantum volume is not guaranteed to reliably predict the performance of wide and shallow circuits, or deep and narrow circuits.

\subsection{Volumetric Benchmarks}
\label{sec:volumetric_benchmarks}
\begin{figure}[t!]
\includegraphics[width=\columnwidth]{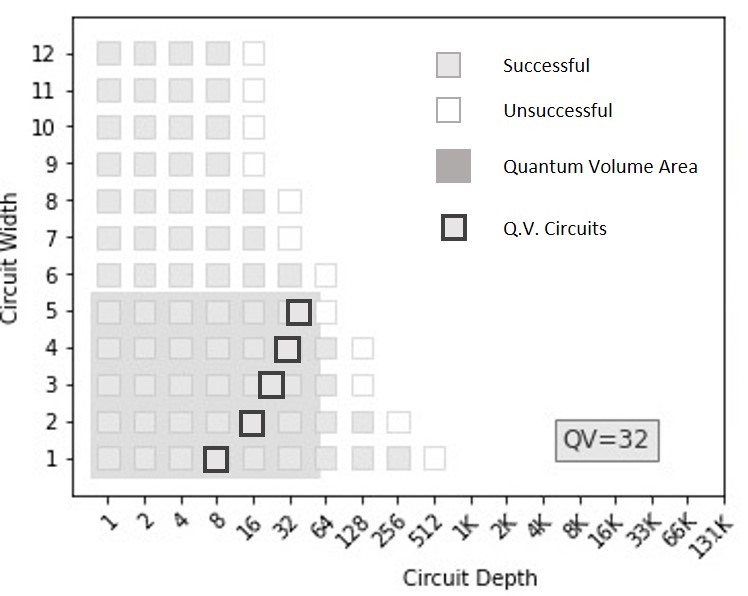}
\caption{\textbf{An example of volumetric benchmarking.} Volumetric benchmarking results for a hypothetical 12-qubit quantum computer, whereby a quantum computer's performance on some family of circuits is mapped out as a function of circuit width and depth. This example uses a binary measure of performance: grey and white squares show the circuit shapes at which the test circuits succeeded and failed, respectively, according to this metric. The location of the quantum volume circuits, that would be used to extract the quantum volume, are shown in bold. Here and throughout, `depth' for all circuits is defined with respect to a particular standardized gate set, and in this gate set the quantum volume circuits are not square.}
\label{fig:vb_intro}
\end{figure}

To present results from our benchmarks we will primarily use the framework of volumetric benchmarking \cite{BlumeKohout2020volumetricframework}, which we now review. Volumetric benchmarking is a framework for constructing and visualizing the results of large, flexible families of benchmarks. It consists of (1) choosing a family of circuits to use as a benchmark, (2) selecting circuits from this family and executing them on the hardware to be benchmarked, and then (3) plotting the observed performance of the hardware on these circuits as a function of the circuits' width and depth. The results are plotted in the depth $\times$ width `volumetric space', as illustrated in Figure \ref{fig:vb_intro}. The data points (here, grey-scale squares) show the result quality for the tested circuits of that shape (width and depth), and in this example the metric of result quality takes one of two values: `success' (grey squares) or `fail' (white squares). Note that Figure~\ref{fig:vb_intro} also contains the location of the quantum volume circuits (black squares) and a quantum volume region (grey rectangle) that we explain in Section~\ref{sec:volumetric_positioning}.

Volumetric benchmarking can be used to estimate and report the performance of both wide shallow circuits and narrow deep circuits, in addition to square circuits like those tested in the quantum volume benchmark. Furthermore, volumetric benchmarking is not restricted to a single kind of circuit (e.g., the specific random circuits of the quantum volume benchmark). This makes volumetric benchmarking an ideal framework within which to place application-oriented benchmarks.

\section{Application-Oriented Benchmarks}
\label{sec:application_oriented_benchmarks}
In this section we introduce our application-oriented benchmarking suite, and we explain how we analyze the data and present the results. A quantum algorithm or application (e.g., Grover's Search) does not itself directly constituent a well-defined benchmark, so in this section we explain both what algorithms and applications are included in our suite and how they are converted into specific benchmarks. Our benchmarking suite consists of four components:
\begin{enumerate}
    \item A set of algorithms or subroutines $\{\mathcal{A}\}$ [e.g., Grover's Search or the Quantum Fourier Transform (QFT)], each with a variable input size $n$.
    \item For each algorithm $\mathcal{A}$, one or more recipes for turning that algorithm into a set of quantum circuits $\{\mathcal{C}\}_n$ for each input problem size $n$. These circuits are always closely related to, but are not necessarily identical to, the circuits that define the algorithm. Note that these circuits are specified using a hardware-agnostic gate set, and they therefore must be compiled into a specific device's native operations before they are run (see Section~\ref{sec:benchmarking_algorithm}).
    \item A procedure for using a given family of circuits $\{\mathcal{C}\}_n$ as a benchmark, including a method for selecting circuits from $\{\mathcal{C}\}_n$ to run (such as sampling from a distribution over $\{\mathcal{C}\}_n$) and, e.g., a method for varying $n$.
    \item Methods for analyzing the data.
\end{enumerate}

Below we introduce each of these components in turn. In Section~\ref{sec:the_benchmarking_suite} we state the algorithms that we have initially included in our benchmarking suite (component 1 above). In Section~\ref{sec:creating_benchmark_circs} we discuss how we turn an algorithm into a specific set of circuits (component 2 above), what should be considered when doing so, and the limitations of some of our current benchmarking circuits. In Section~\ref{sec:benchmarking_algorithm} we introduce the experimental procedure for running a circuit set $\{\mathcal{C}\}_n$ (component 3 above). In Sections~\ref{sec:volumetric_positioning}-\ref{sec:circuit_fidelity} we introduce the methods for analyzing the data (component 4 above). In Section~\ref{sec:volumetric_positioning} we explain how and why we present the results of the benchmarking using volumetric benchmarking plots, and in Section~\ref{sec:vbackgrounds} we explain how we generate `volumetric backgrounds' that are independent predictions for a quantum computer's performance on the application-oriented benchmarks. These plots necessitate defining a circuit's \emph{depth}, and so in Section~\ref{sec:circuit_depth} we discuss three different ways in a which a circuit's depth could be defined---each with strengths and weaknesses---and state the choice that we made in this work. The benchmarking suite currently focuses primarily on result quality (rather than, e.g., time to solution) and this can be quantified in many ways. In Section~\ref{sec:circuit_fidelity} we discuss the merits of different choices for this quality metric, and introduce our choice: a normalized version of classical fidelity. Having introduced our benchmarking suite, in Section~\ref{sec:benchmark_rules} we then discuss what constitutes reasonable usage of the benchmarks, i.e., what are the rules of use?

\subsection{The Algorithms in the Benchmarking Suite} \label{sec:the_benchmarking_suite}
Our benchmarking suite is based on a set of quantum algorithms or applications, i.e., every benchmark in our suite is built from a quantum algorithm. Our intention is that this benchmarking suite will evolve over time (as has, e.g., the SPEC benchmark suite \cite{spec_org, hennessy_patterson_2019_all} for classical computers), and a particularly important part of this evolution will be changes in the applications and algorithms it contains. The initial set of algorithms and applications that we chose is  summarized in Table~\ref{tab:benchmark_suite}. 

\begin{center}
\begin{table}[h]
\begin{tabular}{ |c|c|c| } 
 \hline
 \multicolumn{2}{|c|}{Application-Oriented Benchmarks} \\
 \hline
                & Deutsch-Jozsa \\
 Tutorial       & Bernstein-Vazirani \\
                & Hidden Shift \\
 \hline
                & Quantum Fourier Transform \\ 
 Subroutine     & Phase Estimation \\
                & Amplitude Estimation \\
 \hline
                & Grover's Search \\
                & Hamiltonian Simulation \\ 
 Functional     & Monte Carlo Sampling \\ 
                & Variational Quantum Eigensolver \\
                & Shor's Period Finding \\
 \hline
\end{tabular}
\caption{\label{tab:benchmark_suite}The initial set of quantum algorithms and applications included within our application-oriented benchmarking suite. We anticipate that the algorithms included within our suite will expand and evolve over time, e.g., quantum error correction subroutines are an important quantum computational primitive that are not currently represented in this suite. We find it useful to separate the current suite of algorithms into three informal categories: tutorial, subroutine, and functional.}
\end{table}
\end{center}

\vspace{-0.8cm}     

We refer to our suite as `application-oriented benchmarks' (rather than `application benchmarks') because our suite does not---and is not intended to---only include complete and practically useful algorithms. This is reflected in the three informal categories into which we place the algorithms we have selected: `functional', `subroutines' and `tutorial'. In our suite, we include complete algorithms that are anticipated to be useful within the functional category. It could be argued that an `application-oriented' benchmarking suite should only contain benchmarks based on complete, useful algorithms---but we do not take this position, for two reasons. First, many algorithms use the same subroutines, e.g, the QFT, and so it is valuable to benchmark a quantum computer's performance on those subroutines. This is true even if that subroutine is, on its own, not computationally useful (e.g., the QFT with a product state input and followed immediately by a computational basis measurement can be efficiently simulated classically). This is the motivation for including the algorithms that are within our subroutine category.

Second, the ultimate practical applications of quantum computers are currently uncertain---both the problems that quantum computers will solve and the algorithms that will be used to solve those problem remain unclear. Diversity in a benchmarking suite is therefore important, i.e., the properties of the algorithms and circuits within our benchmarking suite should be diverse. This motivates include simple oracle algorithms (e.g., Deutsch-Jozsa) within our suite, as, unlike most of the other algorithms we have included, they require only shallow (i.e., low depth) circuits. We collect these algorithms in the tutorial category (termed as such because these algorithms are commonly used to introduce concepts in quantum computing). Note that the need for a diverse benchmarking suite is also an important motivation for including many of the other nominally useful algorithms that we selected. This is because, e.g., Shor's algorithm is unlikely to execute with sufficiently low error except when run on error corrected logical qubits---so the main scientific reason for using Shor's algorithm, run directly on physical qubits, to benchmark near-term hardware is to increase the diversity of the tasks with which hardware is tested. We anticipate that as near-term (i.e., NISQ) and long-term (i.e., fault-tolerant error-corrected) applications of quantum computers become clearer, our benchmarking suite will evolve to reflect this.

\subsection{Constructing Benchmarking Circuits}
\label{sec:creating_benchmark_circs}
Each of our benchmarks is based on an algorithm, but an algorithm does not itself define a unique benchmark. For example, most algorithms are defined for a large class of input problems and a quantum computer cannot be benchmarked by exhaustively testing every problem instance. In our benchmarking suite, a specific benchmark $\texttt{A}$ for an algorithm $\mathcal{A}$ is defined by a method for (1) turning that algorithm into a set of circuits $\{\mathcal{C}\}_n$ for each input problem size $n$, (2) selecting $N_{\textrm{circs}}$ circuits to run from $\{\mathcal{C}\}_n$, and (3) selecting $N_{\textrm{params}}$ values for all adjustable parameters (if any) within each selected circuit $\mathcal{C}$. Applying this procedure for a specified $n$ results in $N_{\textrm{circs}} \times N_{\textrm{params}}$ `complete' circuits to be run. A `complete' circuit means that all gates within the circuit are specified (including their parameter values, if any), the circuit begins with all qubits initialized in the $\ket{0}$ state, and the circuit ends with a computational basis measurement of all qubits. In all cases,  the circuit set $\{\mathcal{C}\}_n$ consists of a subset of the circuits that implement the algorithm $\mathcal{A}$ or some simple adaptation of those circuits.

For each algorithm that we chose, we designed one or more benchmarks---defined by the three ingredients above---based on that algorithm. Complete specifications for each benchmark are provided in Appendix \ref{apdx:algorithms_and_applications}. When designing our benchmarks we aimed to satisfy two competing criteria. Firstly, it should be feasible to run and analyze the data produced by a benchmark. For example, ideally a benchmark's circuits should be designed so that the data analysis can be implemented with polynomial (in $n$) time and space classical computations---although note that a benchmark with exponentially scaling classical computation requirements is not necessarily infeasible to run in the near term. Second, a quantum computer's performance on a benchmark $\texttt{A}$ built from an algorithm $\mathcal{A}$ should reflect that computer's performance on $\mathcal{A}$ as closely as possible. Satisfying both criteria is challenging, as they are conflicting. This is because the quantum circuits of a useful quantum algorithm applied to a classically intractable problem instance cannot be efficiently simulated on a classical computer, so we cannot simply compare the output of these circuits to the results of a classical simulation. Any feasible benchmark must circumvent this challenge. 

Many of our initial benchmarks satisfy one but not both of the above criteria. For example, our \texttt{QFT(1)} benchmark inputs a random computational basis state ($\ket{x}$ where $x \in [0..2^n]$) into the standard QFT circuit, adds one to the resultant Fourier basis state (using single qubit gates we map $F\ket{x}\to F\ket{x+1}$ where $x+1$ is modulo $2^n$ arithmetic and $F$ is the QFT unitary), and then applies the inverse QFT. The correct output bit string is the binary representation of $x+1$---which can evidently be efficiently computed---so this benchmark is feasible to implement on any number of qubits. However, this benchmark does not test the action of the device's imperfect implementation of the QFT on entangled input states (although it does test the inverse QFT's action on entangled states), which are the input of the QFT in all its practical applications.  Conversely, some of our other benchmarks require classical simulations that are exponentially expensive in the number of qubits. Improving each of our benchmarks so that they all satisfying the above criteria is important work that is already ongoing (see discussion in Section~\ref{sec:summary-and-conclusions}).

\subsection{The Benchmarking Procedure}
\label{sec:benchmarking_algorithm}
Here we explain the procedure that we use to execute a benchmark \texttt{A}. Executing an application benchmark \texttt{A} consists of (1) choosing a range of problem sizes at which to probe performance, (2) generating a set of circuits for each problem size, which are selected from \texttt{A}'s circuit set $\{\mathcal{C}\}_n$ using the method specified by $\texttt{A}$, and (3) executing the circuits and measuring key components of performance, notably the result fidelity and the execution time. We implement this methodology using Algorithm \ref{alg:benchmark_execution_loop}:


\begin{algorithm}[h!]
    \caption{Benchmark Execution Loop}
    \label{alg:benchmark_execution_loop}
    \begin{algorithmic}[1]
    \State $target \gets backend\_id$
    \State $initialize\_metrics()$
    \For{$problem\_size \gets min\_size, max\_size$}  \Comment{$N_1$}                    
        \For{$circuit\_id \gets 1, max\_circuits$} \Comment{$N_2$}
            \State $execute\_classical()$
            \State $circuit \Leftarrow create\_circuit(circuit\_id)$
            \State $circuit \Leftarrow compile\_load\_circuit(circuit)$
            \For{$params \gets 1, max\_params$}   \Comment{$N_3$}
                \State $execute\_quantum(target, circuit, params, shots)$
            \EndFor
            \State $collect\_circuit\_metrics()$
        \EndFor
        \State $collect\_group\_metrics()$
    \EndFor
    \end{algorithmic}
\end{algorithm}

The algorithm defines three nested loops. The outer loop (labelled $N_1$) is a sweep over the range of problem sizes. The middle loop ($N_2$) represents multiple interleaved executions of classical and quantum execution routines. In all our existing benchmarks the classical routines consist solely of circuit construction and compilation. In this step each of the benchmark's circuits is compiled into the native operations of the hardware under test, using the provider's compilers (our benchmarks therefore test both the performance of the provider's classical compilation algorithms and the performance of the quantum hardware). Future benchmarks may use additional classical computation (e.g., benchmarks based on variational algorithms that include the classical optimization loop).  Each iteration of the middle loop constructs a quantum circuit, which can be a parameterized circuit. The inner loop ($N_3$) consists of executing that circuit, for some range of parameter values if it is a parameterized circuit. Our procedure for implementing a benchmark explicitly contains some user-specified variables---including the number of shots per circuit ($N_{\rm{shots}}$) and the total number of circuits at each $n$ ($N_{\textrm{circs}} \times N_{\textrm{params}}$)---and implicitly contains much implementation freedom (e.g., the circuit compilation algorithm can be chosen by the user). In Section~\ref{sec:benchmark_rules} we discuss a tentative set of rules for what choices for these explicit and implicit user-specified variables are permissible.

\subsection{Volumetric Positioning}
\label{sec:volumetric_positioning}

Our benchmarking suite uses \emph{volumetric benchmarking plots} \cite{BlumeKohout2020volumetricframework} to visualize the results of our quantum application benchmarks. We now explain how we use volumetric benchmarking plots (a necessary component for volumetric benchmarking plots is defining a circuit's depth---we defer this important point to Section~\ref{sec:circuit_depth}). Each of our application-oriented benchmarks is associated with a family of circuits that is parameterized by the input problem size, i.e., each input size $n$ defines a circuit ensemble $\{\mathcal{C}\}_n$. Each application's circuit family has a particular `footprint' that represents the way in which the shape of the circuits---i.e., their width and depth---varies as a function of the input size. We can therefore report the performance of our application benchmarks in the depth $\times$ width `volumetric space'. In this work, for each application benchmark \texttt{A} a fixed input size $n$ always corresponds to a single circuit width $w$. Note that often $w=n$, i.e., a problem instance of size $w$ uses an $n$-qubit circuit, but this is not always the case (e.g., our \texttt{Bernstein-Vazirani(2)} benchmark uses mid-circuit measurements to implement the Bernstein-Vazirani algorithm using $n=2$ qubit circuits for any input problem size $w$). In contrast, a benchmark \texttt{A}'s circuits can have a range of depths for each input size. For example, the algorithms in the `tutorial' category use a varied-depth oracle circuit that encodes the problem instance. We therefore use the circuits' mean depth to define that ensemble's volumetric position, and we report average performance over the circuit ensemble.

To illustrate this idea, we executed our benchmarking suite on a quantum simulator. The lower plot of Figure \ref{fig:transpilation_example_1} shows the result fidelity (see below) for a single benchmark plotted in the volumetric space.
Figure \ref{fig:benchmarks_all_vp_sim_1} shows the result fidelity for each of our benchmarks, plotted in the volumetric space and separated into the three categories of benchmark. Each benchmark consists of running its circuits for a range of problem sizes, which closely correspond to the circuits' widths. Each data point (small rectangles) summarizes the results for the circuits of a particular problem size. The rectangle's location indicates the circuits' width and average depth, and the color indicates the average quality of the result for that circuit shape. We quantify the quality of the result using the average result fidelity, defined in Section \ref{sec:circuit_fidelity}. The width and depth of benchmarking circuits for different applications can be similar, resulting in overlap on the volumetric plots.
Whenever this is the case the result fidelities from the overlapping circuits for different applications are averaged together. We use this averaging in Figure \ref{fig:benchmarks_all_vp_sim_1}, and throughout this paper except where stated.

\subsection{Volumetric Backgrounds}\label{sec:vbackgrounds}
Algorithmic benchmarks directly probe a quantum computer's performance on application-like tasks, but performance on these tasks can also be predicted from a processor's known performance data (e.g., RB error rates). Discrepancies between these predictions and experimental results can reveal the presence and size of interesting and important physical effects, e.g., crosstalk. Throughout this work, we compare the results of our application-oriented benchmarks to a prediction by placing them on top of what we call a `volumetric background', shown by the grey-scale rectangles in Figure \ref{fig:benchmarks_all_vp_sim_1}. A volumetric background means a prediction for the result fidelity, as a function of circuit width and depth, obtained from some method that is independent of our application oriented benchmarks. In Figure \ref{fig:benchmarks_all_vp_sim_1} and throughout most of this paper, we use a volumetric background that is extrapolated from the corresponding device's quantum volume. For visual clarity, we use a binary `success' (grey squares) or `fail' (white squares) background, with `success' corresponding to a result fidelity of at least $\nicefrac{1}{2}$ \footnote{In the quantum volume protocol, success is quantified in terms of the heavy output probability (HOP) rather than result fidelity. The success threshold used for the HOP in the quantum volume protocol is $\nicefrac{2}{3}$, which approximately corresponds to a threshold for result fidelity of $\nicefrac{1}{2}$.}. We use this type of volumetric background throughout this paper, except where stated.

Predicting a circuit's result fidelity from a processor's quantum volume requires an extrapolation. We use the following heuristic  methodology that is arguable the best guess for the result fidelity of a width $w$ and depth $d$ circuit given only a processor's quantum volume. A width $w$ and depth $d$ circuit is predicted to have a result fidelity above $\nicefrac{1}{2}$ if $wd < (\log_2(V_Q))^2$ where $V_Q$ is the quantum volume \cite{Cross_2019}. That is, a circuit is predicted to succeed if its depth $\times$ width is no larger than the largest successful quantum volume circuits. Note that here, as throughout, we use the standardized circuit depth (see Section~\ref{sec:circuit_depth}) and with this definition for the circuit depth quantum volume circuits are not square.

No volumetric background is likely to accurately predict the performance of all circuits, including our application oriented benchmarks. This is because a circuit's performance is not only a function of its shape, i.e., its volumetric position, as it also depends on the type of errors present and whether those errors change with time or with number of qubits. A volumetric background inferred from a quantum volume involves particularly large extrapolations, as discussed in Section~\ref{sec:comparison_to_generic_benchmarks}. For this reason, we also include a `quantum volume region' (the large grey square in Figure \ref{fig:benchmarks_all_vp_sim_1}), alongside any volumetric background extrapolated from a quantum volume. This quantum volume region encompasses all circuit shapes that are smaller than the largest successful quantum volume circuit, and so circuits within this region will have high result fidelity under weaker assumptions.

\begin{figure}[t!]
\includegraphics[width=\columnwidth]{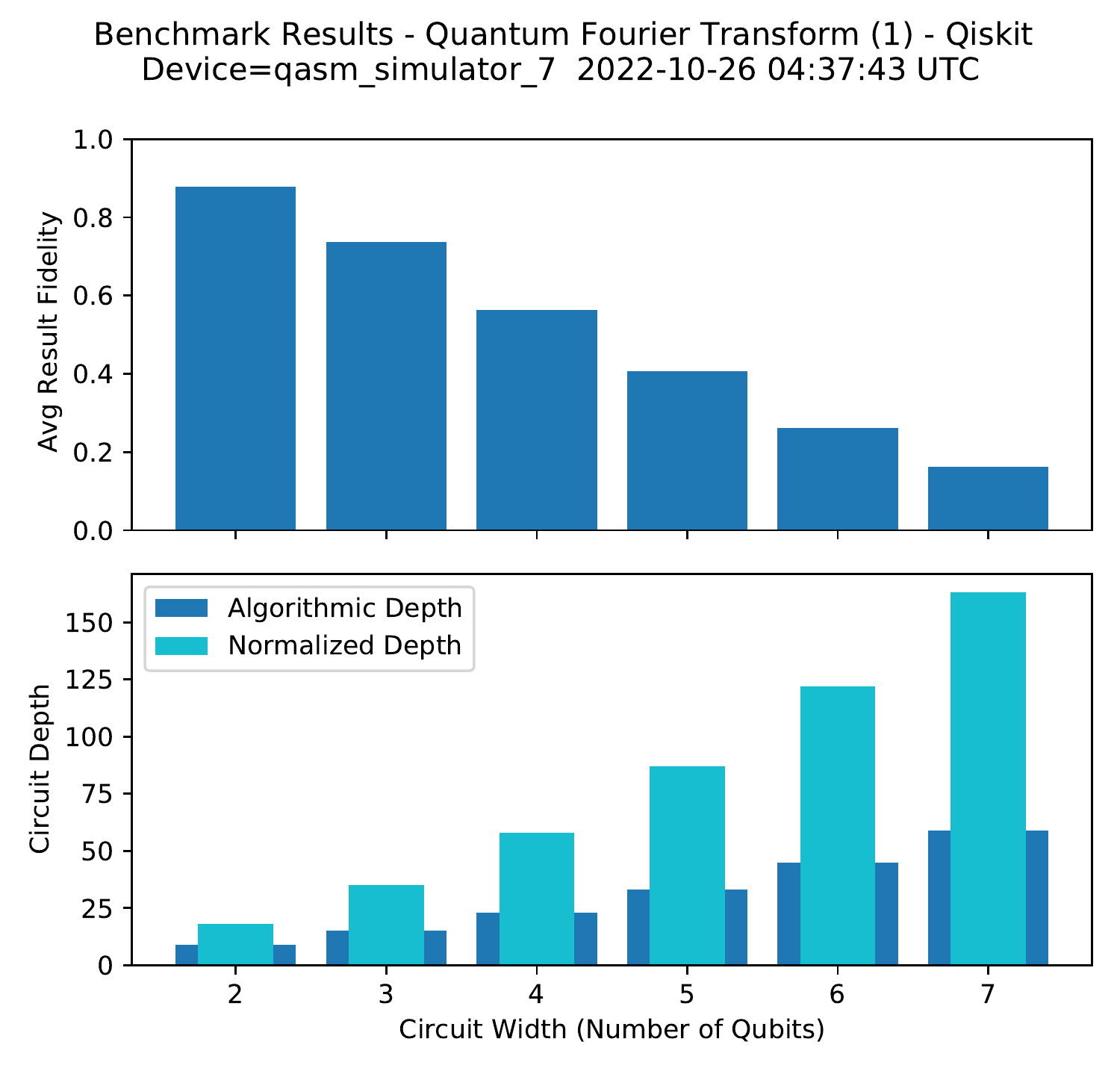}
\includegraphics[width=\columnwidth]{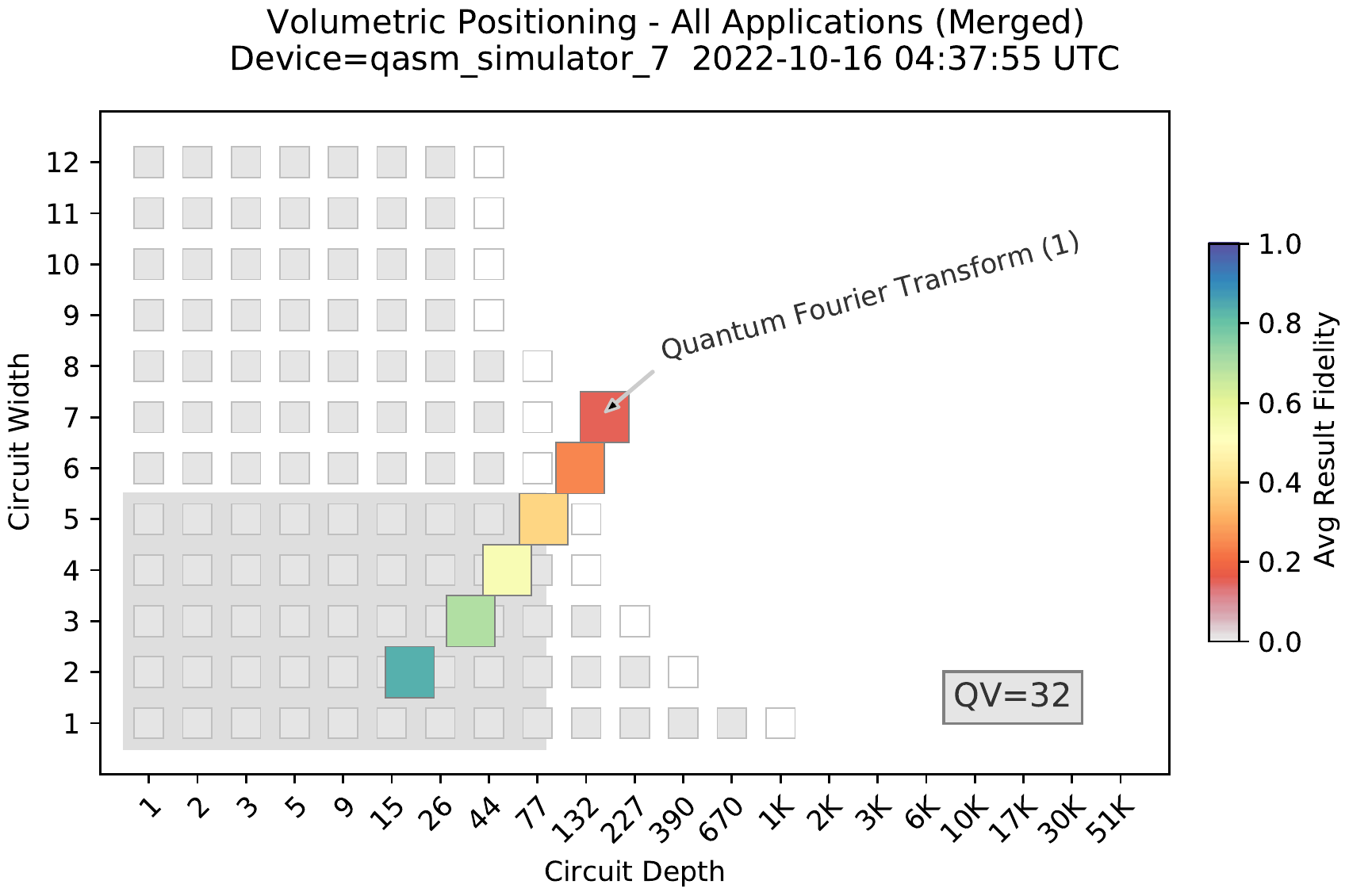}
\caption{\textbf{Demonstrating our methodology.} Demonstrating our methodology by executing our \benchmark{QFT(1)} benchmark up to 7 qubits on a noisy quantum computer simulator. The upper plot shows the average result fidelity as a function of circuit width. The middle plot shows both the circuit's normalized depth (the depth of the circuit after it has been transpiled into our standard gate set) and the circuit's algorithmic depth (the depth when written in the high-level API used to define the benchmarks), as a function of circuit width. The lower plot presents these data on a volumetric background, in order to visualize the `footprint' and result fidelity of the benchmark circuits. In this and all volumetric plots, the circuit depth refers to normalized depth.}
\label{fig:transpilation_example_1}
\end{figure}

\subsection{Circuit Depth}
\label{sec:circuit_depth}
We will primarily present our results in the volumetric space, meaning that a circuit's shape, i.e., its depth and width, specifies where results from that circuit will appear in our plots. Here we introduce and motivate the normalized definition for circuit depth that we use. A circuit's depth is the number of layers of logic gates it contains. However, any specific quantum computation can be implemented with many different circuits, perhaps defined over different gate sets, that have different depths. Our benchmark circuits are defined in terms of a high-level
API (such as Qiskit) and these `high-level circuits' use algorithm-specific gate sets that often contain gates that are not directly implementable on many devices. In order to run a high-level circuit, it  must be converted (transpiled) to the set of primitive quantum gates that the hardware executes. This transpiled circuit's depth---which we call the `physical depth'---is often larger than the depth of the original high-level circuit.
For example, Toffoli gates (multiply-controlled NOT gates) are often used in high-level descriptions of algorithms, but each Toffoli gate must be decomposed into many one- and two-qubit gates in order to run it on most quantum hardware. 

As discussed in detail in Ref.~\cite{BlumeKohout2020volumetricframework}, there are (at least) three useful kinds of definition for a benchmarking circuit's depth:

\begin{itemize}
    \item \emph{Algorithmic depth}. For each algorithm, express its circuits in a canonical form using a natural, algorithm-specific gate set. The depth of these canonical circuits is what we call algorithmic depth. For example, there is a well-known canonical form for the circuits of the QFT. Similarly, Grover's Search consists of repeating two subroutines (diffusion and oracle subroutines) which could be considered the natural gate set for Grover's algorithm.
    \item \emph{Normalized depth}. Choose a basis (gate set) and a canonical programmability constraints---e.g., some set of one- and two-qubit gates with all-to-all connectivity---and some classical compilation algorithm for transpiling a circuit into this gate set. Normalized depth is computed by transpiling a high-level circuit into this canonical gate set, and computing that transpiled circuit's depth.
    \item \emph{Physical depth}. Transpile a high-level circuit into a processor's native operations. The depth of this `low-level' circuit defines the physical depth.
\end{itemize}

These three kinds of definition for a circuit's depth have complementary strengths and weaknesses. Using an algorithm-specific definition for depth (`algorithmic depth') makes it possible to directly connect depth to computationally-relevant quantities, such as the number of queries to the oracle in Grover's algorithm. However, this kind of depth definition makes it challenging to compare the results of different algorithmic benchmarks. At the other extreme, physical depth enables transparent comparison between the results of different algorithmic benchmarks on the same processor (we expect circuits with similar physical depths to have similar result fidelities). However, a particular benchmark circuit's physical depth varies across devices (because, e.g., it depends on a device's connectivity) making cross-device comparisons challenging. Furthermore, physical depth is not always made available by quantum computing providers, and it is not consistently defined as there are ambiguities in what a processor's `native operations' are.

In this work we chose to use a particular version of normalized depth. This is a device- and algorithm-independent definition for depth, obtained by specifying a device-independent basis (a gate set) to which the circuit is transpiled. The middle plot of Figure \ref{fig:transpilation_example_1} shows an example of the depth of circuits from the \benchmark{QFT(1)} benchmark when (1) written in a high-level API (this is `algorithmic depth') and (2) after the circuits have been transpiled to our standard basis, i.e., the normalized depth of the circuit. In this example, and in most of the benchmarks, the normalized depth is larger than the circuit depth when written in a high-level API.

The basis that we selected consists of arbitrary single-qubit rotations around the 3 Pauli axes, a CNOT gate, and a measurement gate. We assume that CNOT gates can be applied between any pair of qubits (i.e., all qubit pairs are connected), and that gates on distinct qubits can be applied in parallel. This gate set, denoted by [`rx', `ry', `rz', `cx'] in Qiskit, is a widely used universal basis \cite{Williams2011} that is hardware-agnostic. Note, however, that this (or any other) basis choice is unavoidably subjective.
Each benchmark circuit is transpiled to this basis to compute a normalized circuit depth. Note that finding a minimal depth compilation into this gate set is not feasible (due to the classical computational complexity of the problem)---so the normalized depth depends on the heuristic classical algorithm used by the transpiler (we use Qiskit's compilation algorithm). This is arguably an undesirable property of normalized definitions for depth. When executed on hardware, the circuit is (necessarily) transpiled to the basis specific to that hardware, which will often include the addition of qubit SWAP gates so as to respect the topology of that hardware. Throughout this paper, `circuit depth' always refers to our normalized definition for circuit depth unless otherwise stated.

\subsection{Quantifying Result Quality with Circuit Fidelity}
\label{sec:circuit_fidelity}
Quantum computing hardware errors are the main roadblock to useful quantum computations, so our benchmarking suite focuses on measuring their impact on each algorithm. Currently our benchmarking suite uses one metric to quantify result quality: a normalized version of classical fidelity, introduced below, that we refer to as `result fidelity'. Like all options for quantifying result quality, this version of fidelity has both strengths and weaknesses, and we are \emph{not} advocating its exclusive adoption. We anticipate that future versions of our benchmarking suite will compute multiple metrics for result quality, as we discuss further below.

Result fidelity is defined in terms of the observed and ideal output probability distributions for a circuit. Each time an $n$-qubit circuit $\mathcal{C}$ is executed on quantum computing hardware we obtain a sample from some distribution $P_{\textrm{output}}$ over $n$-bit strings, where $P_{\textrm{output}}(x)$ is the probability of observing bit string $x$. This will differ from the distribution $P_{\textrm{ideal}}$ that would be sampled from if $\mathcal{C}$ was run on a perfect quantum computer. The error in the implementation of $\mathcal{C}$ can therefore be quantified by the discrepancy between $P_{\textrm{output}}$ and  $P_{\textrm{ideal}}$. There are many complementary ways to quantify the closeness of (or distance between) two probability distributions. The metric we use is a normalized version of the classical fidelity $F_{\rm s}$ \cite{qiskit_org, nielsen2002quantum} (note that classical fidelity is related to the classical Hellinger distance \cite{hellinger1909neue}). The classical fidelity $F_{\rm s}$ is defined by
\begin{equation}
F_{\rm s}(P_{\textrm{ideal}},P_{\textrm{output}}) = \left(\sum_x \sqrt{P_{\textrm{output}}(x)P_{\textrm{ideal}}(x)}\right)^2.
\label{eq:fidelity}
\end{equation}
Our normalized fidelity is given by
\begin{equation}
    F(P_{\textrm{ideal}},P_{\textrm{output}}) = \max\left\{  F_{\textrm{raw}}(P_{\textrm{ideal}},P_{\textrm{output}}) , 0\right\}, \\[10pt]
\label{eq:normalized-fidelity}
\end{equation}
where
\begin{equation}
    F_{\textrm{raw}}(P_{\textrm{ideal}},P_{\textrm{output}}) =  \frac{
    F_{\rm s}(P_{\textrm{ideal}}, P_{\textrm{output}}) - F_{\rm s}(P_{\textrm{ideal}},P_{\textrm{uni}})
    } {1 - F_{\rm s}(P_{\textrm{ideal}},P_{\textrm{uni}})}. \\[10pt]
\label{eq:raw-normalized-fidelity}
\end{equation}

This normalized fidelity is defined for all probability distributions except the uniform distribution $P_{\textrm{uni}}$---in which case the denominator in Eq.~\eqref{eq:raw-normalized-fidelity} is zero---but it is poorly behaved whenever $ F_{\rm s}(P_{\textrm{ideal}},P_{\textrm{uni}}) \approx 1$. However, benchmarking circuits for which $ F_{\rm s}(P_{\textrm{ideal}},P_{\textrm{uni}}) \approx 1$ should be avoided as whenever $P_{\textrm{ideal}}$ is close to $P_{\textrm{uni}}$ it is difficult to observe the effect of errors (as the approximate impact of many kinds of errors, including depolarization, is to cause $P_{\textrm{output}}$ to be a mixture of $P_{\textrm{ideal}}$ and $P_{\textrm{uni}}$). Note that this does not preclude designing benchmarks based on algorithms that, on some practically relevant inputs, are intended to produce coherent superpositions over all computational basis states with uniform (or close to uniform) amplitudes. In such cases, those algorithms should be converted into benchmark circuits that do not have this property. For example, the QFT maps a computational basis state to a uniform superposition of computational basis states, so our \texttt{QFT(1)} benchmark follows the QFT by the inverse QFT, so that these circuits' error-free output state is a computational basis state.

Directly computing $F$ (or $F_{\textrm{s}}$) for a circuit $\mathcal{C}$ requires calculating that circuit's $P_{\textrm{ideal}}$ on a classical computer (meaning simulating $\mathcal{C}$) and estimating $P_{\textrm{output}}$ by running $\mathcal{C}$ $N_{\rm{shots}}$ times on the processor being tested. The classical simulation is exponentially expensive for general $\mathcal{C}$, and the number of experimental shots ($N_{\rm{shots}}$) needed for low-uncertainty estimates of $F$ can also grow quickly with $n$. This is why we aim to design benchmarking circuits for which $P_{\rm{ideal}}$ can be efficiently computed.
 
The reason that we use $F$ instead of the standard fidelity ($F_{\rm s}$) is that $F_{\rm s}$ is non-zero even when the output of the circuit is completely random, i.e., when $P_{\textrm{output}} = P_{\textrm{uni}}$. Moreover, the value of $F_{\rm s}(P_{\textrm{ideal}},P_{\textrm{uni}})$ depends on $P_{\textrm{ideal}}$ (it depends on the sparsity of $P_{\textrm{ideal}}$). This is particularly inconvenient for comparisons between the results of different benchmarks, as the sparsity of $P_{\textrm{ideal}}$ varies. The normalized fidelity has the property that $F(P_{\textrm{ideal}},P_{\textrm{uni}}) = 0$, which is convenient because errors in quantum computing hardware often cause $P_{\rm{output}}$ to converge towards $P_{\rm{uni}}$ as circuits get larger or deeper. We have empirically found that the normalized fidelities of different benchmarks, at similar circuit shapes, are more correlated than the standard fidelities are.

Any metric for result quality has strengths and weaknesses. One inconvenient property of the normalized fidelity is that its `raw' version $F_{\textrm{raw}}$ is negative if $P_{\rm{ideal}}$ is closer to the uniform distribution than to the observed distribution. The maximum in Eq.~\eqref{eq:normalized-fidelity} removes this unwanted effect, but is \emph{ad hoc}. Alternative metrics for quantifying the closeness of $P_{\rm{ideal}}$ and $P_{\rm{output}}$ will be added to our benchmarking suite in the future. Furthermore, note that the quality with which a quantum circuit is implemented can also be quantified in terms of the closeness between the desired and actual quantum evolutions---using metrics like quantum process fidelity and diamond distance. Quantities like process fidelity are not functions of only $P_{\textrm{ideal}}$ and $P_{\textrm{output}}$, as they can only be measured by embedding a circuit $\mathcal{C}$ within multiple different contexts (e.g., ending the circuit with a variety of subroutines that simulate measurements in different bases). We anticipate adding techniques for measuring, e.g., process fidelity, to our benchmarking suite in the future.

Throughout this work, `result fidelity' refers to the normalized fidelity $F$ of Eq.~\eqref{eq:normalized-fidelity}. Typically, we average this fidelity over a circuit ensemble, which we denote by $\bar{F}$, but note that other statistics of the set of circuits' result fidelities could also be computed (Appendix~\ref{apdx:fidelity} provides an example that highlights the limitations of averaging). Figure \ref{fig:transpilation_example_1} shows how we present the average result fidelity of a set of benchmark circuits, along with their widths and normalized depths, using the example of the \benchmark{QFT(1)} benchmark run on a simulation of a noisy quantum computer. The benchmark sweeps over a range of circuit widths and computes the average result fidelity for the circuits of that width, the average depth of the original high-level circuits, and the average normalized depth, i.e., the depth of the circuits after transpilation to the standard basis. This fidelity and depth data are shown, versus circuit width, in the upper two plots of Figure \ref{fig:transpilation_example_1}. The lower plot in Figure \ref{fig:transpilation_example_1} shows the same data plotted on a volumetric background, which is the primary method that we use to visualize our results.

\subsection{Implementation Rules}\label{sec:benchmark_rules}
Our benchmarking suite will likely be used to compare the performance of different devices and competing technologies, so it is important to discuss what constitutes reasonable usage of these benchmarks. Methods for benchmarking quantum computers typically have user-adjustable aspects, which can be both explicit inputs (i.e. adjustable variables) of the benchmarking routine or details of how the benchmarks are implemented that are not formally specified by the benchmarking procedure. Not all choices for these variables and implementation details will necessary constitute reasonable usage of a benchmarking method. 

Quantum computing is still an early-stage technology and we believe that it is important to avoid prescriptive benchmarking suites with inflexible rules. Nevertheless, we think that it is valuable to outline some implementation principles for our benchmarking suite that can be considered a tentative standard practice. Deviations from these suggestions may well be reasonable and useful, but they should be justified. Our benchmarking suite contains many distinct benchmarks, and we anticipate that the number of benchmarks will grow in the future. We consider it to be reasonable for a user to choose which of these benchmarks to run. This is because different benchmarks test performance on different tasks, and a user might not consider all the tasks to be relevant. Note, however, that when the aim is to compare two or more devices we strongly recommend running the same set of benchmarks on all devices whenever feasible (for the experimental results presented herein we did not run all our benchmarks on each tested processor, but these results are explicitly not intended to be used to compare different processors).

Each individual benchmark in our suite has three important adjustable parameters: the values of $n$ for which performance is tested (corresponding to the input problem size, which often also corresponds to the number of qubits in the circuits), the number of circuits ($N_{\rm{circs}}$) that are selected from the benchmarking circuit set $\{\mathcal{C}\}_n$, and the number of times each circuit is run ($N_{\textrm{shots}}$). We consider it to be reasonable for a user to choose the values of $n$ to test, but if two or more processors are to be tested and the results compared either (a) the same values of $n$ should be used for all processors, or (b) a principled methodology should be chosen for selecting the values of $n$ to use (and this methodology should be stated). For example, one reasonable strategy is to increase $n$ until either the result fidelity drops below a threshold value (e.g., 10\%) or the maximum number of qubits available is reached.

We recommend that the number of circuits sampled from $\{\mathcal{C}\}_n$ at each $n$ is at least $N_{\rm{circs}} \geq 10$. Note, however, that small values for $N_{\rm{circs}}$ may sometimes be necessary to keep the total number of circuits that need executing small---as was the case in our experiments, where we fixed $N_{\rm{circs}} = 3$ due to throughput limitations with some hardware (for repeatability, we also ran the same circuits on all devices that we tested, by using a single random seed in each pseudo-random number generator). Whenever the total number of circuits to be run is $\gg 1$, this is not unreasonable as it will simply result in larger statistically fluctuations in the performance observed at a given circuit shape (meaning that there will be more ``speckle'' in a volumetric benchmarking plot) whenever there is significant variation in a processor's result fidelity on circuits within $\{\mathcal{C}\}_n$. In contrast, however, note that it not permissible to discard circuits selected from $\{\mathcal{C}\}_n$, e.g., because a processor’s result fidelity on those circuits is observed to be low (i.e., post-selection of the circuits for high result fidelity). Finally, we recommend that $N_{\rm{shots}} \geq 1000$ (this recommendation was followed in all our experiments \footnote{In all experiments $N_{\rm{shots}} = 1000$, except the \benchmark{VQE(1)} experiments where $N_{\rm{shots}} =4098$ due to a coding oversight.}), although fewer shots per circuit may be reasonable for many purpose (e.g., for computing average result fidelity when $N_{\textrm{circs}} \gg 1$). When small values for $N_{\rm{shots}}$ or $N_{\rm{circs}}$ are used they should be clearly stated and, ideally, statistical uncertainties on all estimated quantities should be computed (currently the benchmarking suite does not automatically compute standard errors).

Our benchmarking suite permits broad implementation freedom. One particularly important freedom is that the benchmarking circuits can be compiled into other circuits that are logical equivalent (i.e., they implement the same unitary evolution). This is necessary if a benchmark’s results are to reflect the performance a user can expect to achieve using a provider’s end-to-end stack. However, for benchmarks whose circuits are, by design, efficiently classically simulatable, arbitrary compilations cannot be permitted---because all such circuits can be compiled into simple quantum circuits (e.g., all processing can be off-loaded to a classical co-processor). Our current benchmarking suite does not directly address these challenges, and compilers designed to game these benchmarks could compress some of our benchmarking circuits into trivial circuits (or even offload all the computation to a classical co-processor). Addressing these challenges without forbidding all compilation is difficult, but judiciously placed ``compilation barriers'' within circuits is a promising solution (this idea is discussed in detail in Ref.~\cite{BlumeKohout2020volumetricframework}). In future versions of our benchmarking suite, we anticipate encoding explicit rules for what kinds of compilation are allowed for each benchmarking circuit---which will only be possible if quantum computing providers facilitate these flexible and user-specifiable compiler restrictions.  

One important role for benchmarking suites is testing and comparing different error mitigation or reduction strategies, e.g., dynamically decoupled gates or zero noise extrapolation. Some error mitigation strategies are not a scalable solution to reducing the errors in a quantum computation, e.g., one strategy for mitigating errors is to just replace experimental data with data obtained by simulating a perfect quantum compute. This is obviously absurd (it is infeasible for any useful computation), and it would be unreasonable for a quantum computing provider to employ this strategy when running a benchmarking suite. However, it is not clear how to forbid unreasonable error mitigation strategies while leaving room for innovation. Instead, we advocate clear and detailed explanations for all error mitigation strategies employed when running any benchmarking suite, including this one. 

\section{Benchmark Results and Analysis} 
\label{sec:benchmark_results}
In this section we present results from executing our application-oriented benchmarks on both a quantum simulator (Section~\ref{simulator_execution_results})
 and a selection of physical devices (Section~\ref{hardware_execution_results}).
All of the benchmarks in our suite have been implemented using the Qiskit API.
Some, but not all, have been implemented using the Cirq and Braket APIs (see discussion in Appendix~\ref{apdx:issues_apis}). 
Throughout, all executions of our benchmarking suite on a quantum simulator are performed using the Qiskit implementation of the benchmarking suite, as it provides rich circuit analysis features.
Execution on all of the hardware was also performed using the Qiskit version of our benchmarks, with the exception of the experiments on Rigetti Computing's hardware, which used the Braket version of our benchmarks.
 
\subsection{Benchmark Results on a Quantum Simulator}
\label{simulator_execution_results}

\begin{figure}[h!]
\includegraphics[width=\columnwidth]{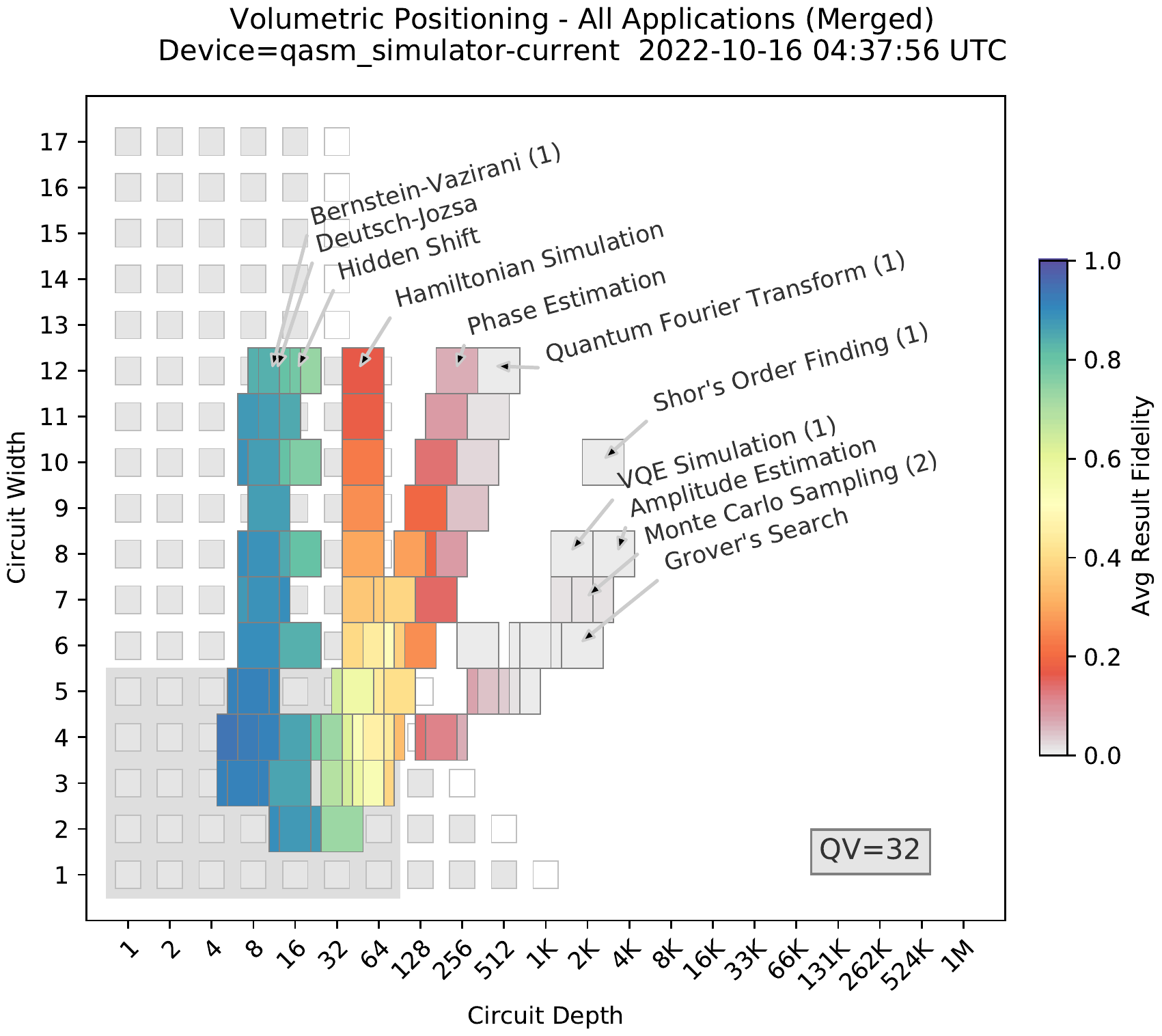}
\includegraphics[width=\columnwidth]{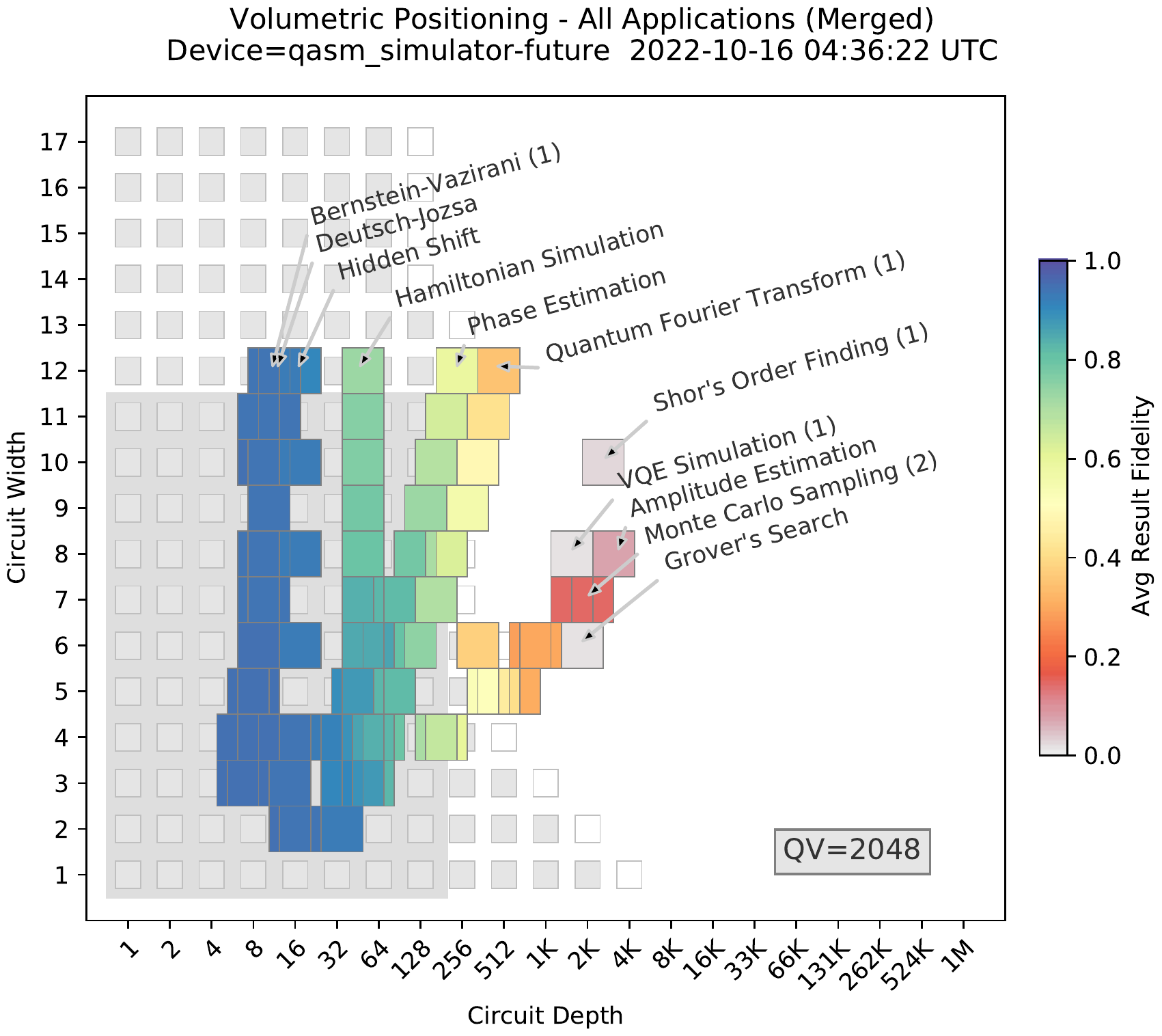}
\caption{\textbf{Demonstrating our benchmarks on a quantum simulator.} Results from running our suite of application-oriented benchmarks on a quantum simulator (colored squares) on top of a volumetric background (grey-scale squares) extrapolated from the quantum volume. The simulator uses all-to-all qubit connectivity and two different error rate scales (details in main text) that result in quantum volumes of 32 and 2048, as shown in each plot. Each volumetric background is extrapolated from the quantum volume using a simple heuristic (see Section~\protect\ref{sec:volumetric_positioning}) that is designed to predict the regions in which a circuit's result fidelity will be above $\nicefrac{1}{2}$.
}
\label{fig:benchmark_results_sim}
\end{figure}

Execution of the entire suite of benchmark programs results in a data file that stores all of the metrics collected for the tested device.
Plotting all of the accumulated average result fidelity data on a single volumetric background produces charts like the ones shown in Figure \ref{fig:benchmark_results_sim}, which were generated by running our benchmarks on the Qiskit Aer simulator. These simulations used all-to-all connectivity between the qubits, and a simple error model in which each $n$-qubit gate was subject to $n$-qubit depolarizing errors, for $n=1,2$. For the simulation that generated the upper plot of Figure \ref{fig:benchmark_results_sim}, the one- and two-qubit gate error rates were 0.003 and 0.03, respectively. These error rates are similar to those found in some current devices (although note that current devices suffer more complex, structured forms of error than pure depolarization). The quantum volume was determined to be 32, by executing the quantum volume protocol directly on this simulator, and this was used to generate the volumetric background (see Section~\ref{sec:volumetric_positioning}). For the lower plot of Figure \ref{fig:benchmark_results_sim}, the one- and two-qubit gate error rates were 0.0005 and 0.005, respectively, resulting in a quantum volume of 2048.

For both plots in Figure~\ref{fig:benchmark_results_sim}, there is close agreement between the average result fidelities of all the application benchmarks. There is also close agreement between the application benchmarks and the volumetric background. This is expected here due to the simplicity of the error model and the use of all-to-all connectivity. For more complex errors (e.g., coherent errors) the exact structure of a circuit can impact its performance, as errors can, e.g., coherently add or cancel \cite{proctor2020measuring}. This means that circuits of the same shape can have very different performance, potentially resulting in significant differences between a device's performance on different applications. Similarly, if there is limited qubit connectivity, after a circuit is transpiled to a device's primitive gates there can be an increase in the circuit depth that is large and strongly dependent on both the application and the device's connectivity. This can cause a sharper decrease in performance with increasing circuit width than anticipated by a volumetric background extrapolated from the quantum volume (using our simple heuristic), and this result fidelity decrease can be application dependent. We observe some of these effects in our experimental results.

\subsection{Benchmark Results on Quantum Hardware}
\label{hardware_execution_results}
In this section we present the results of executing our application-oriented benchmarks on various providers' hardware.
This section is not intended to perform or highlight comparisons between different systems, but rather to demonstrate our benchmarking suite.
Quantum computing hardware is changing rapidly, and we did not have access to the latest hardware from all providers.
Any comparisons would be quickly outdated, and therefore potentially misleading or erroneous.
Here we present a few representative samples of results and highlight several conclusions that we can draw from these results.

\vspace{0.3cm}
\begin{center}
\emph{Rigetti Computing}
\end{center}
\vspace{0.2cm}

One of the first quantum computers available to the public was provided by Rigetti Computing. Rigetti's devices are based on superconducting qubit technology, and they can be accessed using the Amazon Braket Service \cite{amazon_braket}. Here, we show results obtained from executing a subset of our benchmarks on Rigetti Aspen-9, which is a 32-qubit machine \cite{amazon_braket_rigetti}, using the Amazon Braket version of our benchmarks.

\begin{figure}[t!]
\includegraphics[width=\columnwidth]{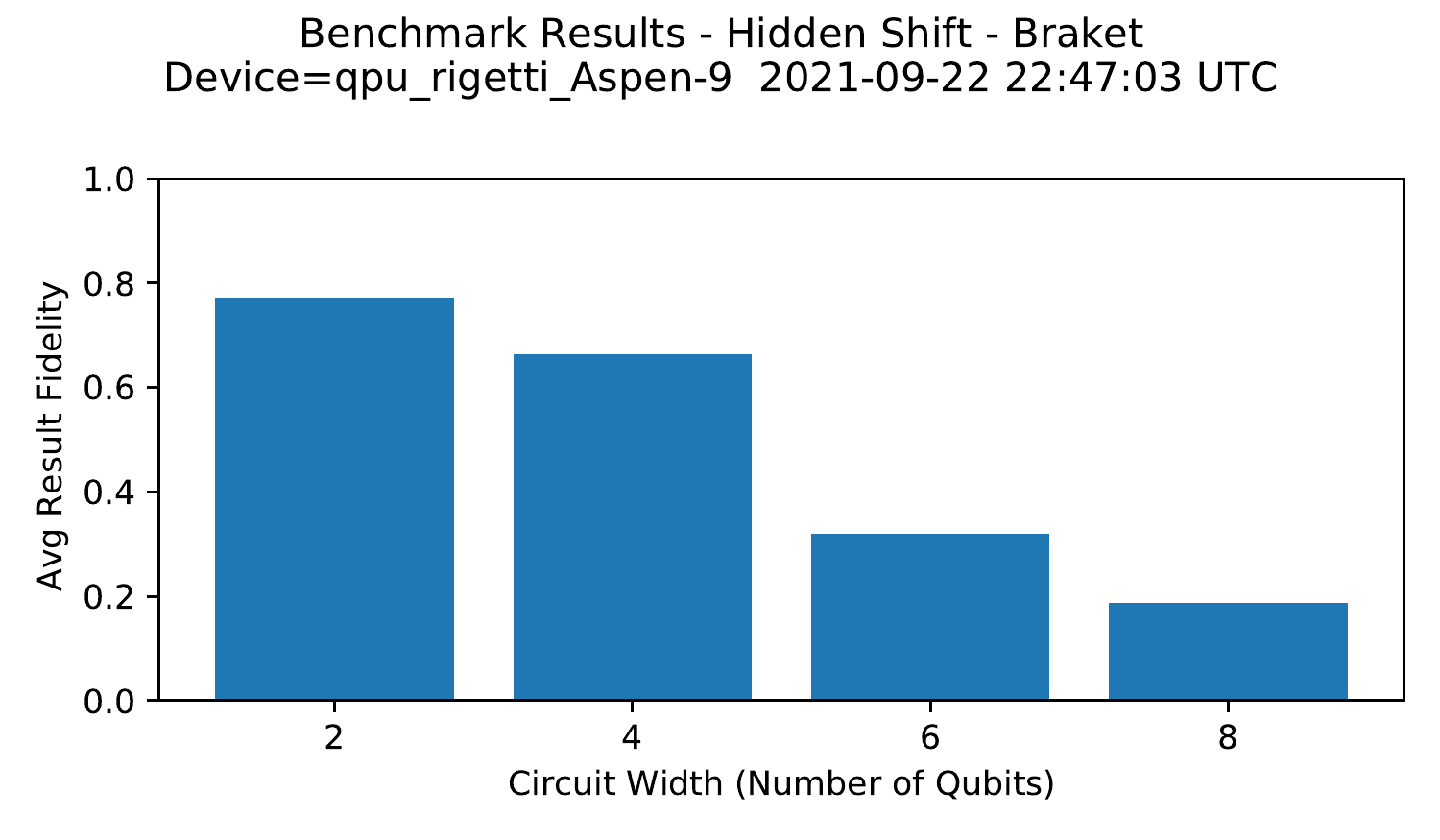}
\includegraphics[width=\columnwidth]{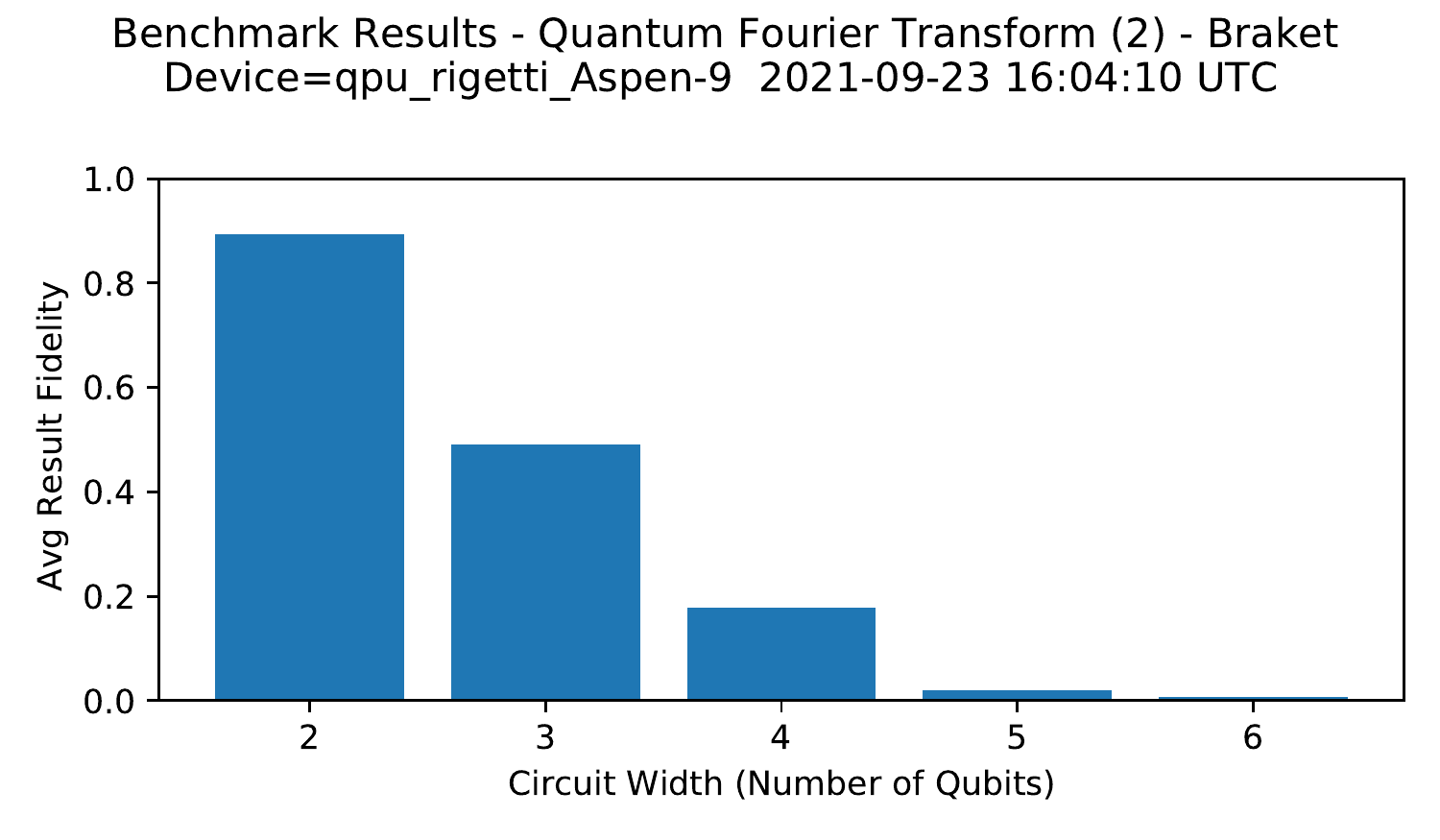}
\caption{\textbf{Demonstrating two benchmarks on Rigetti Aspen-9.} The result fidelity observed when running our  \benchmark{Hidden Shift} benchmark (upper plot) and  \benchmark{QFT(2)} benchmark (lower plot) on Rigetti Aspen-9. For each benchmark, we show the average result fidelity as a function of circuit width (the number of qubits). (\emph{Data collected via cloud service}.)}
\label{fig:rigetti_results}
\end{figure}

Figure \ref{fig:rigetti_results} shows the result fidelity as a function of the circuit width (the number of qubits) obtained when running the \benchmark{Hidden Shift} and \benchmark{QFT(2)} benchmarks on Rigetti Aspen-9. The  \benchmark{Hidden Shift} benchmark was executed on the first 8 qubits of Aspen-9. The  \benchmark{QFT(2)} benchmark was executed on only the first 6 qubits of Aspen-9. This was because, as shown in Figure \ref{fig:rigetti_results}, the result fidelity became negligible at around 5-6 qubits.

The result fidelity decreases as the circuit width increases, as expected.
There are numerous factors that contribute to the precise form of the result fidelity decay that we observe, including the qubit connectivity. This is because, in general, these benchmarks use controlled rotation gates between non-connected qubits, which must be implemented using multiple two-qubit gates between pairs of connected qubits.

This is a fundamental consequence of limited qubit connectivity, but note that none of our benchmarks are tailored to utilize the subset of qubits in a device that have the best connectivity for the given application.
Instead, the results represent what a user would experience when executing applications using the qubits $1$ through $n$, for an $n$-qubit circuit---although note that a provider's compiler \emph{can} employ intelligent qubit selection strategies (see discussion in Section~\ref{sec:impact_compiler_opts}). 

\vspace{0.7cm}       
\vspace{0.3cm}
\begin{center}
\emph{IBM Quantum}
\end{center}
\vspace{0.2cm}

IBM Quantum provides a large selection of quantum computing devices through IBM Quantum Services \cite{ibmq2021}. IBM Q's devices are based on superconducting qubit technology.
Below we present results from several of those machines.
Note that there are larger and more recent IBM Q systems, including systems that have larger quantum volumes than those that we tested.

\begin{figure}[t!]
\includegraphics[width=\columnwidth]{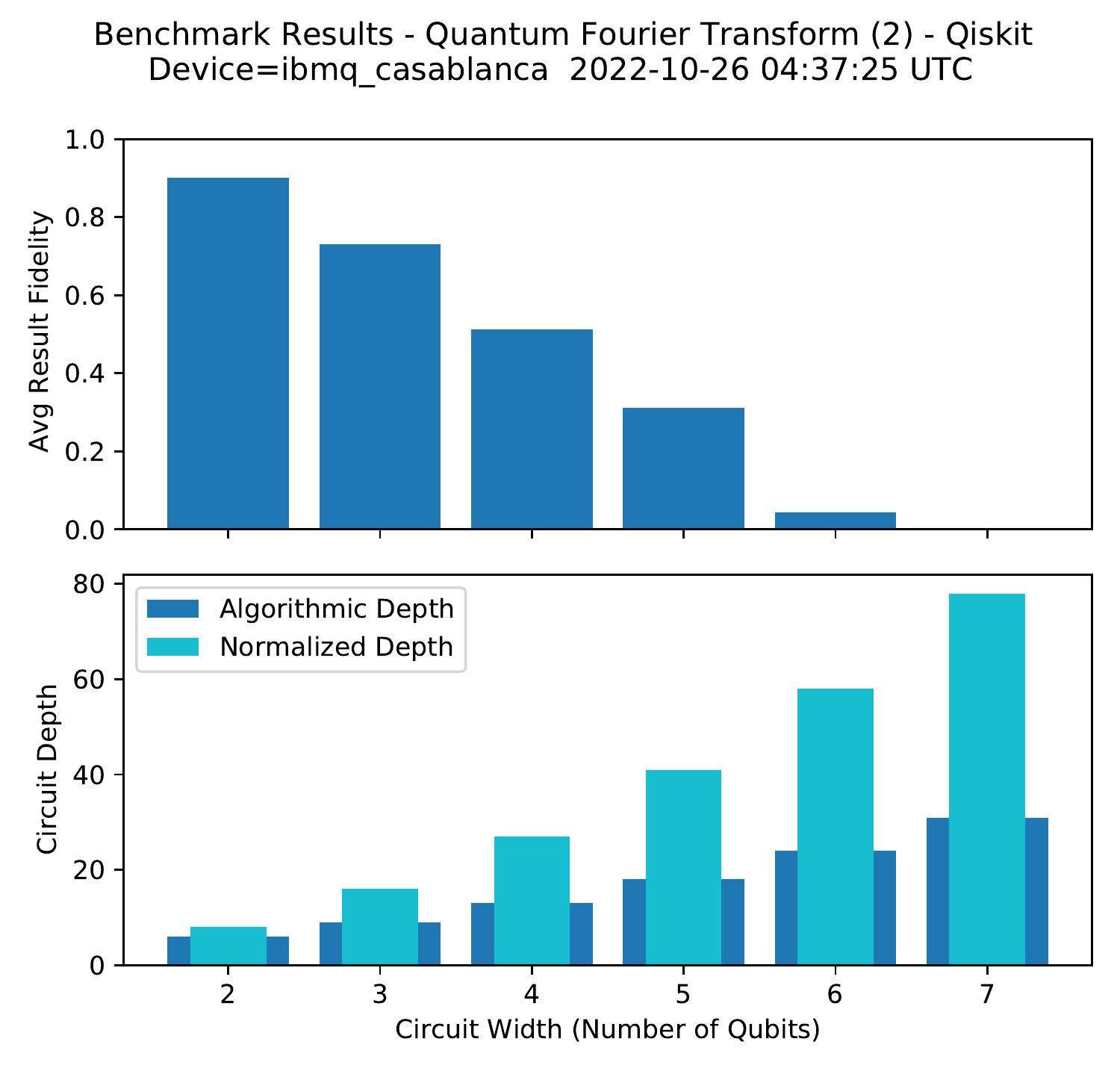}
\caption{\textbf{Demonstrating a QFT-based benchmark on IBM Q Casablanca.} Results from running the  \benchmark{QFT(2)} benchmark, the second of our two QFT-based benchmarks, on IBM Q Casablanca. The upper and lower plots show the average result fidelity and circuit depth, respectively, as a function of circuit width (the number of qubits). (\emph{Data collected via cloud service}.)}
\label{fig:ibmq_casablanca_results_1}
\end{figure}

Figure \ref{fig:ibmq_casablanca_results_1} shows results from running the  \benchmark{QFT(2)} benchmark on IBM Q Casablanca, which is a 7 qubit device. We show the average result fidelity and average circuit depth as a function of the circuit width (the number of qubits). The lower plot of the figure shows both the normalized circuit depth (which is what `circuit depth' refers to throughout this paper unless otherwise stated) and the depth of the circuits when written in the high-level API used to define the benchmark (see Section~\ref{sec:application_oriented_benchmarks}). Note that neither of these depths correspond to the depth of the circuits after they have been transpiled to the device's native operations (i.e., the circuit's physical depth), which will typically be larger.

\begin{figure}[t!]
\includegraphics[width=\columnwidth]{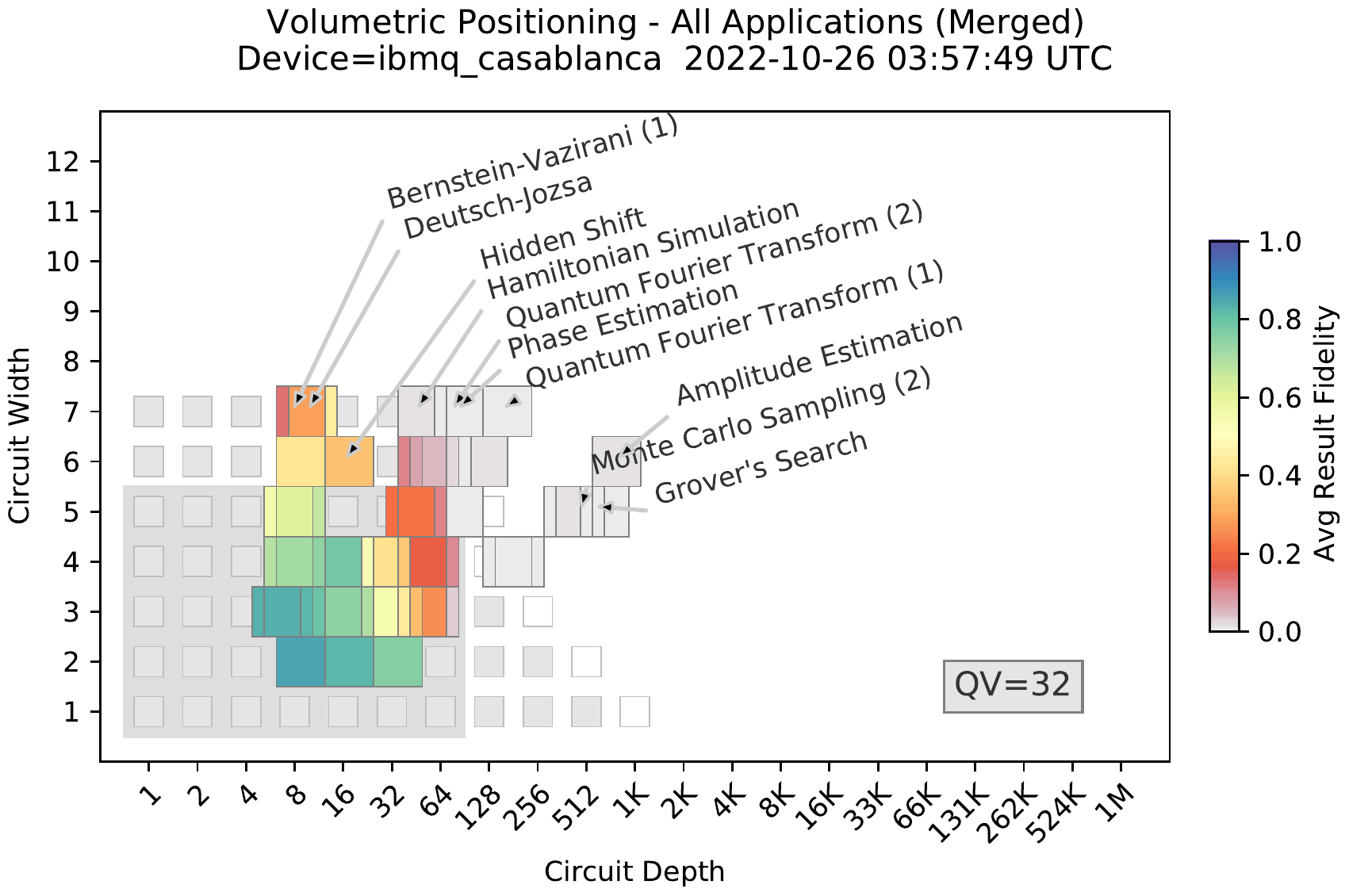}
\includegraphics[width=\columnwidth]{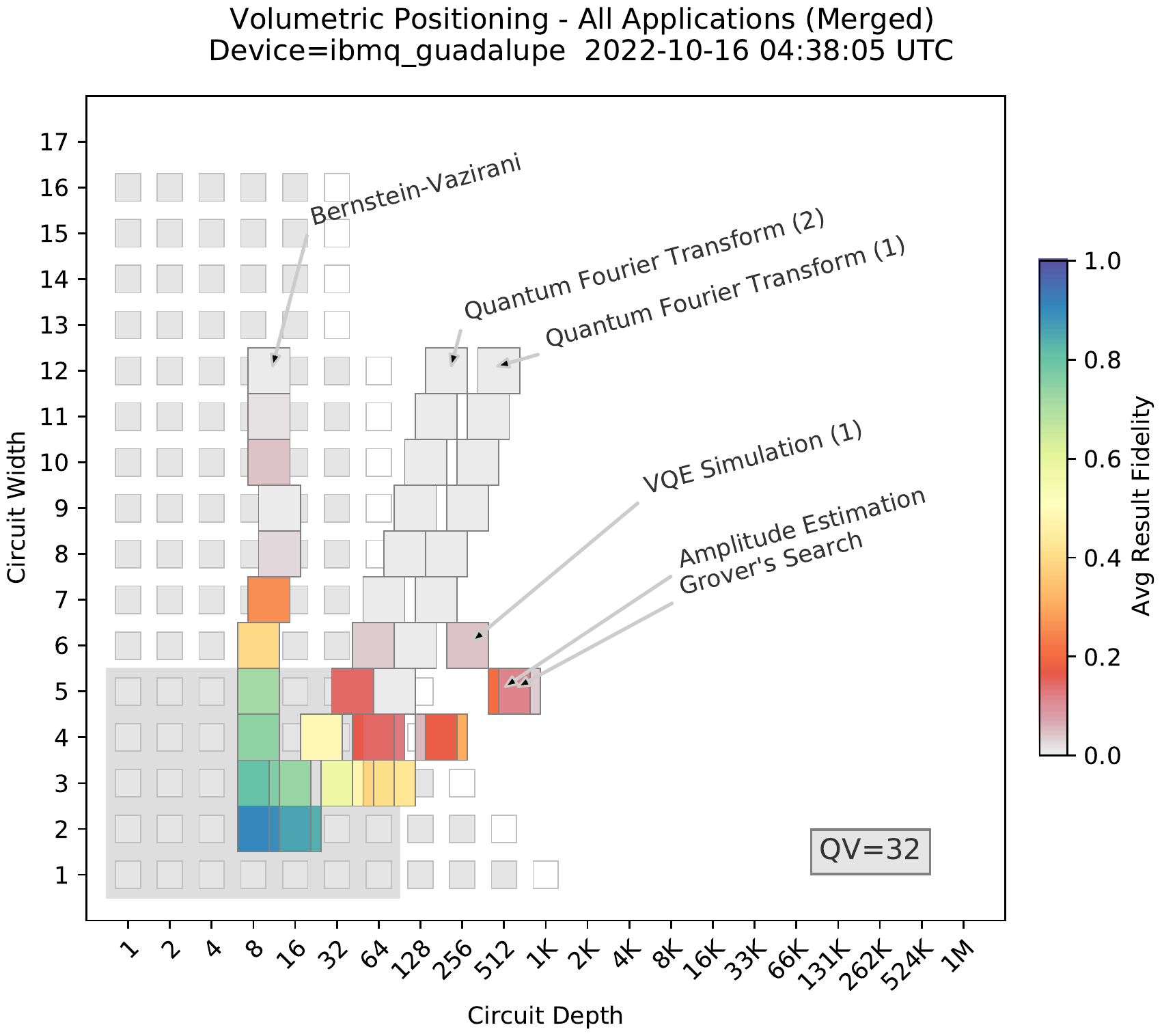}
\caption{\textbf{Benchmarking results on IBM Q Casablanca and Guadalupe.} The result fidelities obtained when running subsets of our benchmarking suite on IBM Q Casablanca and IBM Q Guadalupe. The volumetric backgrounds (grey squares) extrapolated from each device's quantum volume (32 in both cases) do not accurately predict the performance of wide circuits on IBM Q Guadalupe. (\emph{Data collected via cloud service}.)}
\label{fig:ibmq_guadalupe_results_all_subset}
\end{figure}

Figure \ref{fig:ibmq_guadalupe_results_all_subset} shows the results of running a subset of the benchmarking suite on IBM Q Casablanca and IBM Q Guadalupe. IBM Q Casablanca and IBM Q Guadalupe are 7 and 16 qubit devices, respectively, that both have a quantum volume of 32. Some of these benchmarks use circuits with a depth of well over 1000 (e.g., \benchmark{Grover's Search} benchmark), and we observed low result fidelities for all such circuits. All application circuits whose volumetric positions are in the region predicted to have low result fidelity by the volumetric background (white squares and blank region) do have low result fidelities. However, there are application circuits with low result fidelities (e.g., red squares) whose volumetric positions fall within the high result fidelity region of the volumetric background (grey squares). We observe that performance drops off with increasing width much more quickly than predicted by the volumetric background. This is likely primarily because this volumetric background is extrapolated from the quantum volume using a simple heuristic that does not account for a device's connectivity. This could potentially be addressed with a more sophisticated heuristic, or by instead using measured volumetric backgrounds (see Section~\ref{sec:comparison_to_generic_benchmarks}).

\begin{figure}[t!]
\includegraphics[width=\columnwidth]{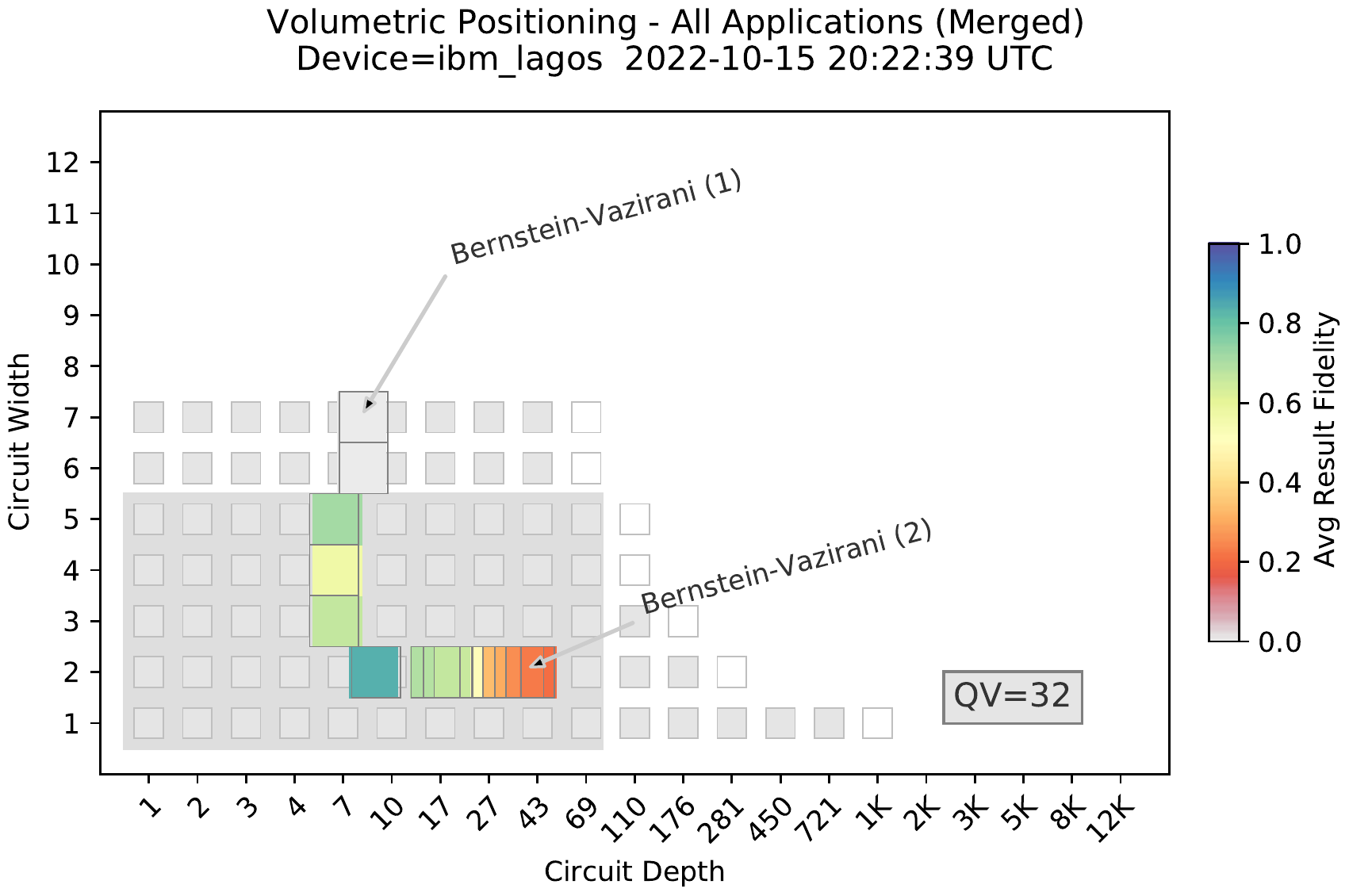}
\caption{\textbf{Benchmarks that use mid-circuit measurements.} Results from running the \benchmark{Bernstein-Vazirani(1)} and  \benchmark{Bernstein-Vazirani(2)} benchmarks on IBM Q Lagos. These benchmarks are based on two different implementations of the Bernstein-Vazirani algorithm: \benchmark{Bernstein-Vazirani(1)} uses $n+1$ qubit circuits for $n$-bit inputs, whereas \benchmark{Bernstein-Vazirani(2)} uses mid-circuit measurements to reuse one of the qubits, and therefore needs only 2 qubits for any size input. Both benchmarks use the same random 2-6 bit algorithm inputs. We observe higher result fidelity when using mid-circuit measurements to reduce the width of the circuits. (\emph{Data collected via cloud service}.)}
\label{fig:ibm_lagos_results_bv_bms}
\end{figure}

Mid-circuit measurement and qubit reset to make qubits reusable is a recent quantum computing enhancement, and IBM has introduced this feature for some of their systems. Our benchmark suite has been designed so that it can be adapted to include circuits that utilize mid-circuit measurements, as we now briefly demonstrate. As we review in Appendix \ref{apdx:mid_circuit_measurements}, 
the effect of this advance is that certain applications can be implemented with circuits over fewer qubits, as qubits can be reused. Figure \ref{fig:ibm_lagos_results_bv_bms} shows the results of executing two benchmarks, constructed from the Bernstein-Vazirani algorithm, on IBM Q Lagos, which is a 7 qubit device. In both versions of the benchmark the algorithm's input is an $n$ bit integer, and on IBM Q Lagos $n$ ranges from 2 to 6.
In the  \benchmark{Bernstein-Vazirani(1)} benchmark, an $n$ bit integer input uses an $n+1$ qubit circuit, so the circuit widths range from 3 to 7. In the  \benchmark{Bernstein-Vazirani(2)} benchmark, an $n$ bit integer input uses only 2 qubits for all $n$, one of which is repeatedly reset and reused with the aid of mid-circuit measurements. This results in width 2 circuits for all $n$.
In Figure \ref{fig:ibm_lagos_results_bv_bms} we observe that the result fidelity drops precipitously when the width of the circuit exceeds 5, which places it outside the quantum volume region (the grey rectangle). In the  \benchmark{Bernstein-Vazirani(2)} benchmark, the circuits' widths do not exceed 2, and we observe a larger result fidelity for the same problem instances.
This illustrates the potential value of mid-circuit measurements for increasing an algorithm's result fidelity. 

\vspace{0.3cm}
\begin{center}
\emph{Quantinuum}
\end{center}
\vspace{0.2cm}

Quantinuum System Model H1.1 was released commercially in 2020~\cite{honeywell_2020}. This quantum computing system currently encodes up to 12 qubits in trapped atomic ions using the quantum charged coupled architecture~\cite{pino_dreiling_figgatt_gaebler_moses_allman_baldwin_foss-feig_hayes_mayer_etal_2021}. The qubits are fully connected by using ion transport operations to rearrange the ions to interact any pair of qubits. The system has an average single-qubit gate fidelity of 99.99(1)\% and average two-qubit gate fidelity of 99.70(5)\%, as measured using standard randomized benchmarking~\cite{honeywell_2021}. The system is also capable of applying high-fidelity mid-circuit measurement and reset to any qubit~\cite{gaebler2021suppression}.

\begin{figure}[t!]
\includegraphics[width=\columnwidth]{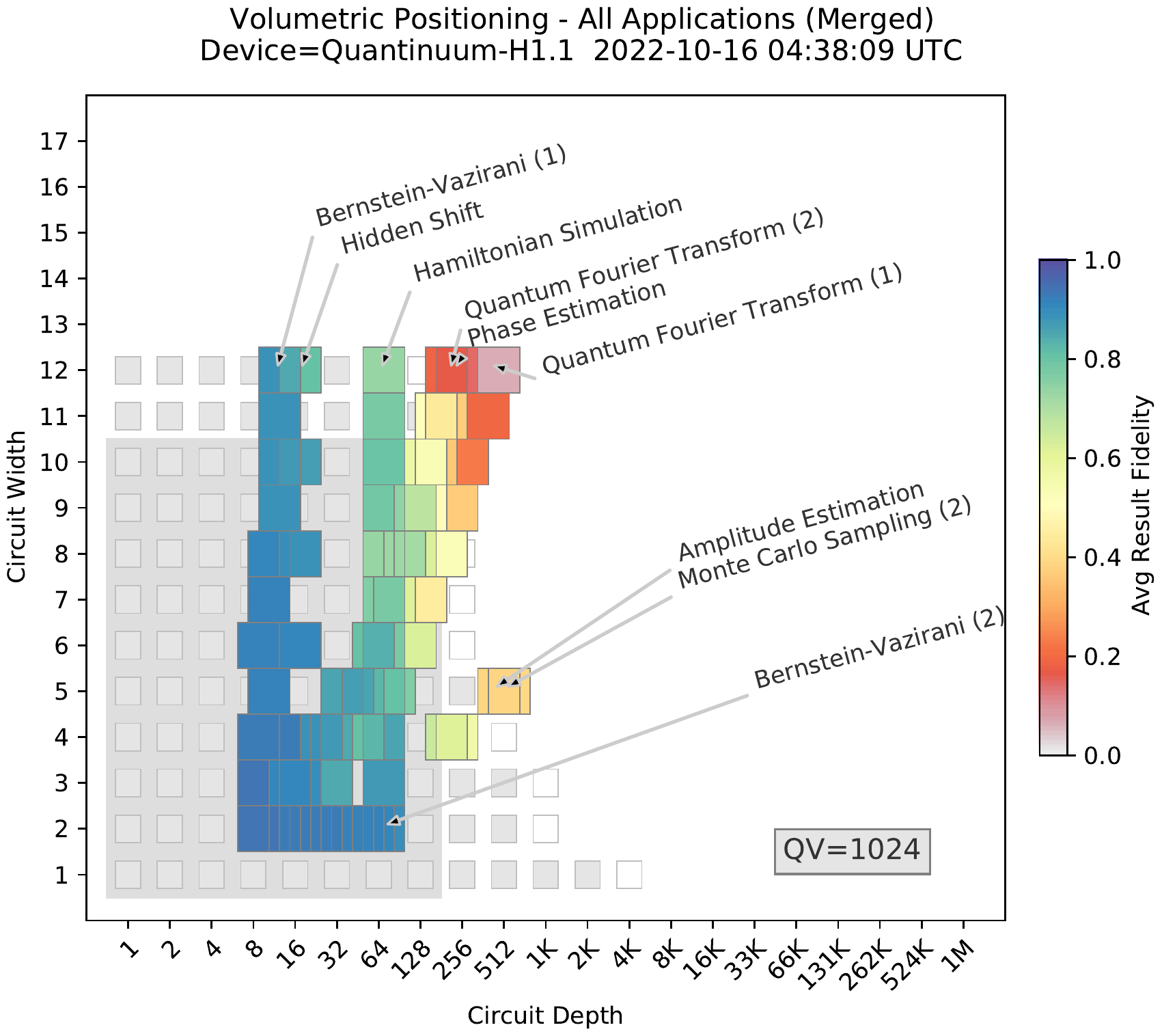}
\caption{\textbf{Benchmark results on Quantinuum's H1.1.} The result fidelities obtained when running our benchmarking suite on H1.1, which is a 12-qubit device. Each benchmark was run with 500 shots and at most five random circuits per qubit number. For all wide and shallow circuits the result fidelity is high. Quantinuum has measured the quantum volume of this device to be 1024~\protect\cite{honeywell_new_2021}, and this value is used to construct the volumetric background region. The volumetric background is broadly predictive of the performance of our algorithmic benchmarks. (\emph{Data collected by Quantinuum in-house}.)}
\label{fig:quantinuum_ALL-vplot-2}
\end{figure}

Figure \ref{fig:quantinuum_ALL-vplot-2} shows the results of executing a subset of our application-oriented benchmarks on the Quantinuum H1-1. For these tests the code was slightly edited to more easily run with the Quantinuum API but the qiskit version is also supported. These results show that the Quantinuum system can run 12-qubit, moderate-depth circuits that contain interactions between many pairs of qubits and still obtain moderate to high fidelity results. Furthermore, these results are broadly consistent with the component-level fidelities measured by one- and two-qubit randomized benchmarking.
The \benchmark{Bernstein-Vazirani(1)},  \benchmark{Hidden-Shift}, and  \benchmark{Hamiltonian Simulation} benchmarks all return high-fidelity results ($\bar{F} \geq 0.77$) for all qubit numbers up to the 12 qubit system size limit. Benchmarks that require deeper circuits, such as the  \benchmark{QFT(1)} and \benchmark{QFT(2)} benchmarks and the \benchmark{Phase Estimation} benchmark, still produce results that are distinguishable from random outputs. The smallest fidelity for the \benchmark{QFT (1)} was $\bar{F} = 0.07$, which was obtained for the case of $n=12$. Overall, we observe that the volumetric background extrapolated from a quantum volume of 1024  \cite{honeywell_new_2021} provides a broadly, but not entirely, accurate prediction of the performance of our algorithmic benchmarks.

The H1.1 can implement mid-circuit measurement and reset operations, which we benchmarked using the \benchmark{Bernstein-Vazirani(2)} benchmark. We see that while both our benchmarks based on the Bernstein-Vazirani algorithm achieve high fidelity results ($\bar{F} >0.929$), \benchmark{Bernstein-Vazirani(2)} performs slightly better than \benchmark{Bernstein-Vazirani(1)} for larger register size. The \benchmark{Bernstein-Vazirani(2)} benchmark can also be implemented for larger algorithm input sizes ($n>12$) as it only uses two qubits for any input, although we did not do this in our experiments, whereas the \benchmark{Bernstein-Vazirani(1)} benchmark is limited to $n\leq 12$ due to the machine's current restriction to 12 qubits.

\vspace{0.3cm}
\begin{center}
\emph{IonQ}
\end{center}
\vspace{0.2cm}

IonQ is a provider of quantum computers that are available via Amazon Braket \cite{amazon_braket_ionq}, Microsoft Azure Quantum \cite{microsoft_qsharp} and Google Cloud \cite{google_cloud_ionq}. IonQ systems support most leading quantum programming environments, and we executed our benchmarks on the IonQ system using the Qiskit provider module supplied by IonQ. Their hardware utilizes trapped ytterbium ions where qubits are encoded in two states of the ground hyperfine manifold. The IonQ system currently on the cloud consists of 11 qubits that are fully connected.

\begin{figure}[t!]
\includegraphics[width=\columnwidth]{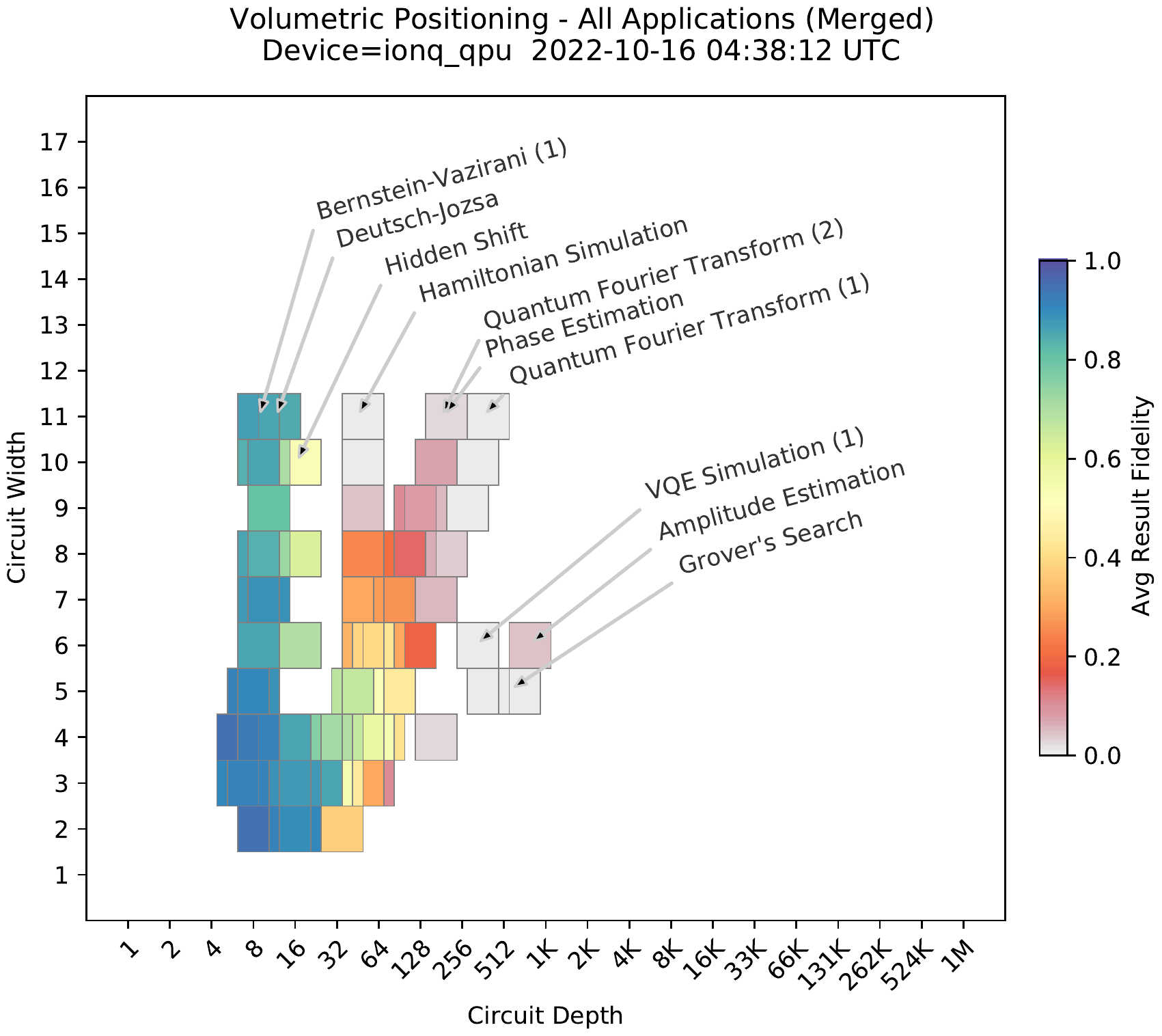}
\caption{\textbf{Benchmark results on IonQ's cloud-accessible system.} The result fidelities obtained when running our benchmarking suite on IonQ's Harmony cloud-accessible system, which features 11 qubits. For all wide and shallow circuits the result fidelity is high. IonQ does not publish a quantum volume for this device, and we did not measure it directly, so we do not include a volumetric background in this plot. (\emph{Data collected via cloud service}.)}
\label{fig:ionq_qpu_ALL-vplot-2}
\end{figure}

\begin{figure}[t!]
\includegraphics[width=\columnwidth]{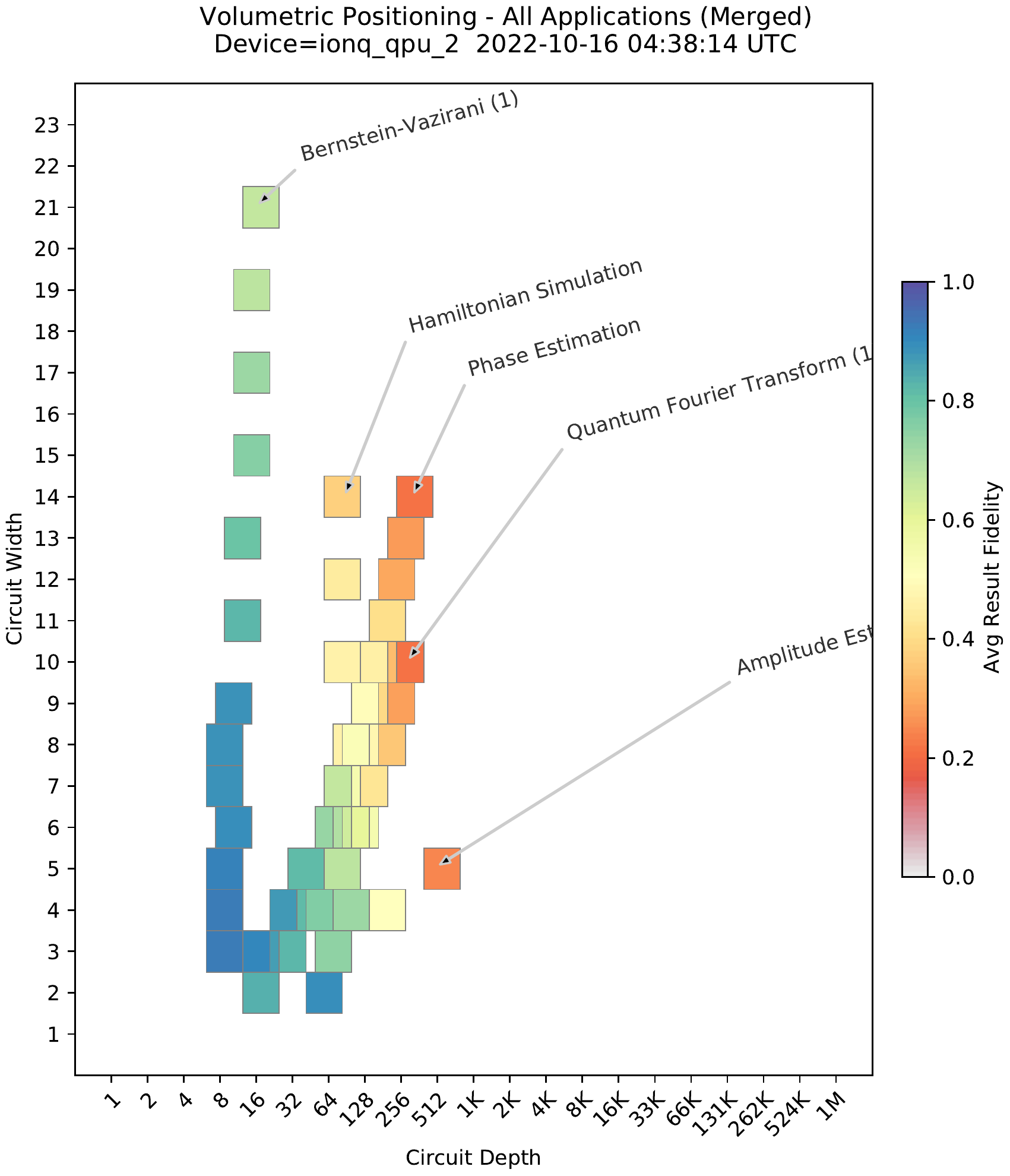}
\caption{\textbf{Benchmark results on IonQ's Aria system.} The result fidelities obtained when running our benchmarking suite on up to 21 qubits of IonQ's Aria system. We observe a significant performance improvement in comparison to the previous generation of IonQ hardware. IonQ does not publish a quantum volume for this device, and we did not measure it directly, so we do not include a volumetric background in this plot. (\emph{Data collected by IonQ in-house}.)}
\label{fig:ionq_qpu2_ALL-vplot-new}
\end{figure}

Figure \ref{fig:ionq_qpu_ALL-vplot-2} shows results from executing a subset of the application-oriented benchmarks on the IonQ system. IonQ does not publish a quantum volume metric for this device, and we chose not to run the quantum volume circuits on any of the tested hardware. In Figure \ref{fig:ionq_qpu_ALL-vplot-2} we observe high-fidelity results for the \benchmark{Bernstein-Vazirani(1)}, \benchmark{Deutsch-Jozsa} and \benchmark{Hidden Shift} benchmarks for all qubit numbers, up to the current system limit of $n=11$. For benchmarks that use deeper circuits we observe a steeper decrease in result fidelity with $n$, and the observed performance is broadly, but not entirely, consistent with the predictions of the volumetric background.

IonQ has recently introduced its next-generation quantum hardware \cite{zhu2021generative}. This hardware is purported to feature an order of magnitude reduction in gate error rates in comparison to IonQ's current system on the cloud. Figure \ref{fig:ionq_qpu2_ALL-vplot-new} shows the results from running a subset of our benchmarking suite on this device, including benchmarks from all 3 categories in Table \ref{tab:benchmark_suite}. We observe a significant increase in performance in comparison to the previous generation hardware, as is apparent by comparing Figs.~\ref{fig:ionq_qpu_ALL-vplot-2} and \ref{fig:ionq_qpu2_ALL-vplot-new}. For instance, the \benchmark{Bernstein-Vazirani(1)} benchmark is run for up to 21 qubits, and the result fidelity for the largest circuit is 70\%. This can be compared with the result fidelity for the largest circuit for the \benchmark{Bernstein Vazirani(1)} benchmark on the cloud quantum computer, which is 78\% for 11 qubits. Note that the maximum width and depth of the \benchmark{Hamiltonian Simulation} and \benchmark{Phase Estimation} benchmark circuits here exceeds that in all other IonQ experiments, as here we tested up to 14 qubits.

\vspace{0.3cm}
\begin{center}
\emph{Compatible Hardware}
\end{center}
\vspace{0.2cm}

Our application-oriented benchmarking suite is designed to be applicable to any gate-based quantum computer, and our implementation of that suite is applicable to any platform that can be interfaced with Qiskit, Cirq, Braket, or Q\#. Table~\ref{tab:compatible} contains a non-exhaustive list of hardware that that we did not test in this work but that is compatible with our implementation of our benchmarking suite. Note that support for different quantum hardware devices by these programming environments will evolve over time.

\begin{center}
\begin{table}[h]
\begin{tabular}{ |c|c|c| } 
 \hline
 Hardware     & Programming Environment \\
 \hline
 Google QCS     & Cirq \\
 \hline
 Alpine Quantum          & Qiskit \\
 Technologies            & Cirq \\
 \hline
\end{tabular}
\caption{\label{tab:compatible}The Application-Oriented Benchmark Suite introduced in this work can be executed on other quantum computing hardware, using the programming environments specified.}
\end{table}
\end{center}

\subsection{Comparison to Generic Benchmarks}
\label{sec:comparison_to_generic_benchmarks}
In this work we have compared results from our application-oriented benchmarks with a volumetric background. As introduced in Section~\ref{sec:volumetric_positioning}, a volumetric background is a prediction for the result fidelity of circuits, as a function of circuit shape (i.e., volumetric position), that is constructed from the results of some independent benchmark or performance data. So far in this paper, we have considered only volumetric backgrounds that are extrapolated from a single performance metric: the quantum volume.

Our simulated and experimental results show that this extrapolated volumetric background can provide reasonably accurate predictions of application performance in some instances (particularly with all-to-all connectivity devices) but not always. There are many reasons why we should not expect our volumetric background extrapolated from the quantum volume to always accurately predict performance, including: (1) the extrapolation uses a very simple heuristic that does not account for device connectivity (see Section~\ref{sec:volumetric_positioning}), and (2) we often extrapolate the quantum volume to predict the performance of circuits containing significantly more qubits than were used in any of the quantum volume circuits.

\begin{figure}[t!]
\includegraphics[width=\columnwidth]{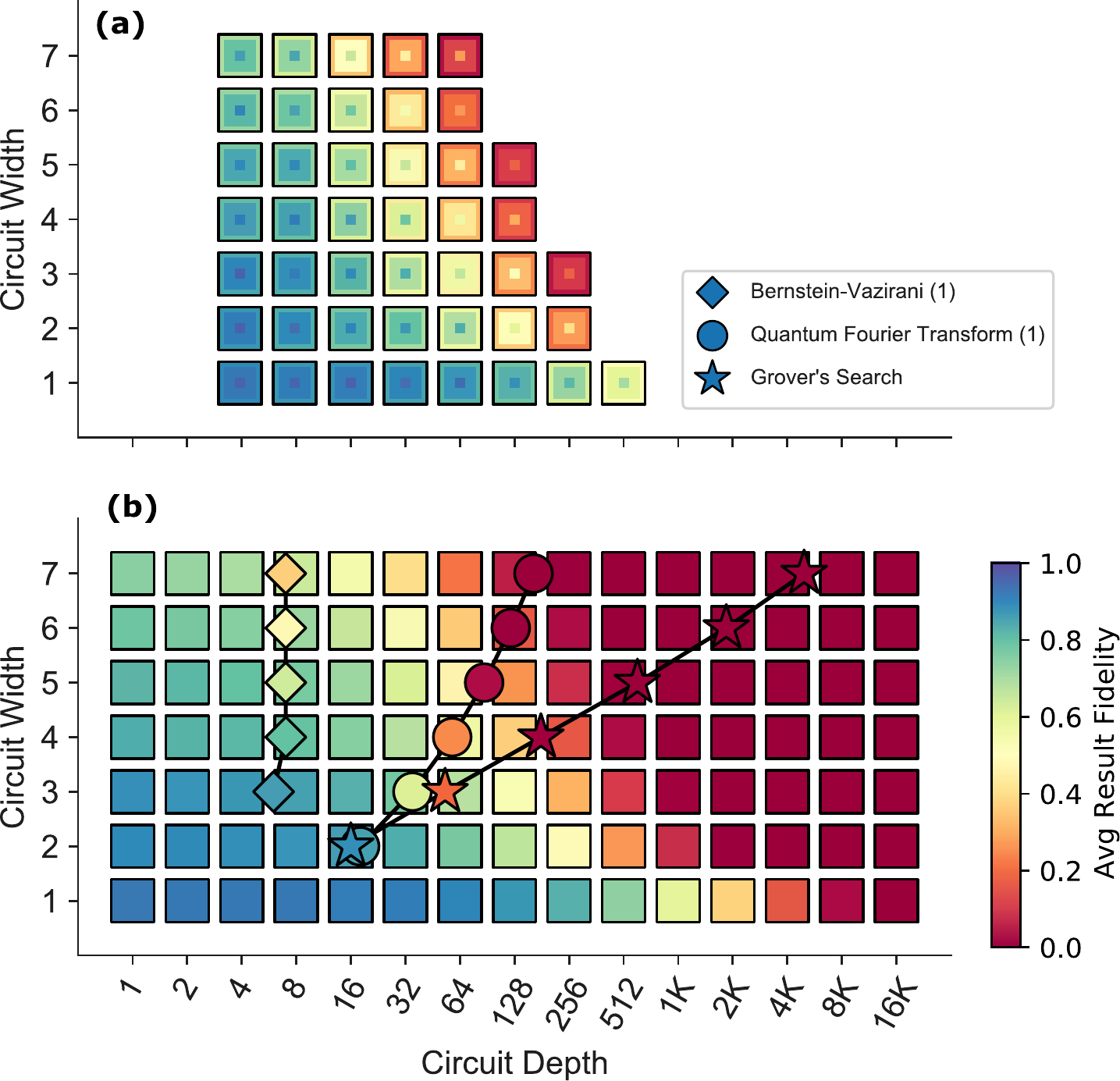}
\caption{\textbf{Comparing mirror circuit and application-oriented benchmarks.} \textbf{(a)} Results from running randomized mirror circuits \protect\cite{proctor2020measuring} on IBM Q Casablanca. At each circuit shape 40 circuits were sampled and run. The maximum, average, and minimum result fidelity that was observed at each circuit shape is shown, in concentric squares. \textbf{(b)} The average result fidelity of the mirror circuits is extrapolated to arbitrary depths to create a volumetric background (squares), and compared to the results from running a subset of the application benchmarks on IBM Q Casablanca (diamonds, circles and stars). We observe good correlation between the different benchmarks at low width, but an increasing discrepancy between the volumetric background and the applications as the circuit width increases. This is likely primarily due to the different frequency with which gates between non-connected qubits are used in the four different benchmarks. Mirror circuit benchmarks that mimic the frequency with which a particular algorithms uses gates between non-connected qubits might reduce this discrepancy, which would result in a randomized benchmark that is no longer algorithm agnostic, but this is not explored herein.}
\label{fig:all_vplots_casablanca}
\end{figure}

Predictions of the performance of a suite of algorithmic benchmarks from a single number, like the quantum volume, are unlikely to be consistently accurate---due to the complexities of quantum computer's errors and their interactions with other characteristics of a processor (e.g., connectivity). Furthermore, it is difficult for any generic benchmark to reliably predict the performance of a diverse suite of algorithms. To understand why, below we briefly explore one alternative to volumetric backgrounds based on the quantum volume: a volumetric background created by running `randomized mirror circuits' \cite{proctor2020measuring}. For this illustrative comparison, we focus on a single device, IBM Q Casablanca, and the particular class of randomized mirror circuits used in Ref.~\cite{proctor2020measuring}. Figure \ref{fig:all_vplots_casablanca} (a) shows the result fidelities obtained when running randomized mirror circuits on IBM Q Casablanca.

We use the data in Figure \ref{fig:all_vplots_casablanca} (a) to construct a volumetric background by (1) using the average result fidelity for randomized mirror circuits of a given shape as our predictor for the performance of other circuits of this shape, and (2) extrapolating this to all circuit depths using the theory for randomized mirror circuits presented in \cite{proctor2021scalable}. This volumetric background is shown in Figure \ref{fig:all_vplots_casablanca}, alongside the average result fidelities of three of our application benchmarks: the \benchmark{Bernstein-Vazirani(1)} benchmark, the \benchmark{QFT(1)} benchmark, and the \benchmark{Grover's Search} benchmark. To facilitate a more precise comparison to the results of the application-oriented benchmarks, here we (1) do not average together the results for application-oriented benchmarks with similar volumetric positions, and (2) we use a color-scale volumetric background. The general trend of all three benchmarks and the volumetric background are unsurprisingly in agreement: result fidelity decreases with increasing circuit width and depth. However, there are substantial differences between the rates at which the performance of each type of circuit drops off with increasing circuit size.

There are multiple possible reasons for the observed discrepancy between the result fidelity of randomized mirror circuits and algorithmic benchmarks. Perhaps the most important is that each of the four circuit families (the three algorithms and the randomized mirror circuits) requires a different number of physical two-qubit gates at a fixed normalized circuit depth. When transpiled into the standardized gate set with all-to-all connectivity (which is used to define normalized circuit depth) each circuit family will use a different average number of two-qubit gates at fixed depth $d$. Furthermore, the rate that each pair of qubits are coupled in these circuits will vary, and so each circuit family will incur a different CNOT overhead when it is transpiled to the physical gate set of IBM Q Casablanca (which has limited connectivity). The randomized mirror circuits incur no CNOT overhead at this transpilation stage, as they were designed to contain only CNOT gates between connected qubits. This is a compelling explanation for why the predictions from the randomized mirror circuits are systematically over-optimistic. The physical CNOT count discrepancy will typically increase with the number of qubits (i.e., circuit width), which explains the faster result fidelity decrease with circuit width that we observe for the application benchmarks versus the randomized mirror circuit. It is possible to design randomized mirror circuits that use the same number of physical CNOT gates as any given application benchmark (on average and at a given depth $d$), but this physical CNOT density varies between algorithmic benchmarks---so any randomized benchmark that is designed to match this density is no longer algorithm agnostic. 

The discrepancies between the results of our application-oriented benchmarks and both the volumetric backgrounds inferred from randomized mirror circuit data and from a device's quantum volume are indicative of the challenges associated with predicting algorithm performance from generic benchmarks. One way to potentially remove the discrepancy we observe in Figure \ref{fig:all_vplots_casablanca} is to use physical depth to define a circuit's volumetric position. This would mean that an application's volumetric `footprint' will be device-dependent (something we choose to avoid in this work). We leave an investigation of this to future work.

\subsection{Impact of Compiler Optimization Techniques}
\label{sec:impact_compiler_opts}
In our experiments on hardware, the benchmarks were intentionally run using the provider's default settings for the transpilation and execution of the circuits. This is how many users work with a programming system, even a quantum one. Many of quantum computing programming packages offer compilation options that will modify the circuits before execution to account for device characteristics, such as qubit connectivity or timing parameters. While using these parameters can result in better performance, it is often difficult for users to know which parameters should be used or what values to choose for each target device. Providers also may elect to deliberately exclude automatic optimization techniques to avoid interfering with users' custom optimizations. The result can be that users fail to obtain the best possible performance from a vendor's hardware by default.

\begin{figure}[t!]
\includegraphics[width=\columnwidth]{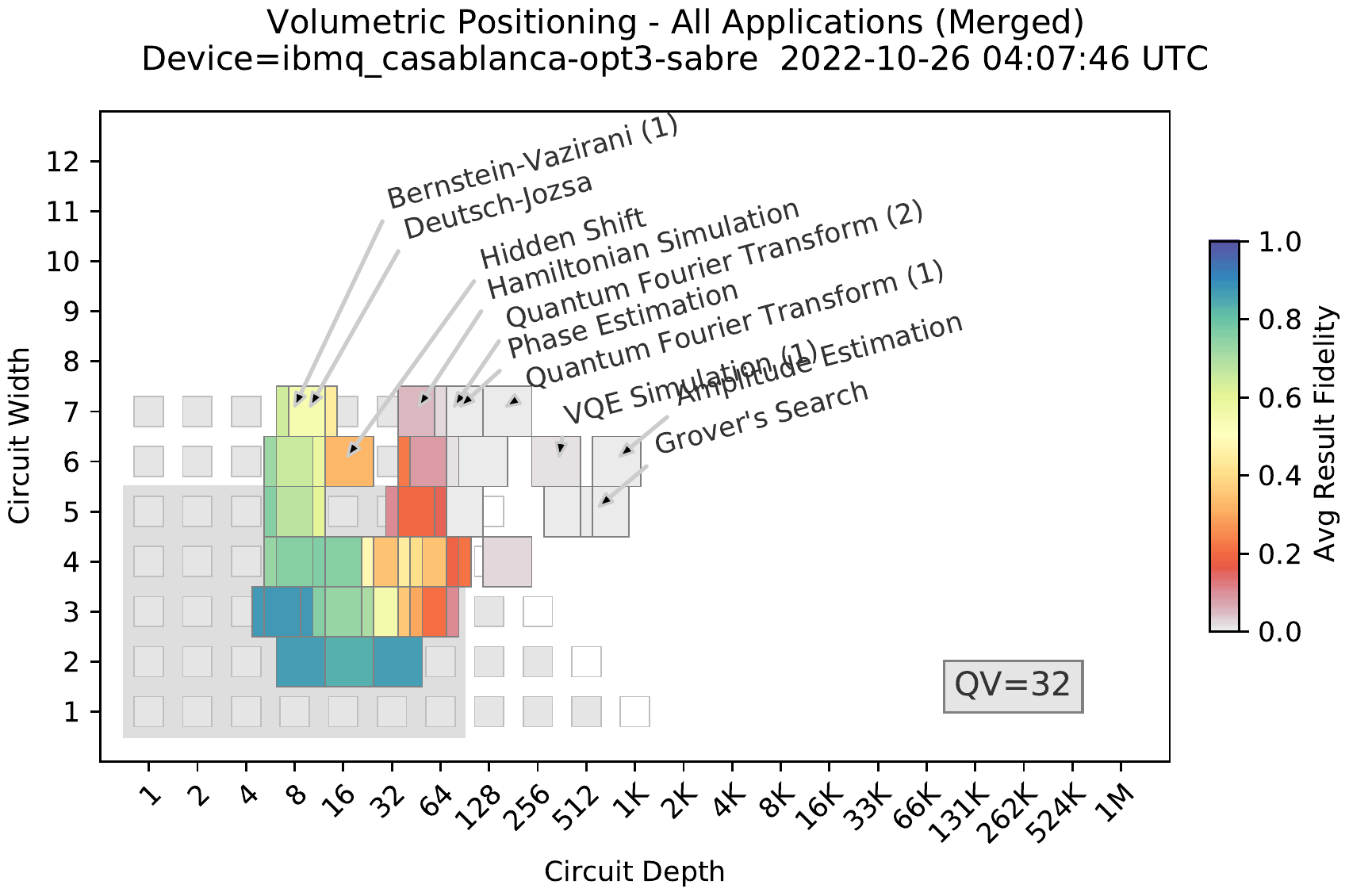}
\includegraphics[width=\columnwidth]{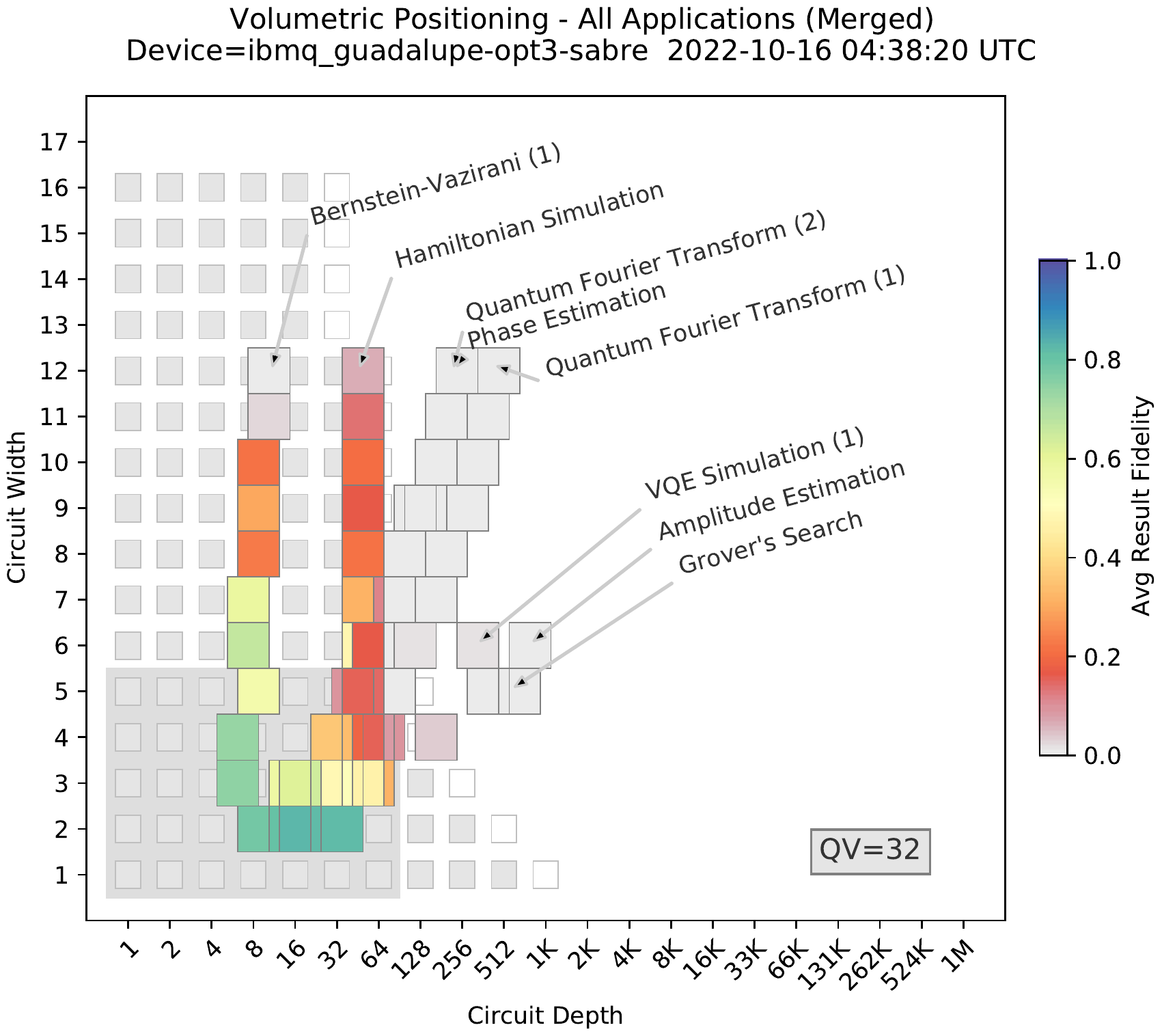}
\caption{\textbf{Benchmarking results on IBM Q Casablanca and Guadalupe with compiler optimizations.} The result fidelities obtained when running a subset of our benchmarking suite on the IBM Q Casablanca and Guadalupe systems using the compiler optimization strategy defined in the text. There is an improvement in result fidelity for wide circuits, as compared with the results shown in Figure \ref{fig:ibmq_guadalupe_results_all_subset}. (\emph{Data collected via cloud service}.)}
\label{fig:ibmq_guadalupe_results_opt3_sabre}
\end{figure}

Figure~\ref{fig:ibmq_guadalupe_results_opt3_sabre} illustrates the impact of a simple compiler optimization and layout strategy as applied to execution on the IBM Q Guadalupe device. Note that the fidelity obtained for circuits that are wide and shallow is significantly improved using this strategy because it optimizes the layout of the circuit for the qubit connectivity of the device. In this case, the Qiskit execution method was configured to use $optimization\_level=3$, which provides the highest circuit optimization but at the expense of longer execution time for the classical circuit optimizer \cite{qiskit_org}. Additionally, the execution was configured to use $layout\_method=`sabre' $ and $routing\_method=`sabre'$, which apply information about the connectivity of the target device to layout the circuit and route interactions between qubits optimally.

\begin{figure}[t!]
\includegraphics[width=\columnwidth]{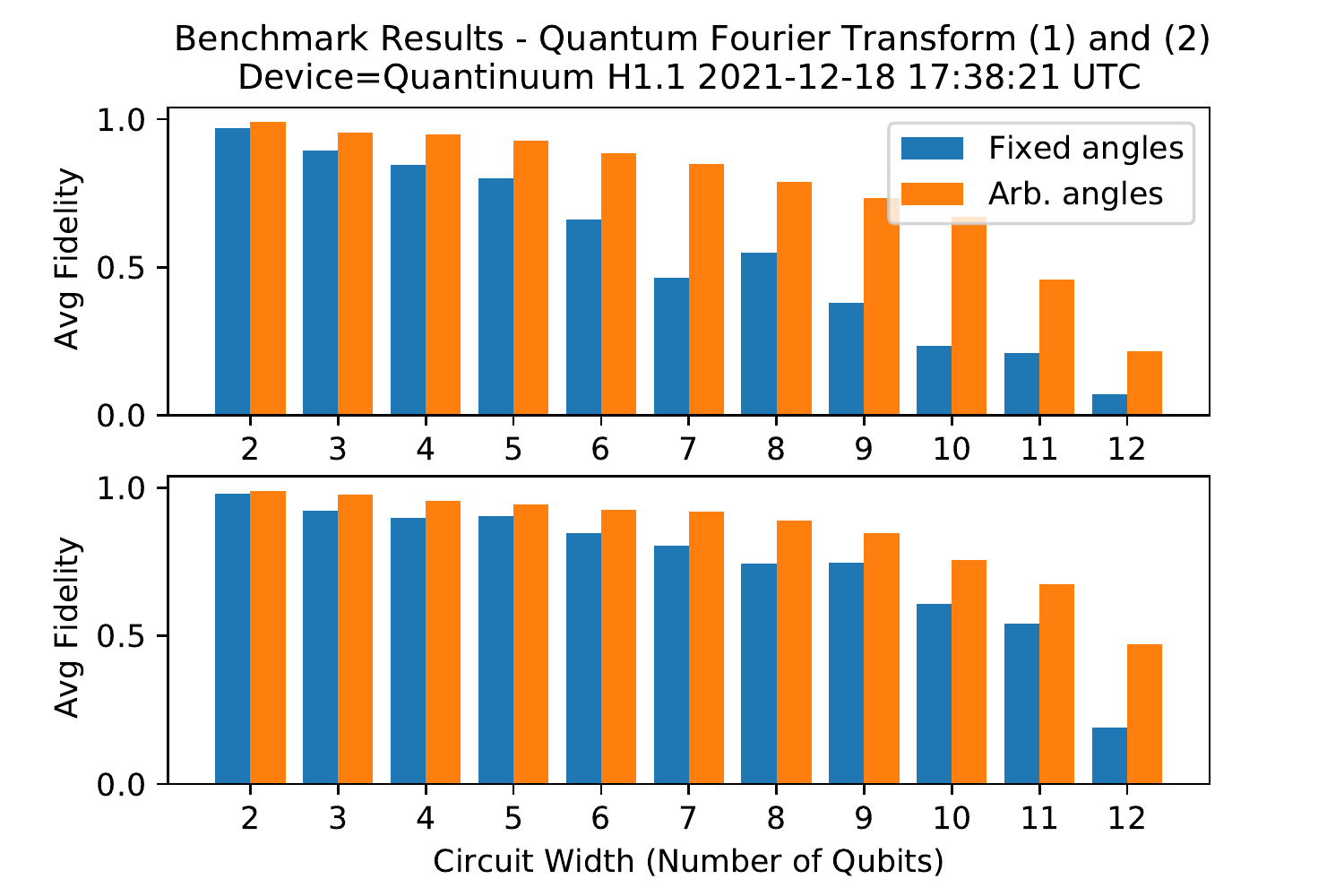}
\caption{\textbf{Benchmarking results on Quantinuum's H1.1 with compiler optimizations.} The result fidelities obtained when running a subset of our benchmarking suite on H1.1 using arbitrary angles. For both QFT-based benchmarks the number of two-qubit interactions is roughly cut in half and the errors are further reduced by decreasing the angle of two-qubit rotations. (\emph{Data collected by Quantinuum in-house}.)}
\label{fig:quantinuum_2_opt}
\end{figure}

On the H1.1 system we experimented with using arbitrary angle two-qubit gates to implement the \benchmark{QFT(1)} and \benchmark{QFT(2)} benchmarks. As shown in Fig.~\ref{fig:quantinuum_2_opt}, these circuit optimizations increased the results fidelity. In trapped-ion systems, two-qubit gates are typically implemented with the M{\o}lmer-S{\o}renson interaction. In most cases, the resulting gate is produced with a fixed laser power that ideally results in the unitary $\exp[-i(\pi/4) XX]$, which is equivalent to CNOT up to local unitaries. A controlled-$Z(\theta)$ operation, which appears in many of the application oriented benchmarks, requires two CNOTs. The M{\o}lmer-S{\o}renson gate can also be implemented with an arbitrary angle $\exp[-i(\phi/2) XX]$ by varying the laser power as a function of $\phi$. The resulting gate is equivalent to a controlled-$Z(\theta)$ gate up to local unitaries. This modification halves the number of two-qubit interactions required to implement controlled-$Z(\theta)$ (for most $\theta$) and may further decrease two-qubit errors that scale with $\phi$. From the \benchmark{QFT(1)} and \benchmark{QFT(2)} benchmarks plotted in Fig.~\ref{fig:quantinuum_2_opt}, we can verify that the arbitrary angle gates are yielding a significant improvement. For example, for \benchmark{QFT(1)} without arbitrary angle gates the fidelity for $n=12$ qubits was 0.07 but with the arbitrary angle gates it had a 3$\times$ improvement to 0.216.

\section{Measuring Execution Time}
\label{sec:execution_time}
Most existing quantum computing benchmarks focus on measuring result quality, but the time taken to run an application is also an important metric of performance \cite{johnson_faro_2021, Bertels_2020, M_ller_2017, cao_hirzel_2020}. In particular, the time to solution is a crucial metric when comparing quantum and classical computations, e.g., comparing quantum and classical compute times was central to recent demonstrations of quantum advantage \cite{Arute2019-mk, Zhong2020-rk}. Our benchmarking suite is designed to quantify various aspects of the time to solution, as we explain below. In Section~\ref{sec:execution_pipeline} we discuss how to measure and subdivide the total execution time. In Section~\ref{sec:quantum_execution_time} we then turn to the component of execution time that our benchmarking suite currently focuses on: the time taken to execute the quantum circuits.

\subsection{Total Execution Time}
\label{sec:execution_pipeline}
The total time required to execute a complete quantum application is evidently a critical aspect of a quantum computer's performance. In order to understand a quantum computer's performance in detail, the total execution time should ideally be divided into the times taken for distinct sub-tasks (e.g., circuit compilation), and an ideal benchmarking suite would measure each such time. In general, a quantum application is a hybrid quantum-classical computation consisting of running one or more quantum circuits interleaved with classical processing. We therefore suggest that the following non-exhaustive set of sub-tasks and associated compute times will be important to measure:
\begin{enumerate}
    \item Compilation time ($t_{\rm compile}$)---the time spent compiling an application's circuits, written in a high-level language (e.g., Q\# or Qiskit), into the hardware's native operations and language (e.g., OpenQASM, or perhaps the pulse-level description of the circuit).
    \item Classical computation time ($t_{\rm classical}$)---the time spent implementing classical computations to, e.g., process data from quantum circuits already run (as in variational algorithms  \cite{Peruzzo2014, McClean_2016, farhi2014quantum}).
    \item Quantum execution time ($t_{\rm quantum}(N)$)---the time required for $N$ shots of a circuit. This is the only time that we report on in this work.
\end{enumerate}
An additional component in the total execution time when using shared-usage cloud-access quantum hardware is the queuing time, i.e., the time that circuits spend in a queue waiting to be executed. This is an artifact of shared-usage access models, and arguably its main importance here is that it can complicate reliable measurements of other aspects of the total execution time.

To illustrate the potential utility of measuring different components of total execution time, we consider one aspect of running variational quantum algorithms. 
These algorithms involve repeated execution of a parameterized circuit for different parameter values. There are two distinct ways in which these circuits, that differ only in the parameter values, can be implemented. Each such circuit can be recompiled from a high-level language description (incurring the $t_{\rm compile}$ cost each time), or a single parameterized circuit can be compiled with its parameters set and updated using low-level classical controls. These two methods for running parameterized circuits will result in different compute times, and avoiding recompilation could potentially result in substantial compute time savings. Our \benchmark{Variational Quantum Eigensolver} benchmark has been designed to enable this comparison, by looping over multiple instances of a circuit with differing parameter values. Furthermore, we suggest that benchmarks that can reveal and quantify other potential improvements in compute time will be important tools for testing and improving quantum computers.

\begin{figure}[t!]
\includegraphics[width=\columnwidth]{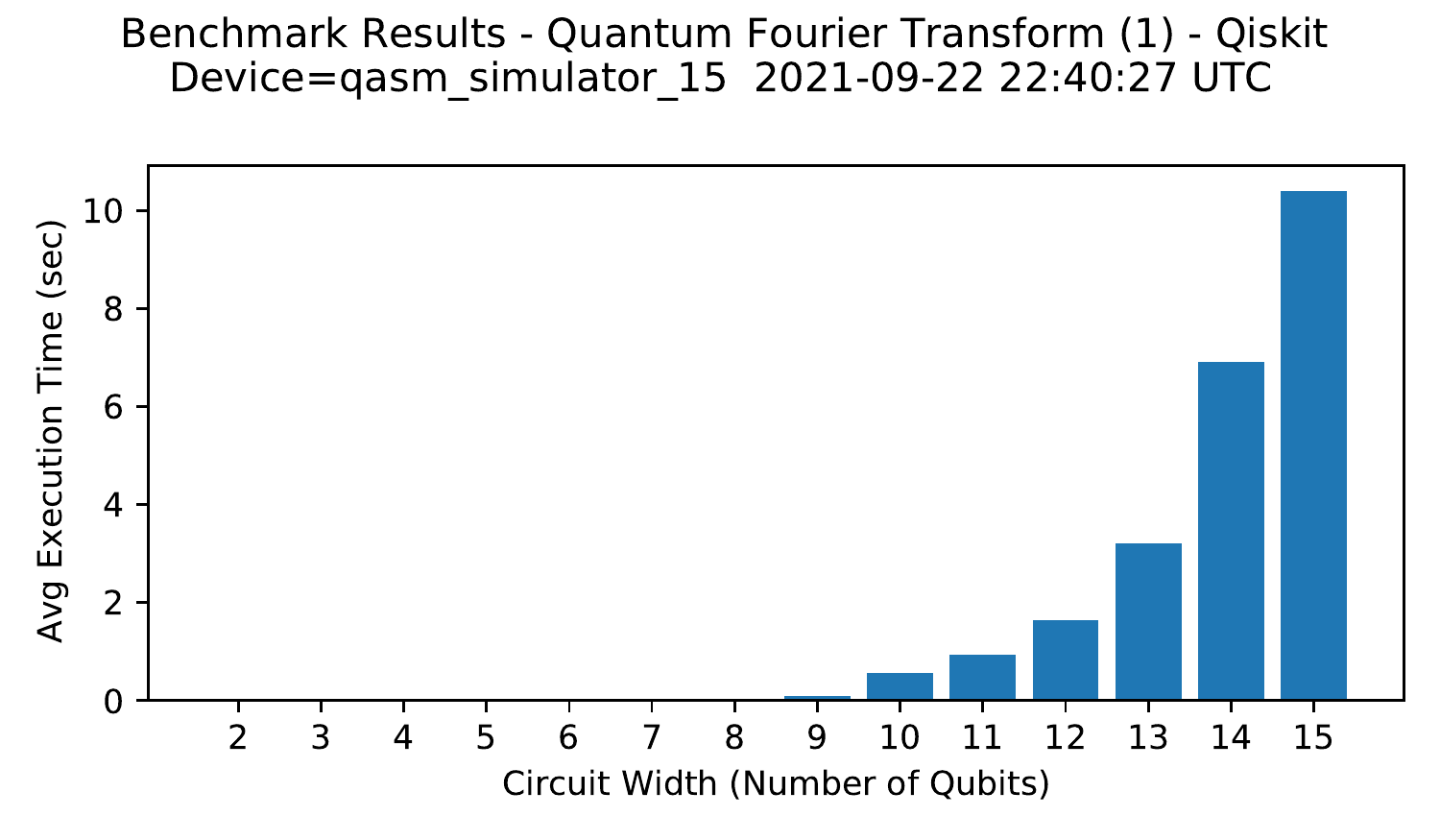}
\includegraphics[width=\columnwidth]{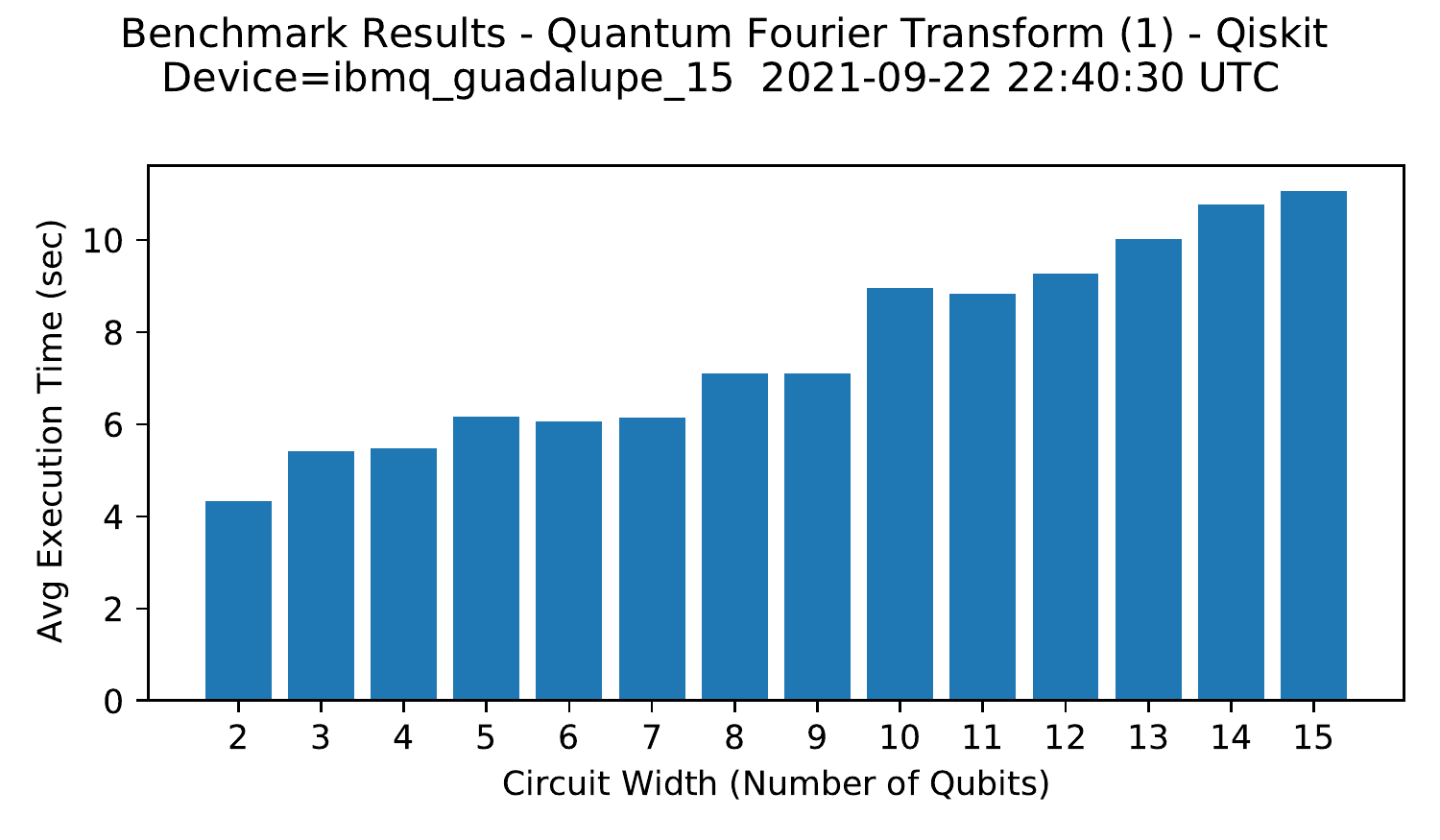}
\caption{\textbf{Measuring quantum execution time.} The average quantum execution time observed when running the \benchmark{QFT(1)} benchmark up to 15 qubits, with $N=1000$ shots, on a Qiskit Aer simulation of a noisy quantum computer (upper plot) and IBM Q Guadalupe (lower plot). (\emph{Data collected via cloud service}.)}
\label{fig:qft_1_15_exec_times}
\end{figure}

\subsection{Quantum Execution Time}
\label{sec:quantum_execution_time}
The aspect of execution time that our current benchmarking suite focuses on is 
quantum execution time $t_{\rm quantum}(N)$. This is the time taken to run $N$ `shots' of a quantum circuit (as the results from quantum circuits are probabilistic in nature, they are each typically run many times, with an individual circuit execution referred to as a `shot'). The quantum execution time depends on both the circuit (e.g., it will increase with the circuit depth), and on the number of shots $N$. In our experiments we use $N=1000$, and we do not study the dependence of $t_{\rm quantum}$ on $N$. Note that is important to use the same value of $N$ when comparing different devices.

The quantum execution time is not directly accessible to a quantum computer user, but cloud quantum computers typically return a quantity that is analogous to $t_{\rm quantum}(N)$, following the execution of the quantum circuit. This time is recorded by our benchmarking suite, and it is what we report. Note, however, that the precise meaning of the timing information returned by a provider's API is not always clear to the end user (and is possibly different for different providers). This should be kept in mind when comparing quantum execution time results between platforms.

\begin{figure}[t!]
\includegraphics[width=\columnwidth]{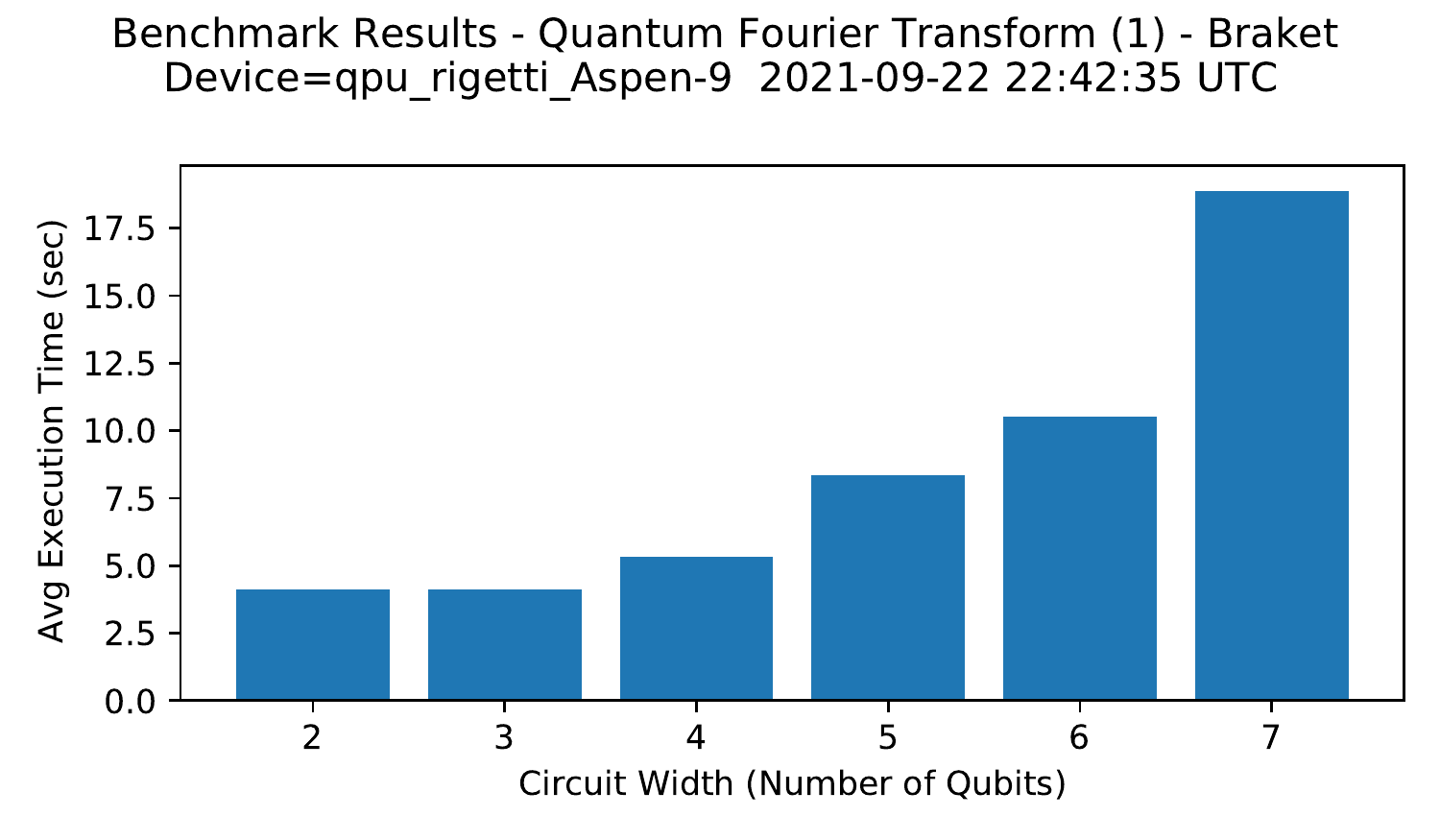}
\includegraphics[width=\columnwidth]{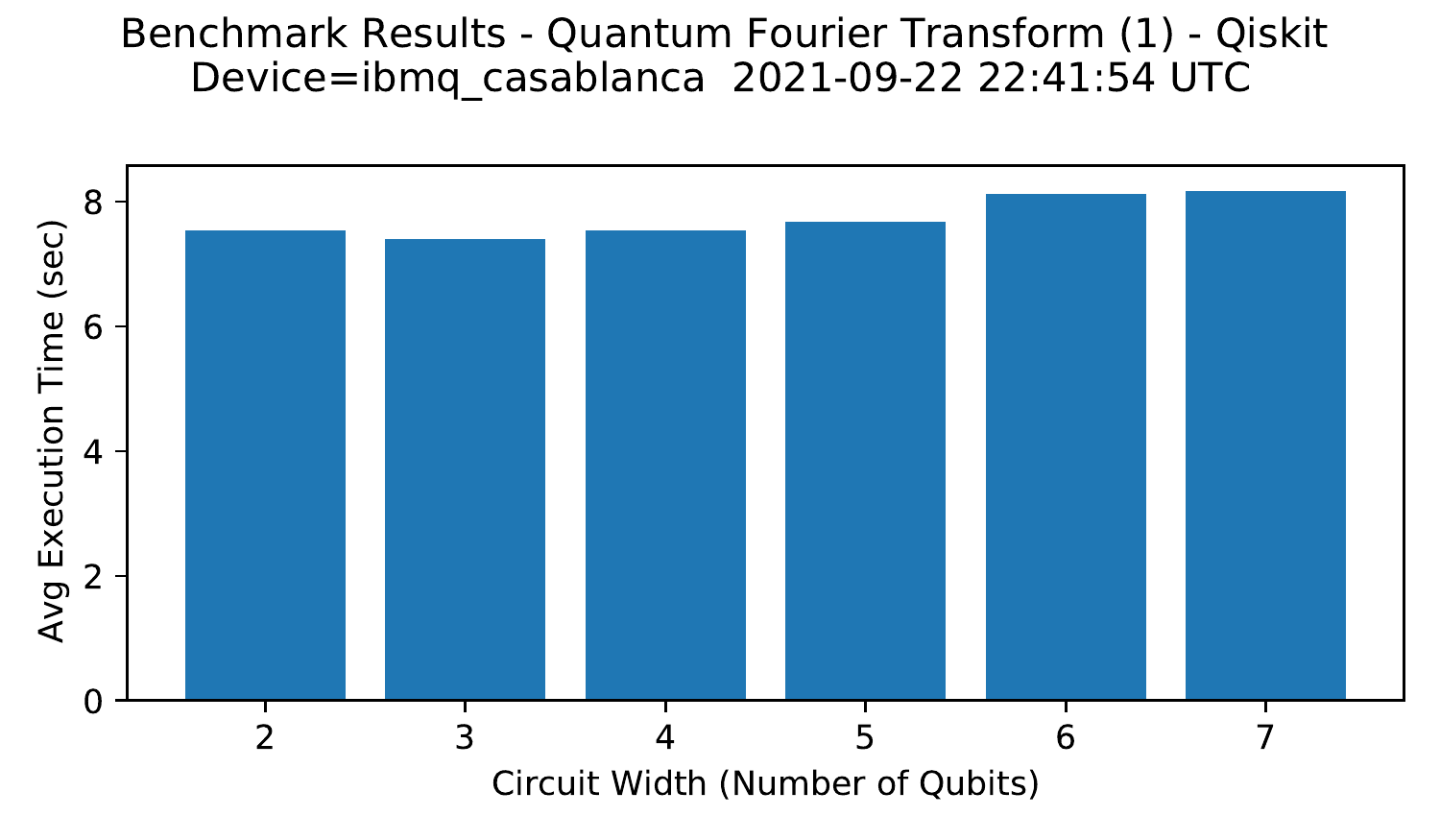}
\caption{\textbf{System differences in quantum execution time.} We observe significant differences in the quantum execution time between different systems. This example shows the average quantum execution time observed when running the \benchmark{QFT(1)} benchmark on Rigetti Aspen-9 (upper plot) and IBM Q Casablanca (lower plot). Note the different minimum execution times and the different rates of increase with circuit width. (\emph{Data collected via cloud service}.)}
\label{fig:ibmq_rigetti_execution_time}
\end{figure}

Figure \ref{fig:qft_1_15_exec_times} shows the average time to execute the circuits of the \benchmark{QFT(1)} benchmark, as a function of circuit width, for a quantum simulator and for IBM Q Guadalupe. The execution time increases exponentially on the simulator, and linearly on this quantum hardware. Note that the exponential growth in the time to simulate these circuits does not imply that these circuits are classically hard to simulate \footnote{The QFT acting on standard input states can be efficiently simulated on a classical computer using a special-purpose simulation algorithm.}, only that this classical simulator exhibits exponential simulation time scaling on these circuits.

Figure \ref{fig:ibmq_rigetti_execution_time} shows results from running the same \benchmark{QFT(1)} benchmark on Rigetti Aspen-9 and IBM Q Casablanca. The average execution time is shown as a function of the circuit width (the number of qubits) up to 7 qubits.
On IBM Q Casablanca, the execution time is approximately independent of circuit width, suggesting a minimal execution time of $\sim 7$ seconds for any circuit.
The increase in execution time is small as circuit width and depth increase (although note that this trend should not be naively extrapolated to larger widths and depths). In contrast, on Rigetti Aspen-9 the minimum execution time is about 4 seconds and there is a steep increase in execution time as circuit width increases (we conjecture that this is because all gates are serialized on Aspen-9, meaning that gates on distinct qubits that could be applied in parallel are applied in serial).
We found that there is significant variation in the execution time profile across the many different devices that we tested.

\section{Summary and Conclusions}
\label{sec:summary-and-conclusions}
Quantum computing is an emerging technology, and existing quantum computers experience a wide range of complex errors. To understand the capabilities of current and near-term quantum hardware the research community needs a robust set of performance benchmarks and metrics.
In this work we introduced a suite of quantum application-oriented performance benchmarks and demonstrated them on a selection of cloud-accessible quantum computers. Our benchmarking suite is based on a variety of well-known quantum algorithms, and it is designed to probe the effectiveness of quantum hardware at executing applications. Each of our application benchmarks tests a quantum computer's performance on variants of that application's quantum circuits. They are designed to map out performance as a function of the input problem's size. We measure performance both in terms of result fidelity and the quantum execution time, providing users with proxies for both the quality of and the time to solution. 

The initial version of our benchmarking suite, reported here, has a variety of limitations that we have highlighted throughout. Our intention is that our benchmarking suite will evolve to address both these limitations and new benchmarking needs---as has been the case with successful classical computing benchmarking suites such as SPEC \cite{spec_org, hennessy_patterson_2019_all}.
For this reason, we expect that the first version of our suite of benchmark programs, that we have presented in this paper, will be enhanced substantially over time. Indeed, since the initial version of this paper was released on arXiv there has been broad community engagement to enhance our benchmarking suite, and we anticipate reporting on these expansions and improvements to our suite in future publications.

One important limitation of some of our benchmarks is that they require exponentially expensive classical computations to implement, and they will be infeasible for $O(100)$ qubits. Our benchmarks that circumvent this problem do so by using only circuits that can be efficiently classically simulated. However, such benchmarking circuits must be designed with care so that a quantum computer's performance on the benchmark is indicative of its performance on the algorithm on which it is based when applied to practically useful problems. A general-purpose solution to this benchmark design problem was recently introduced \cite{proctor2022establishing} that enables efficiently measurement of the process fidelity of any algorithmic circuit. Integrating this technique with our algorithmic benchmarks would ensure that all our benchmarks are scalable and robust, making them practical to run on hundreds or thousands of qubits. Integrating this technique \cite{proctor2022establishing} and related methods \cite{proctor2020measuring, Ferracin2019-ou} into our benchmarking suite is a promising avenue for future research. 

Another limitation of our benchmarking suite is that our measurements of execution time are currently rudimentary. We primarily quantify compute time in terms of the time required to execute each quantum circuit. This is rudimentary because the definition of this time can vary by provider, and it neglects the classical co-processing inherent in quantum applications (e.g., circuit compilation time) that must be quantified for an accurate measure of the total compute time. We suggest that further research is required into how to delineate and consistently measure each aspect of total compute time.

Our primary method for summarizing a quantum computer's result quality on our benchmarks is volumetric benchmarking \cite{BlumeKohout2020volumetricframework, proctor2020measuring} plots, which display performance as a function of the problem input size (circuit width) and the quantum computation's length (circuit depth). This framework enables transparent comparisons between applications, as well as comparisons to other performance metrics. In this work we compared performance on our benchmarks to predictions extrapolated from the quantum volume \cite{Cross_2019} metric. Our experimental results demonstrate that, on the tested devices, the quantum volume is broadly predictive of application performance. However, the reliability of extrapolating the quantum volume to predict application performance varies between devices. A more quantitative and comprehensive comparison between performance on application-oriented benchmarks and application-agnostic benchmarks (e.g., the quantum volume benchmark \cite{Cross_2019} or mirror circuit benchmarks \cite{proctor2020measuring}) would constitute interesting future work.

The performance of the hardware that we tested suggests that substantial advances are required for quantum computers to outperform their classical counterparts on useful tasks. For example, our volumetric plots clearly communicate that an improvement in the quantum volume from, e.g. 32 to 2048, will not make it possible to successfully execute Shor's algorithms on useful problem instances---which, as is well-known, will almost certainly require large-scale quantum computers running fault tolerant quantum error correction. However, the potential for advances in algorithms, in quantum computing hardware, and in the software stack mean that quantum advantage on some problem might be achievable in the near-term. As new algorithms are developed they can be added to our suite, and our benchmarks can quantify the performance gains provide by technological advances. Therefore, our suite is well-suited to tracking progress towards useful quantum advantage.

Cloud-accessible quantum computers are attracting a wide audience of potential users, but the challenges in understanding their capabilities is a significant barrier to adoption of quantum computing. Our benchmarking suite makes it easy for new users to quickly assess a machine's ability to implement applications, and our volumetric visualization of the results is designed to be simple and intuitive. We therefore anticipate that our benchmarking suite will encourage adoption of quantum computing, and further economic development within the industry.

\section*{Data and Code Availability}
\label{sec:data_and_code_availability}
The code for the benchmark suite introduced in this work is available at \href{https://github.com/SRI-International/QC-App-Oriented-Benchmarks}{https://github.com/SRI-International/QC-App-Oriented-Benchmarks}. Detailed instructions are provided in the repository. The source code for quantum volume and volumetric benchmarking are contained in their respective public repositories \cite{qiskit_measuring_quantum_volume,github_pygsti}. All data and code required to reproduce all plots shown herein are available at  https://doi.org/10.5281/zenodo.6972744.

\section*{Acknowledgement}
We acknowledge the use of IBM Quantum services for this work. The views expressed are those of the authors, and do not reflect the official policy or position of IBM or the IBM Quantum team.
IBM Quantum. https://quantum-computing.ibm.com/, 2021. We acknowledge IonQ for the contribution of results from hardware not yet generally available. We acknowledge Quantinuum for contributing the results from their commercial H1.1 hardware. Timothy Proctor's contribution to this work was supported in part by the U.S. Department of Energy, Office of Science, Office of Advanced Scientific Computing Research through the Quantum Testbed Program. Sandia National Laboratories is a multi-program laboratory managed and operated by National Technology and Engineering Solutions of Sandia, LLC., a wholly owned subsidiary of Honeywell International, Inc., for the U.S. Department of Energy’s National Nuclear Security Administration under contract DE-NA-0003525.  Publisher acknowledges the U.S. Government license to provide public access under the DOE Public Access Plan (https://www.energy.gov/downloads/doe-public-access-plan). All statements of fact, opinion or conclusions contained herein are those of the authors and should not be construed as representing the official views or policies of the U.S. Department of Energy, or the U.S. Government.

\clearpage
\appendix
\clearpage

\section{Selected Algorithms and Applications}
\label{apdx:algorithms_and_applications}
In this appendix we examine the quantum programs that we selected for use as benchmarks, and describe how we turned those programs into benchmarks. Complete details for every benchmark can be found at \href{https://github.com/SRI-International/QC-App-Oriented-Benchmarks}{https://github.com/SRI-International/QC-App-Oriented-Benchmarks}. In this appendix, all of numerical results presented are based on execution of the application circuits on a noisy simulator with a specific noise model. The noise model consists of one- and two-qubit depolarizing errors with error rates of 0.003 and 0.03, respectively.

\subsection{Shallow Oracle-Based Algorithms}
\label{sec:oracle_algorithms}
The first group of quantum programs that we converted into benchmarks consist of three `oracle' algorithms.
\begin{itemize}
\item Deutsch-Jozsa Algorithm \cite{Deutsch-Jozsa_1992}
\item Bernstein-Vazirani Algorithm \cite{Bernstein-Vazirani_1997,  Wright_2019,  Linke_2017}
\item Hidden Shift Problem \cite{Rotteler_2010, Wright_2019, Linke_2017}
\end{itemize}

All of these algorithms share a simple structure in which the primary qubits of the circuit are initialized into a quantum superposition and a `secret' integer value is embedded into the circuit using a special `oracle' operation.
The oracle uses quantum gates to manipulate the relative phases of the qubits to effectively `hide' the integer, encoded as a string of ones and zeros, in the resulting quantum state.
At the end of the circuit, the quantum state is transformed back into the computational basis and the secret integer value is revealed.

\begin{figure}[h!]
\includegraphics[width=\columnwidth]{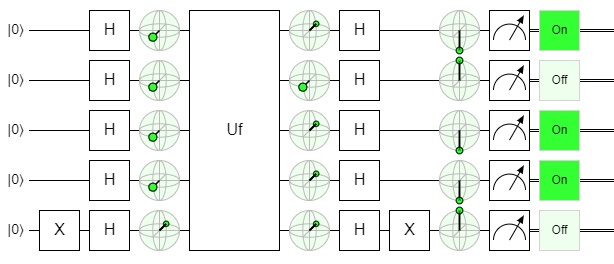}
\caption{An example of a circuit that implements the Bernstein-Vazirani algorithm, with illustrative Bloch spheres that denote the states of the qubits throughout the circuit.}
\label{fig:bm_bernstein_vazirani_5}
\end{figure}

\begin{figure}[h!]
\includegraphics[width=0.20\textwidth]{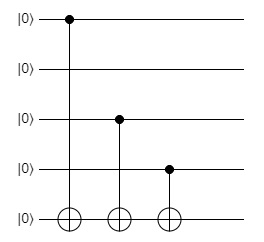}
\caption{An example of the circuit that implements the Bernstein-Vazirani Oracle function $ U_f $. This example encodes the integer 13.}
\label{fig:bm_bernstein_vazirani_uf_5}
\end{figure}

Figure \ref{fig:bm_bernstein_vazirani_5} shows a circuit diagram for the Bernstein-Vazirani algorithm (BV) on 5 qubits, one of which is used as an ancilla or accessory qubit.
An example of an oracle circuit $ U_f $ encoding the integer value $ 13 $ (bit string `1101') is shown in Figure \ref{fig:bm_bernstein_vazirani_uf_5}.
Each `1' bit is encoded using a controlled NOT gate on the 5\textsuperscript{th} ancilla qubit.

To illustrate how these circuits work, we have inserted small Bloch sphere icons into the main circuit diagram before and after the oracle and again before the measurements. The oracle operates on the uniform superposition generated by Hadamard gates on the first 4 qubits to encode the secret integer into a quantum state. Each qubit that performs a CNOT operation will experience phase kickback that rotates the phase of its quantum state by $\nicefrac{\pi}{2}$. The final set of Hadamard gates transforms the qubits back into the computational basis and the value encoded in the oracle is revealed upon measurement.

The main circuit of the Deutsch-Jozsa algorithm (DJ) is identical to the Bernstein-Vazirani algorithm, but it uses a slightly more complex oracle that encodes either a `constant' or `balanced' function.
The oracle is very similar to the Bernstein-Vazirani---it also uses a sequence of CNOT gates---so it is not shown here.

\begin{figure}[h!]
\includegraphics[width=0.45\textwidth]{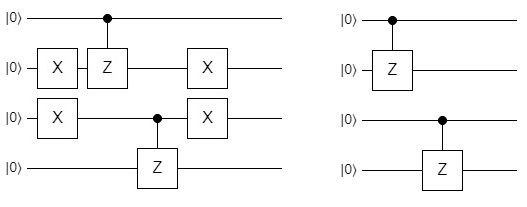}
\caption{Examples of circuits that implement the oracle functions for the Hidden Shift algorithm: $ U_f $ (left) and $ U_g $ (right).}
\label{fig:bm_hidden_shift_uf_4}
\end{figure}

The Hidden Shift algorithm (HS) applies two oracle functions in a row, which contrasts with the DJ and BV algorithms.
The first application of the oracle is identical to the second application, except that it includes a hidden `shift'---a secret integer that encodes a difference between the values returned from the two oracles (shown in Figure \ref{fig:bm_hidden_shift_uf_4}).
The two oracles together result in a circuit depth for the HS that is nearly double that of the BV and DJ algorithms.
Due to the similarities between these algorithms, we group them together as the `tutorial' category.

\begin{figure}[t!]
\includegraphics[width=\columnwidth]{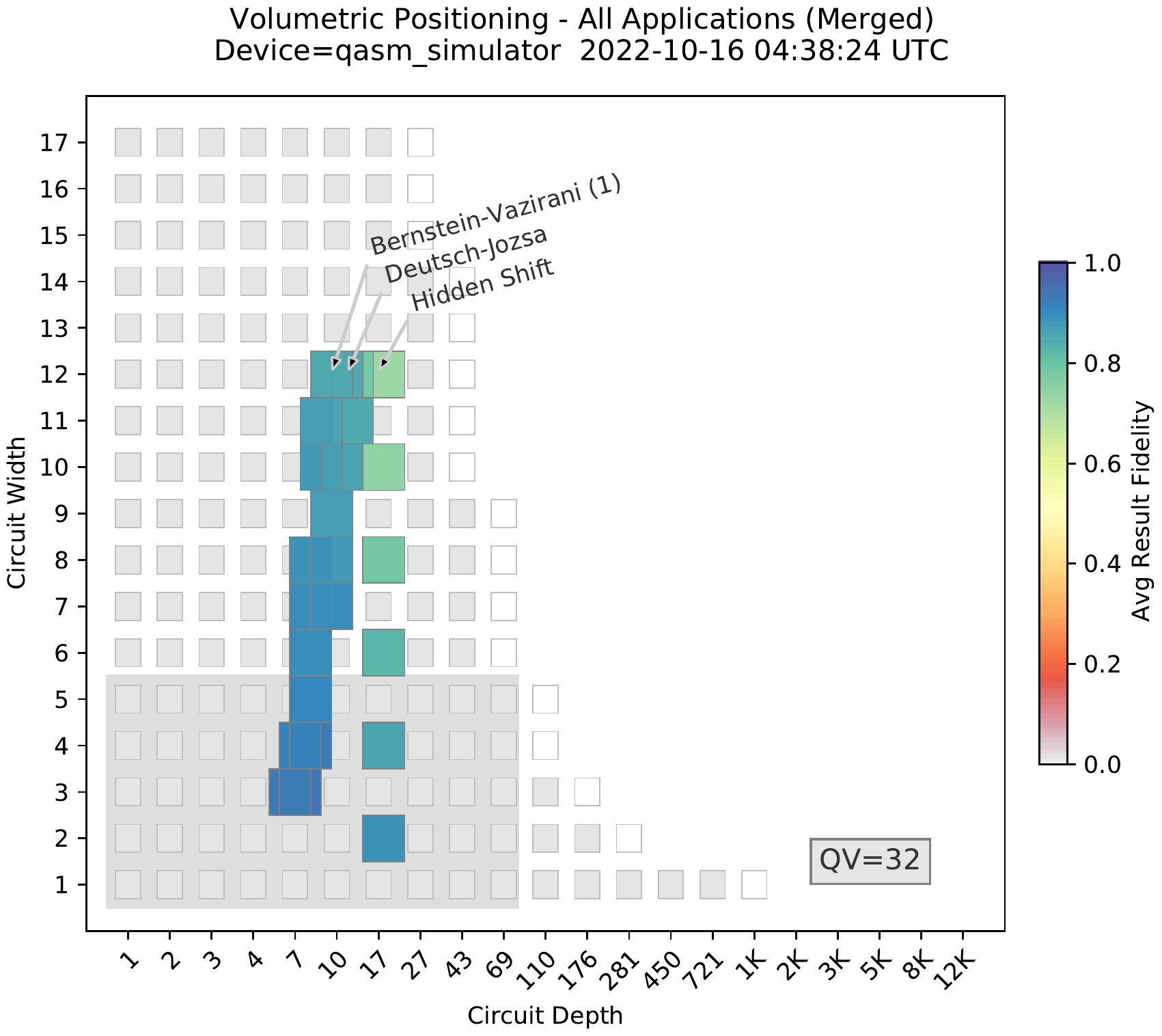}
\caption{Results from running the three oracle algorithm benchmarks on a simulator.}
\label{fig:shallow_circuits_vp_1}
\end{figure}

To construct benchmarks out of each of these algorithms, each circuit of a specific width is executed multiple times with a random integer encoded in the oracle. Figure \ref{fig:shallow_circuits_vp_1} shows the results obtained from executing these 3 benchmarks on the noisy simulator, for between 2 and 12 qubits.

\subsection{Quantum Fourier Transform}
\begin{figure}[t!]
\includegraphics[width=\columnwidth]{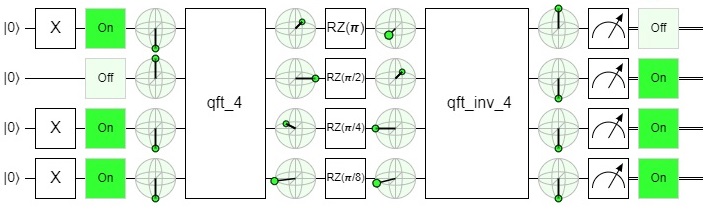}
\caption{An example of the circuits used in the \benchmark{QFT(1)} benchmark.}
\label{fig:bm_qft_1_4}
\end{figure}

The QFT \cite{Coppersmith1994AnAF} and its inverse are used extensively as components of many well-known quantum applications, e.g., Shor's algorithm. We constructed two benchmarks based on the QFT. The first of these benchmarks, which we refer to as the \benchmark{QFT(1)} benchmark, is shown in Figure \ref{fig:bm_qft_1_4} and is as follows. The $n$ qubits are initialized to encode a random integer $x$ ($x=13$ in the example shown in Figure \ref{fig:bm_qft_1_4}). This is followed by the standard QFT circuit, which encodes the input integer value into the Fourier basis, through a series of Hadamard gates and controlled Z rotations. Then a set of simple rotation gates are applied that add 1 modulo $2^n$, in the Fourier basis, to the input state. This is then followed by the standard circuit for implementing the inverse QFT, which converts the Fourier basis state back into the computational basis, and then all the qubits are measured. The resulting bit string should be the $x + 1$ modulo $2^n$ (in the example of Figure \ref{fig:bm_qft_1_4}, 14, rather than the initial value of 13).
This benchmark can be thought of as a type of Loschmidt echo, or a simple form of mirror circuit \cite{proctor2020measuring}, in which the QFT is undone by the inverse QFT that follows it. The inclusion of the $+1$ operation in the middle of the circuit prevents a transpiler from easily compiling this circuit down to an identity (although note that this current implementation is does not strictly prevent compilation down a simpler operation). Additionally, it adds sensitivity to errors that would be canceled by a perfect Loschmidt echo (a forwards and backwards evolution experiment) but that do impact the performance of the QFT. Note, however, that a randomly selected operation in the center of the circuit is required to guarantee sensitivity to all errors \cite{proctor2020measuring}---and this benchmark will likely be updated in the future to include this random operation.

\begin{figure}[t!]
\includegraphics[width=0.35\textwidth]{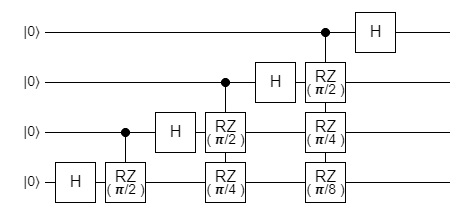}
\caption{The standard circuit that implements the $n=4$ qubit QFT. The standard $n$-qubit QFT circuit is used in the first of our QFT-based benchmarks.}
\label{fig:bm_qft_1_qft_4}
\end{figure}

The QFT sub-circuit used in the \benchmark{QFT(1)} benchmark is displayed in Figure \ref{fig:bm_qft_1_qft_4}.
The inverse QFT sub-circuit is not shown as it is a mirror image of the QFT circuit (with gates replaced with their inverses, meaning that the rotation angles are negated and the Hadmard gates are unchanged).
Note that this QFT implementation does not use SWAP gates (so it effectively reverses the order of the qubits) which simplifies its use in the benchmark programs.

\begin{figure}[h!]
\includegraphics[width=\columnwidth]{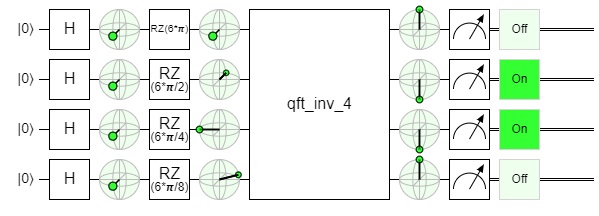}
\caption{An example of the circuits used in the \benchmark{QFT(2)} benchmark.}
\label{fig:bm_qft_2_4}
\end{figure}

Our second QFT benchmark [\benchmark{QFT(2)}] executes only a standard circuit for the inverse QFT. In the \benchmark{QFT(2)} benchmark, the inverse QFT operates on a quantum state that is prepared in a Fourier basis state. 
In this method, instead of creating that quantum state using the QFT (which converts any computational basis state into a Fourier basis state), we use a set of one-qubit gates---specifically, a series of Hadamard gates and $Z$ rotations, to encode a specific integer value $x$ in the Fourier basis. Figure~\ref{fig:bm_qft_2_4} demonstrates this circuit, for the case of encoding the integer value 6. The inverse QFT reverses that operation, so if there are no errors the measurement result should be the bit string encoding $x$.

The circuits of the \benchmark{QFT(2)} benchmark are similar to the second half of circuits of the \benchmark{QFT(1)} benchmark, so the circuit depth in \benchmark{QFT(2)} is approximately half that in \benchmark{QFT(1)} (for fixed $n$).
A useful consequence of this is that the benchmarks can be used to cross-validate each other: if both benchmarks are a good quantification of a quantum computer's performance on the QFT and its inverse, and the QFT and its inverse have approximately the same error, then the result fidelity from \benchmark{QFT(1)} will be approximately the square of the result fidelity from \benchmark{QFT(2)}.

\begin{figure}[t!]
\includegraphics[width=\columnwidth]{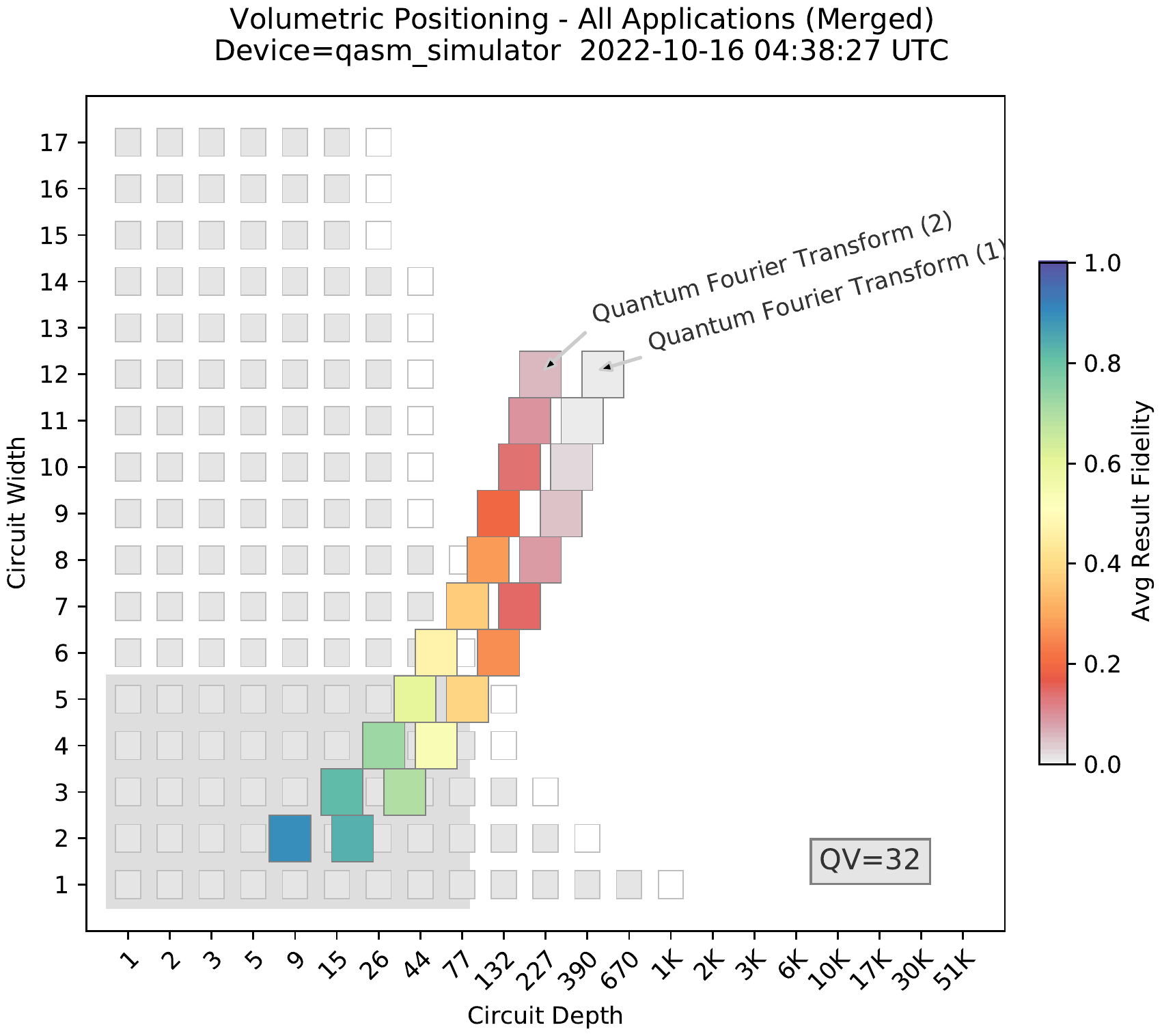}
\caption{Results from running the two QFT-based benchmarks on a simulator.}
\label{fig:qft_circuits_vp_1}
\end{figure}

Figure \ref{fig:qft_circuits_vp_1} shows the results obtained from executing the \benchmark{QFT(1)} and \benchmark{QFT(2)} benchmarks. We see the expected relationship between the fidelities of the two benchmarks (note that this simple relationship can be proven to hold for simple depolarizing errors like those used in this simulation).

\subsection{Grover's Search Algorithm}
Grover's algorithm \cite{Grovers_1996}, also referred to as the quantum search algorithm, is one of the most well known quantum algorithms due to its quadratic run-time speedup over the best known classical algorithm.
This algorithm aims to find the correct item from an un-ordered list.
It achieves a quantum speedup by starting in a uniform superposition and using repeated calls to a quantum oracle to amplify the probability of measuring the correct state.
The amplification used in Grover's algorithm was then generalized into amplitude amplification \cite{Brassard_2002}, which is used in the Amplitude Estimation and Monte Carlo algorithms.

\begin{figure}[ht]
\includegraphics[width=\columnwidth]{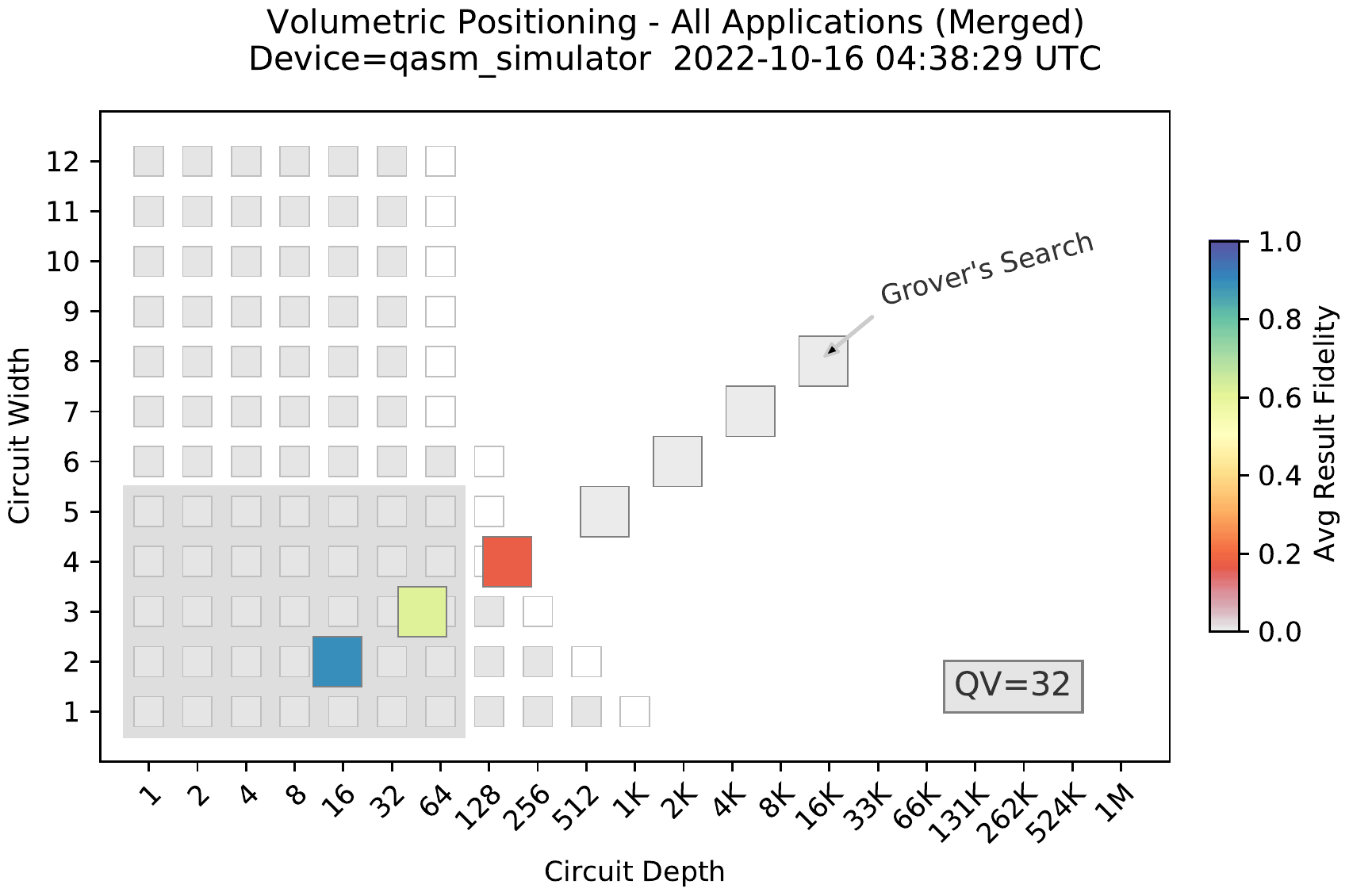}
\caption{Results from running the \benchmark{Grover's Search} Benchmark on a simulator.}
\label{fig:bm_grovers_vp}
\end{figure}

Our benchmark based on Grover's Search algorithm is designed so that the oracle marks a given bit string with a phase of $-1$. At each circuit width, multiple circuits are constructed by generating random bit strings of length equal to the width of the circuit. Figure \ref{fig:bm_grovers_vp} shows the results of executing the \benchmark{Grover's Search} benchmark on a simulator.
This volumetric plot highlights the fast increase in circuit depth with increasing circuit width for this algorithm---which causes a rapid drop in the result fidelity as the circuit width grows. This occurs because Grover's algorithm for $n$-qubits uses many $n$-qubit Toffoli gates.
These gates need to be decomposed into one- and two-qubit primitive gates, which substantially increases the depth. Note that shallower decompositions of $n$-qubit Toffoli gates are possible using ancilla qubits, which we do not do here.

\subsection{Phase and Amplitude Estimation}
Quantum Phase Estimation \cite{Kitaev1996QuantumMA} is another important quantum subroutine that is fundamental to many quantum algorithms, including Shor's algorithm \cite{Shor_1997}.
The goal of the algorithm is to estimate the eigenvalues of a unitary operator by calculating the corresponding phases.
A generalization of phase estimation is amplitude estimation \cite{Brassard_2002}, where the goal is to estimate the amplitudes of a quantum state. Due to the importance of phase estimation and amplitude estimation as subroutines for more complicated quantum algorithms, benchmarking based on these routines will provide insight into the capabilities of a quantum computer.

The \benchmark{Phase Estimation} benchmark is based on picking a unitary whose phase is to be measured. We choose a phase gate with eigenvalues $\exp(\pm i \theta)$, where $\theta$ is defined to a precision $2^{-k}$, where $k$ is the number of qubits in the register that stores the measured phase. The benchmark selects random values for $\theta$ of the form $\frac{n}{2^k}$, where $k$ represents the total number of measured qubits and $n$ is an integer between 0 and $2^k-1$.

The \benchmark{Amplitude Estimation} benchmark contains two quantum subroutines: quantum circuits that implement quantum amplitude amplification and quantum phase estimation.
The quantum amplitude amplification is used to encode the amplitudes of a target quantum state into a phase which can be extracted using phase estimation. This benchmark performs the simplest case of quantum amplitude estimation where, similar to phase estimation, we chose specific circuits such that the amplitudes are specified to a precision which can be exactly encoded in the available qubits.

\begin{figure}[t!]
\includegraphics[width=\columnwidth]{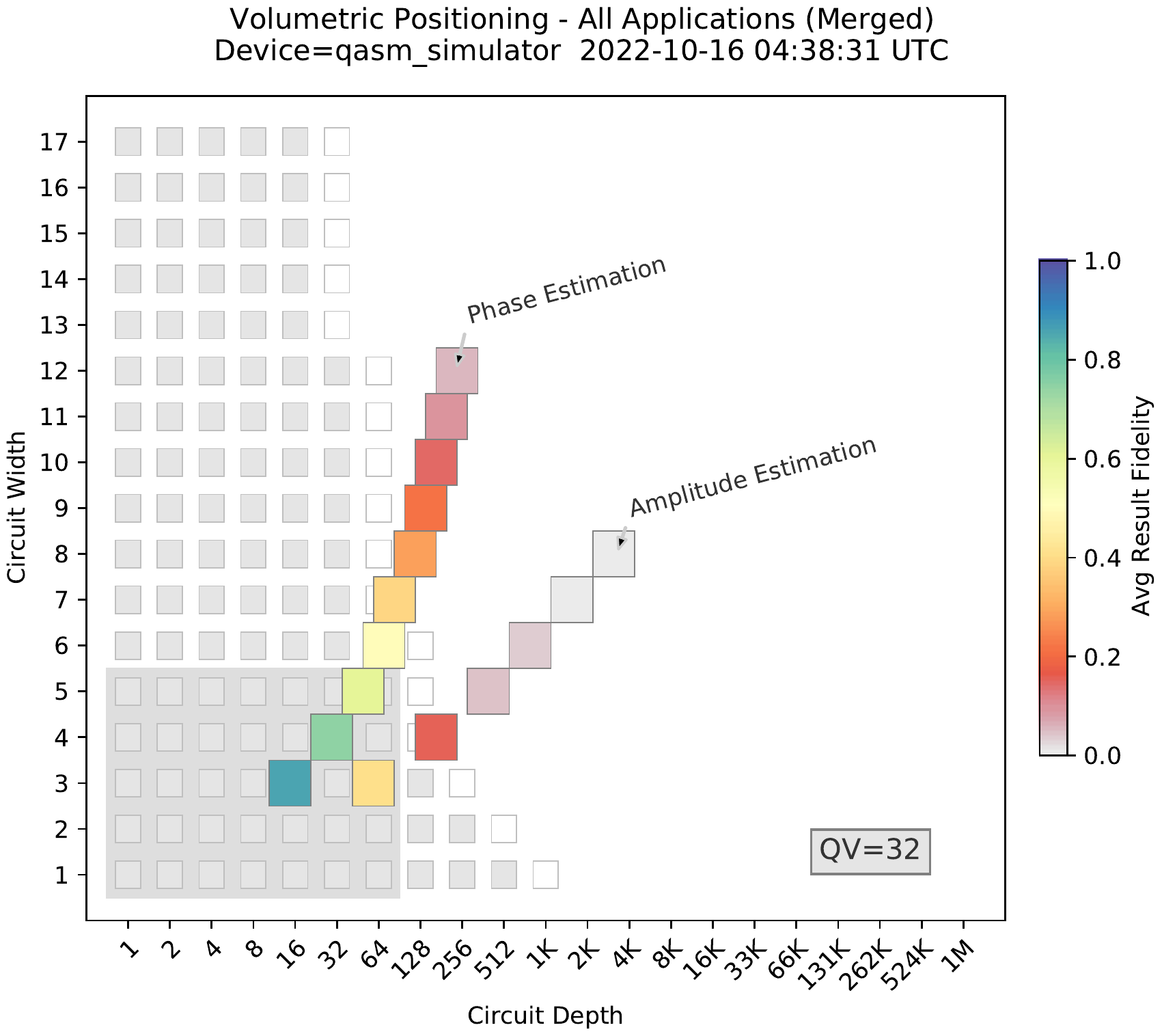}
\caption{Results from running the \benchmark{Phase Estimation} and \benchmark{Amplitude Estimation} benchmarks on a simulator.}
\label{fig:bm_estimations_vp_1}
\end{figure}

Figure \ref{fig:bm_estimations_vp_1} shows the results obtained when executing our \benchmark{Phase Estimation} and \benchmark{Amplitude Estimation} benchmarks. Note the depth of the \benchmark{Phase Estimation} benchmark's circuits scale similarly to the depth of the circuits in the two QFT-based benchmarks. This is because the QFT is the subroutine of phase estimation that uses the deepest circuits. The \benchmark{Amplitude Estimation} benchmark, on the other hand, involves deeper circuits due to the quantum amplitude amplification protocol. This figure shows that a quantum computer with a quantum volume of 32 is incapable of high fidelity implementations of the  Amplitude Estimation algorithm (at least when implemented in its standard form), even for tiny problem instances.

\subsection{Hamiltonian Simulation}
\label{alg:ham-sim}
One of the most promising applications for quantum computers is in the simulation of quantum systems \cite{Feynman1982}. Many simulation problem require simulating the evolution of a Hamiltonian, so we designed a benchmark based on Hamiltonian evolution.
The Hamiltonian we choose is the anti-ferromagnetic Heisenberg chain of 1-D spins with disordered x- and z-fields and an open boundary condition:
\begin{equation}
    H=J\sum_{i=0}^{n-2}\vec{\sigma}_i\cdot\vec{\sigma}_{i+1}+w\sum_{i=0}^{n-1}(h_{x,i}\sigma^x_i+h_{z,i}\sigma^z_i),
\label{eq:hamiltonian} 
\end{equation}
where $J$ and $w$ are the strength of the interactions and the disordered fields respectively, $h_{x,i}$ and $h_{z,i}$ give the strength of the $X$ and $Z$ disordered fields at site $i$, $n$ is the total number of qubits, and $\sigma^{\{x,y,z\}}$ are the usual Pauli operators.
In this benchmark, $J=1$, $w=10$, and $h_{x,i}$ and $h_{z,i}$ are randomly generated in the range $(-1,1)$.
The algorithm initializes the qubits in a product state and evolves them according to a Trotterized version of the Hamiltonian \cite{Suzuki1976}. The Trotterization implements the term $\exp(-iJ\vec{\sigma}_i\cdot\vec{\sigma}_{i+1})$ exactly using a construction from \citet{optimal_circuits_2004}. We use a fixed number of Trotter steps.

\begin{figure}[t!]
\includegraphics[width=\columnwidth]{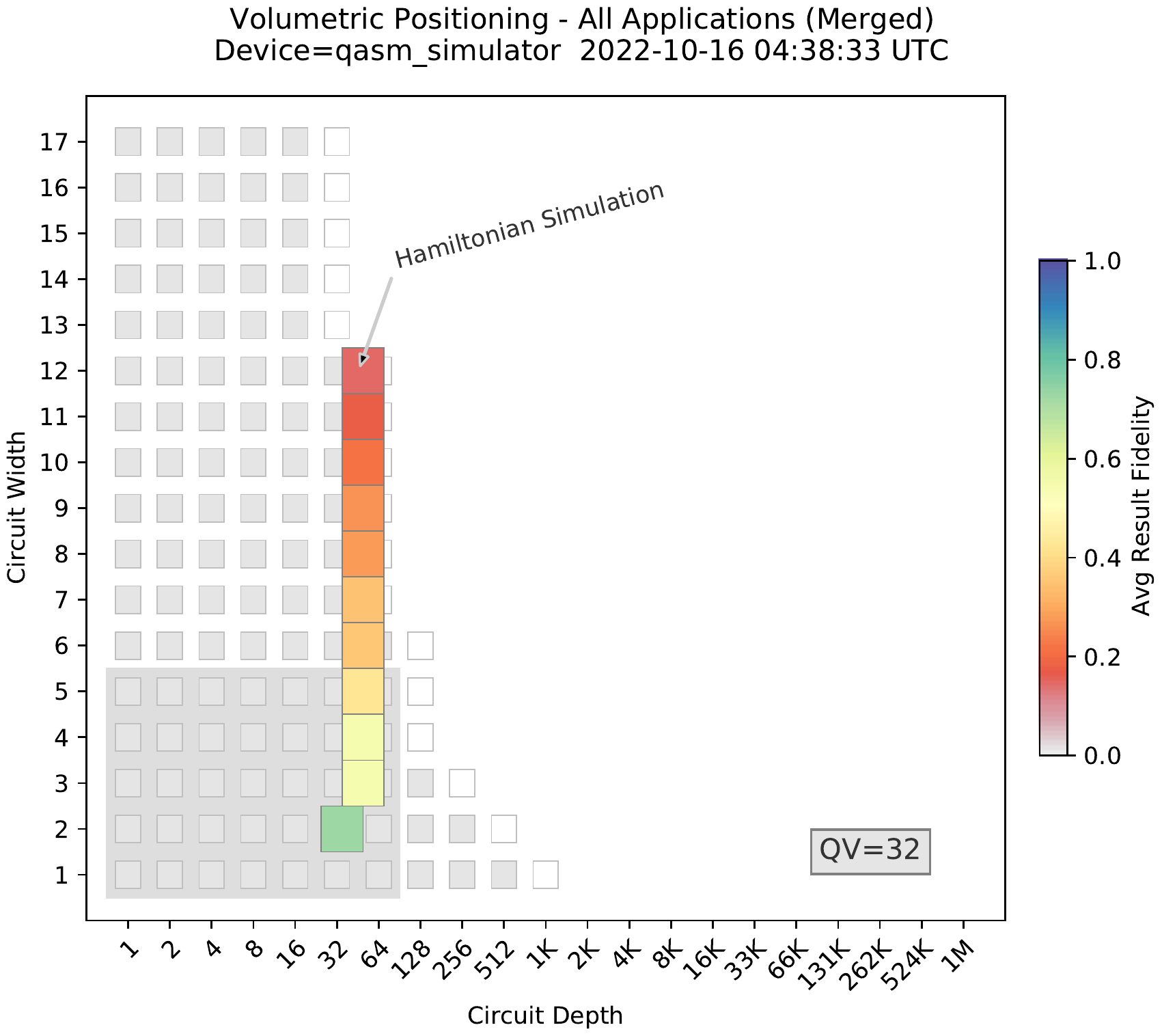}
\caption{Results from running the \benchmark{Hamiltonian Simulation} benchmark on a simulator.}
\label{fig:bm_hamiltonians_vp_1}
\end{figure}

Figure \ref{fig:bm_hamiltonians_vp_1} shows the results of running the \benchmark{Hamiltonian Simulation} benchmark on a simulator. Note that the depth of the circuits does not increase with increasing width, due to the form of the Hamiltonian that we choose. This benchmark uses exponentially scaling classical computations, as the error-free circuit output is computed using a general-purpose circuit simulation.

One way to construct a Hamiltonian simulation benchmark that does not require expoentially expensive classical computations is to use a phenomenon called many-body localization \cite{Childs2018, zhu2021}, where evolution under a random Hamiltonian such as the one in Eq. \ref{eq:hamiltonian} preserves certain quantities.
However, long evolution times are required for many body localization, resulting in circuits containing thousands of circuit layers. A benchmark based on this method would therefore arguably not be useful for current hardware.

\subsection{Monte Carlo Sampling}
Classical Monte Carlo algorithms use random processes to solve problems numerically where other methods might be infeasible. A quantum algorithm which has a quadratic speed up over classical Monte Carlo algorithms was proposed by \citet{Woerner2019}, which we have used here as the basis for our \benchmark{Monte Carlo Sampling} benchmark. The Monte Carlo sampling algorithm aims to compute the expected value of some function $f(X)$, where $X$ is a random variable that $p(X)$ distributed. The algorithms uses two oracles. One of these oracles, $\mathcal{R}$, is defined by
\begin{equation}
    \mathcal{R}\left|0\right>_n\left|0\right>=\sum_i{\sqrt{p(X=i)}\left|i\right>_n\left|0\right>}.
\end{equation}
This encodes the probability distribution $p(X)$ into the amplitudes of $n$ qubit states. The other oracle, $\mathcal{F}$, is defined by
\begin{equation}
    \mathcal{F}\left|i\right>_n\left|0\right>=\left|i\right>_n\left(\sqrt{1-f(i)}\left|0\right>+\sqrt{f(i)}\left|1\right>\right).
\end{equation}
This encodes the function value $f(i)$ into the amplitude of a single auxiliary qubit. By applying $\mathcal{R}$ and $\mathcal{F}$ on a $n+1$ qubit system initialized to the zero state, we obtain
\begin{equation}
    \begin{split}
        &\mathcal{F}\mathcal{R}\left|0\right>_n\left|0\right>= \\
        &\sum_i{\left|i\right>_n\left(\sqrt{p(X=i)}\sqrt{1-f(i)}\left|0\right>+\sqrt{p(X=i)}\sqrt{f(i)}\left|1\right>\right)}.
    \end{split}
\end{equation}
Recall that for the amplitude estimation, given the operator $\mathcal{A}$,
\begin{equation}
    \mathcal{A}\left|0\right>_{n+1}=\sqrt{1-a}\left|\psi_0\right>\left|0\right>+\sqrt{a}\left|\psi_1\right>\left|1\right>,
\end{equation}
the amplitude estimation algorithm yields $a$. The Monte Carlo Sampling algorithm works by applying
\begin{equation}
    \mathcal{A}=\mathcal{F}\mathcal{R}
\end{equation}
and then perform amplitude estimation on the resultant state to produce
\begin{equation}
    a=\sum_i{p(X=i)f(i)}=E[f(X)],
\end{equation}
which is the expectation value of the $f$ function. 

\begin{figure}[t!]
\includegraphics[width=\columnwidth]{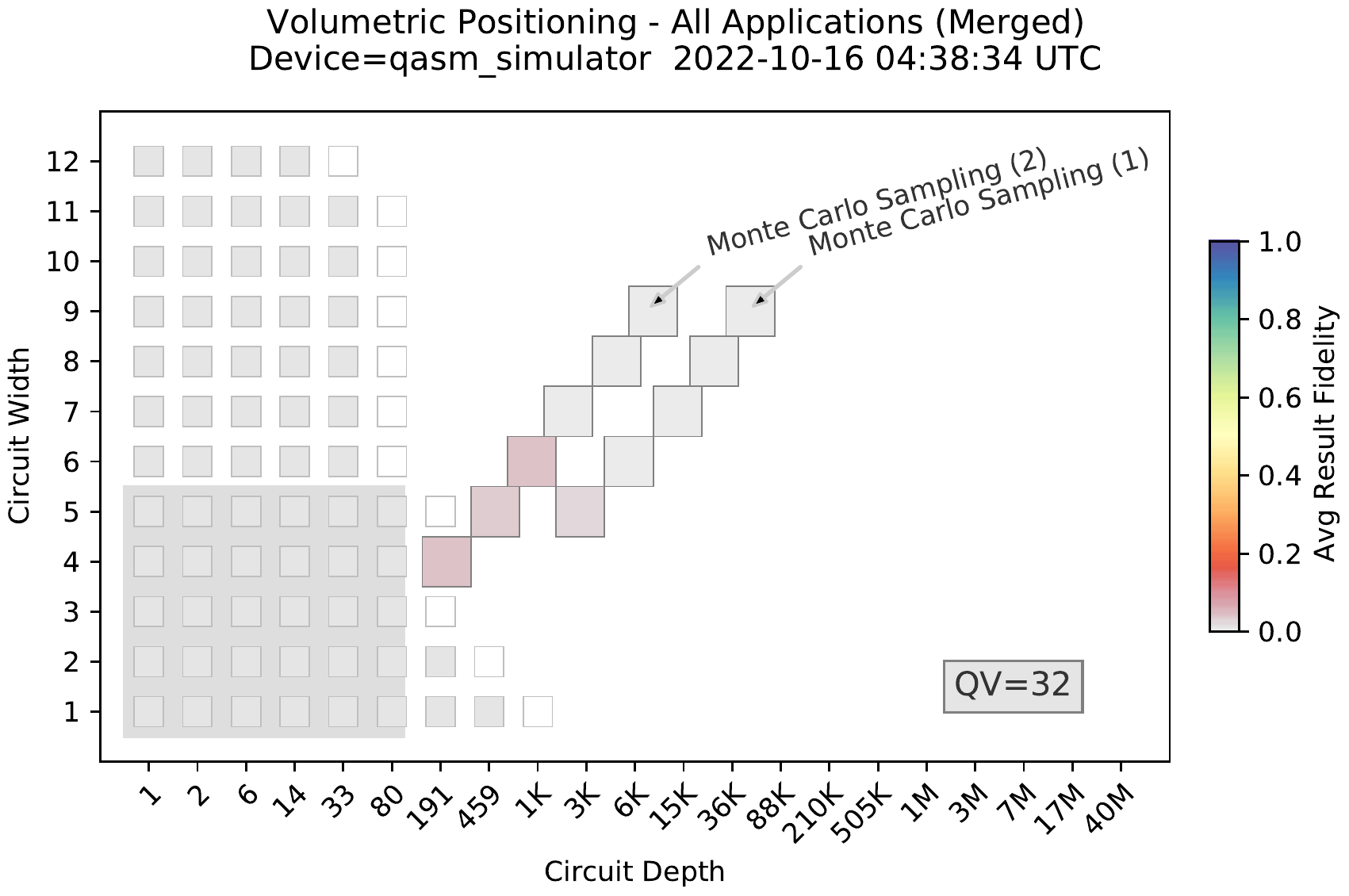}
\caption{Results from running the two benchmarks based on the Monte Carlo Sampling algorithm on a simulator.}
\label{fig:mc_circuits_vp_1}
\end{figure}

We implemented two benchmarks based on running the Monte Carlo Sampling algorithm for different choices of $f$ and $p$. \benchmark{Monto Carlo Samping(1)} enables the user to 
choose $f$ and $p$. For general distributions and functions, the circuits of this benchmark are very deep. \benchmark{Monto Carlo Samping(2)} fixes $p$ to the uniform distribution (this $\mathcal{R}$ oracle can then be implemented using only Hadamard gates), and the fixes $f$ to the parity function [$f(i)=1$ if $i$ has an odd number of 1's in its binary expansions and $f(i)=0$ otherwise], which means that the $\mathcal{F}$ oracle can be implement using only CNOT gates. Figure~\ref{fig:mc_circuits_vp_1} shows results from both versions of our benchmark.

\subsection{Variational Quantum Eigensolver}
Solving the electronic structure problem is one of the most promising applications for quantum computers, with potential contributions to drug discovery \cite{Yuan20_015003}, developing new materials and catalysts for more efficient light-harvesting and CO$_2$ reduction \cite{Troyer21_033055}, understanding high-temperature super conductivity \cite{Yuan20_015003}, and other important use cases \cite{Troyer17_7555}.
In the NISQ era, one of the most promising approaches is the variational quantum eigensolver (VQE) \cite{Peruzzo2014, McClean_2016}.

The VQE algorithm implements a parameterized circuit for a chosen wavefunction ansatz.
It then measures the energy expectation value of the wave function and optimizes the circuit parameters towards a direction that lowers the energy expectation value. The 
VQE algorithm is based on the variational principle \cite{Szabo-Ostland}, 
\begin{equation}
    E=\frac{\left<\Psi(\vec{\theta})\big|H\big|\Psi(\vec{\theta})\right>}{\left<\Psi(\vec{\theta})\big|\Psi(\vec{\theta})\right>}\geq E_0,
\end{equation}
in which the energy expectation value of an parameterized wave function $\Psi(\vec{\theta})$ is lower bounded 
by the exact ground state energy $E_0$. In other words, the optimal parameter
$\vec{\theta}^\ast$ that minimizes the energy expectation value corresponds to 
the closest description to the exact ground state within the flexibility of the 
wave function ansatz. Herein, $H$ is the electronic Hamiltonian defined as 
\begin{equation}
    H=\sum_{pq}{h_{pq}a_p^\dagger{a}_q}+\frac{1}{2}\sum_{pqrs}{g_{pqrs}a^\dagger_pa^\dagger_qa_sa_r},
\end{equation}
in which $h_{pq},g_{pqrs}$ are the one- and two-electron integrals, and 
$a, a^\dagger$ are the Fermionic annihilation and creation operators.

An important aspect of the VQE algorithm is the choice of the wave function ansatz. Many choices exist, and for our VQE-based benchmark we choose to use the unitary Coupled Cluster with singles and doubles ansatz (unitary-CCSD). This ansatz is defined by \cite{Cao_2019}
\begin{equation}
    |\Psi\rangle = e^{T-T^\dagger}|\Phi\rangle
\end{equation}
where $T$ is the cluster operator and $|\Phi\rangle$ is the reference state, which is chosen to be the Hartree-Fock (HF) state.
The cluster operator is defined by
\begin{equation}
    T=\sum_{i,a}t^a_ia_ia^\dagger_a+\sum_{i>j,a>b}t^{ab}_{ij}a_ia^\dagger_aa_ja^\dagger_b
\end{equation}
in which $i, j$ indicate molecular orbitals that are occupied in the HF state, and $a, b$ indicate molecular orbitals that are empty in the HF state. $t_j^a$ and $t^{ab}_{ij}$ are the wave function parameters associated with single and double excitations.

The unitary CCSD ansatz is a variant of the widely used traditional CCSD ansatz. 
Unlike the traditional CCSD ansatz, which fails to model strongly correlated systems,
the unitary CCSD ansatz has been shown to produce highly accurate results in both weakly and strongly correlated systems. 
However, there is no known classical method that can evaluate the energy expectation value of the unitary CCSD ansatz with polynomial cost. In contrast, its unitary nature means that it is natural to implement on a quantum computer. 

Both the electronic Hamiltonian and the unitary CCSD ansatz are defined with Fermionic 
operators. In order to implement them on a quantum computer, one needs to first transform
them to Pauli operators. Our \benchmark{VQE} benchmarks use the Jordan-Wigner transformation 
\cite{Jordan1928}: 
\begin{equation}
    \begin{split}
            a_p^\dagger\rightarrow I^{\otimes p-1}\otimes[\frac{1}{2}(\sigma^x_p-i\sigma^y_p)]\otimes\sigma^{z{\otimes N-p}}, \\
            a_p\rightarrow I^{\otimes p-1}\otimes[\frac{1}{2}(\sigma^x_p+i\sigma^y_p)]\otimes \sigma^{z\otimes N-p},
    \end{split}
\end{equation}
where $I, X, Y$ and $Z$ are Pauli matrices and $N$ is the total number of qubits. 
Upon the Jordan-Wigner transformation, both the Hamiltonian and the cluster operator
becomes a weighted sum of Pauli words 
\begin{equation}
    H(T)=\sum_i^M{g_i(t_i)P^i},
\end{equation}
where $P^i$ is a tensor product of Pauli matrices,
\begin{equation}
    P^i=\{I,\sigma^x,\sigma^y,\sigma^z\}^{\otimes N}.
\end{equation}
The VQE algorithm proceeds by iteratively evaluating and optimizing the energy by varying the wave function parameters, until the minimum is found.

The full VQE algorithm includes both a quantum and a classical part. The former is used to implement the ansatz and measure the energy, and the latter is applied to optimize the ansatz parameters. Our initial VQE-based benchmark, \benchmark{VQE(1)}, only implements the quantum part of the algorithm. For \benchmark{VQE(1)}, we randomly sample the wave function parameters and choose a fixed measurement basis (varying the measurement basis is generally necessary to measure the expectation value of the Hamiltonian $H(T)$).

\begin{figure}[t!]
\includegraphics[width=\columnwidth]{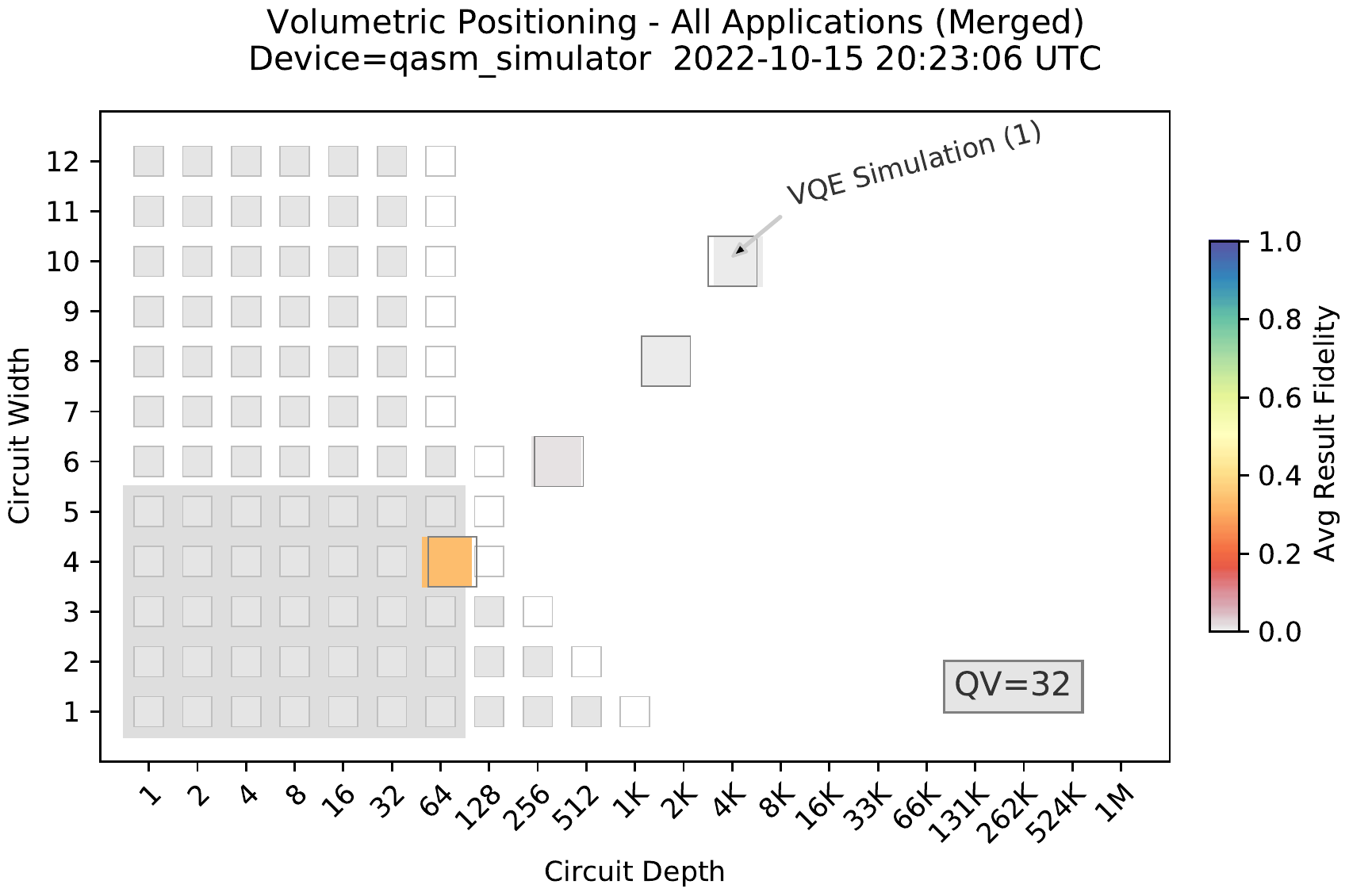}
\caption{Results from running \benchmark{VQE(1)}, one of our two VQE-based benchmarks, on a simulator.}
\label{fig:bm_vqe_vp}
\end{figure}

In our benchmarks, the Hamiltonian and the wave function is implemented for the NaH molecule with a bond length of 1.9 Angstrom with the STO-6G basis\cite{Szabo-Ostland}. By changing the number of orbitals to simulate, we vary the number of qubits. The smallest case is $n=4$ qubits, which corresponds to 2 electrons and 4 spin orbitals. Figure \ref{fig:bm_vqe_vp} shows the results of executing \benchmark{VQE(1)} on a quantum simulator up to 10 qubits.

\subsection{Shor's Order Finding}
Shor's factoring algorithm \cite{Shor_1997} is one of the most well-known applications of quantum computing. Shor's factoring algorithm includes both a quantum and classical subroutine.
The quantum subroutine finds the smallest integer period, or the order, of a periodic function while the classical subroutine uses this order to determine the factors of a number in polynomial time.
This process may need to be repeated multiple times, in case an invalid order is generated.
Since the order finding routine is the bottleneck of integer factorization, our focus is on benchmarking the quantum subroutine of order finding.

The algorithm is designed for finding the order of functions that take the form $f(x) = a^x \mod N$ where $a < N$. The order $r$ of this function is the $r$ satisfying $a^r \mod N = 1$. If $r$ is known, a classical computer can compute the prime factors of $N$ in polynomial time. Our benchmarks based on Shor's order finding algorithm generate random values of $r$ and $N$ from which a corresponding base value $a$ can be analytically calculated. The quantum algorithm for order finding is actually a variation of quantum phase estimation where the chosen unitary operator performs modular exponentiation. This modular exponentiation encodes the order in the phases of a quantum state. 

\begin{figure}[b!]
\includegraphics[width=\columnwidth]{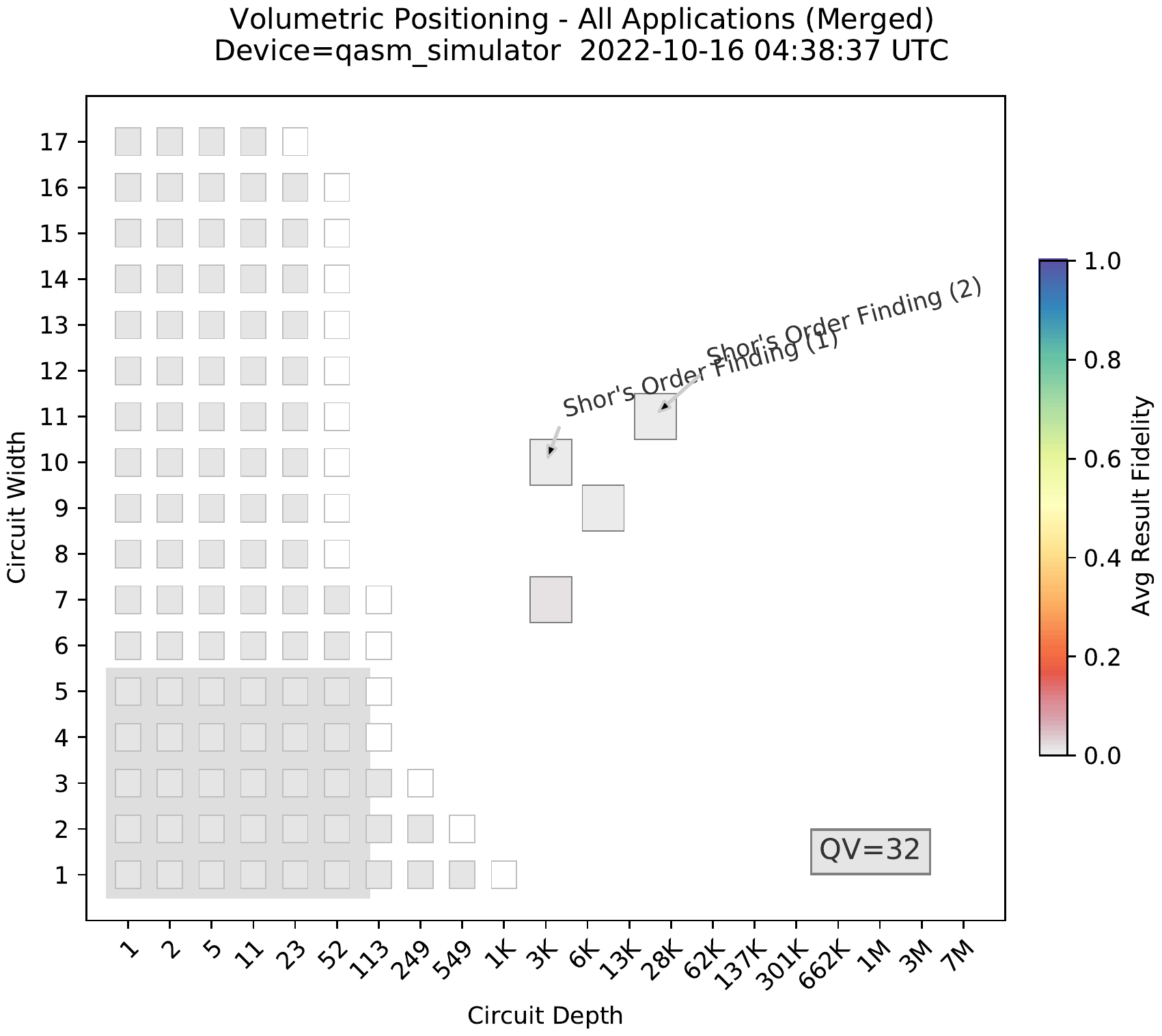}
\caption{Shor's Order Finding Algorithms  }
\label{fig:shors_vp_1}
\end{figure}

We constructed two benchmarks based on Shor's order finding algorithm, \benchmark{Shor's Order Finding(1)} and \benchmark{Shor's Order Finding(2)}, based on variants of Shor's algorithm introduced by Stephane Bearuegard \cite{Beauregard_2003}.
\benchmark{Shor's Order Finding(1)} is based on the standard formulation of Shor's algorithm, and it requires $4n + 2$ qubits where $n$ is the problem instance size (meaning that $N$ is an $n$ bit integer).
\benchmark{Shor's Order Finding(2)} uses a version of Shor's algorithm that utilizes mid-circuit measurements to reduce the number of qubits. This method uses $2n + 3$ qubits. Figure \ref{fig:shors_vp_1} shows the results obtained executing both these benchmarks on a simulator.

\section{Limitations of the Average Result Fidelity}\label{apdx:fidelity}
In this appendix we illustrate some of the limitations inherent to using the average result fidelity to quantify performance, using an example. This example illustrates the limitations of both fidelity and of averaging this fidelity over multiple circuits. We use the  \benchmark{Deutsch-Jozsa} benchmark are our example. The Deutsch-Jozsa algorithm is an $n+1$ qubit oracle algorithm that computes whether $f:\{0,1\}^n \to \{0,1\}$ is a balanced or constant function using a single call to an oracle that implements $f$. If the function is constant the first $n$ qubits should all return 0, and otherwise at least one of the qubits should return 1. In our benchmark we randomly choose the oracle: with 50\% probability $f$ is a balanced function, and with 50\% probability $f$ is constant. The balanced function that we use is the parity function, in which case all the qubits should return 1. Now consider two quantum computers, $Q_A$ and $Q_B$, that always return the bit strings $0000\dots$ and $1000\dots$, respectively. In both cases, if we run the Deutsch-Jozsa algorithm we always get the correct \emph{algorithm} output for one of the two inputs, and the incorrect output for the other, i.e., arguably $Q_A$ and $Q_B$ perform equally badly on this algorithm. But we find that $\bar{F}_{\rm s}=0.5$ for $Q_A$ and $\bar{F}_{\rm s}=0$ for $Q_B$, where $\bar{F}_{\rm s}$ is $F_{\rm s}$ averaged over the balanced and constant oracle. This is because for $Q_B$, $F_{\rm s}(P_{\rm output}, P_{\rm ideal}) = 0$ for both inputs (as $P_{\rm ideal}(111\dots)=1$ for the balanced function) \footnote{Here we have used $F_{\rm s}$ rather than the normalized fidelity $F$ for simplicity; the result for $F$ differs only by a small $n$-dependent factor.}. 

This example illustrates two points. First, the fidelity between $P_{\rm ideal}$ and $P_{\rm output}$ for a particular algorithm input does not directly correspond to the probability of obtaining the correct result on that input. In this example, $Q_B$ always returns the correct algorithm result on the balanced function input, even though $F_{\rm s}(P_{\rm ideal},P_{\rm output})=0$. To avoid this effect we would need an algorithm-specific metric for result quality. We suggest that the result fidelity should be complemented (but not replaced) with such metrics. Second, quantifying performance using only an average over problem instances can obscure important performance information. We suggest that it will be important to also study, e.g., worst-case performance over problem instances. 

\begin{figure}[t!]
\includegraphics[width=\columnwidth]{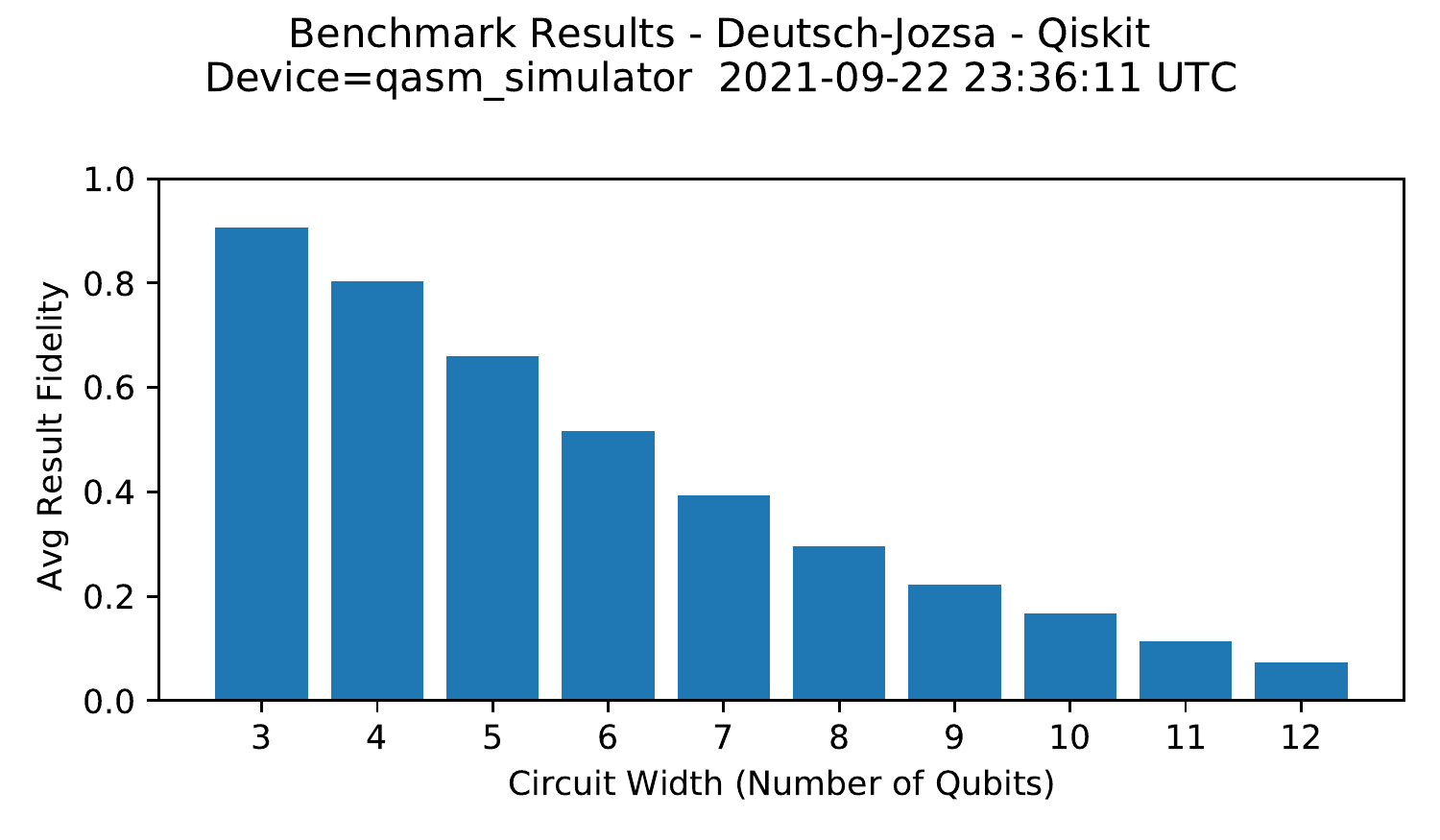}
\includegraphics[width=\columnwidth]{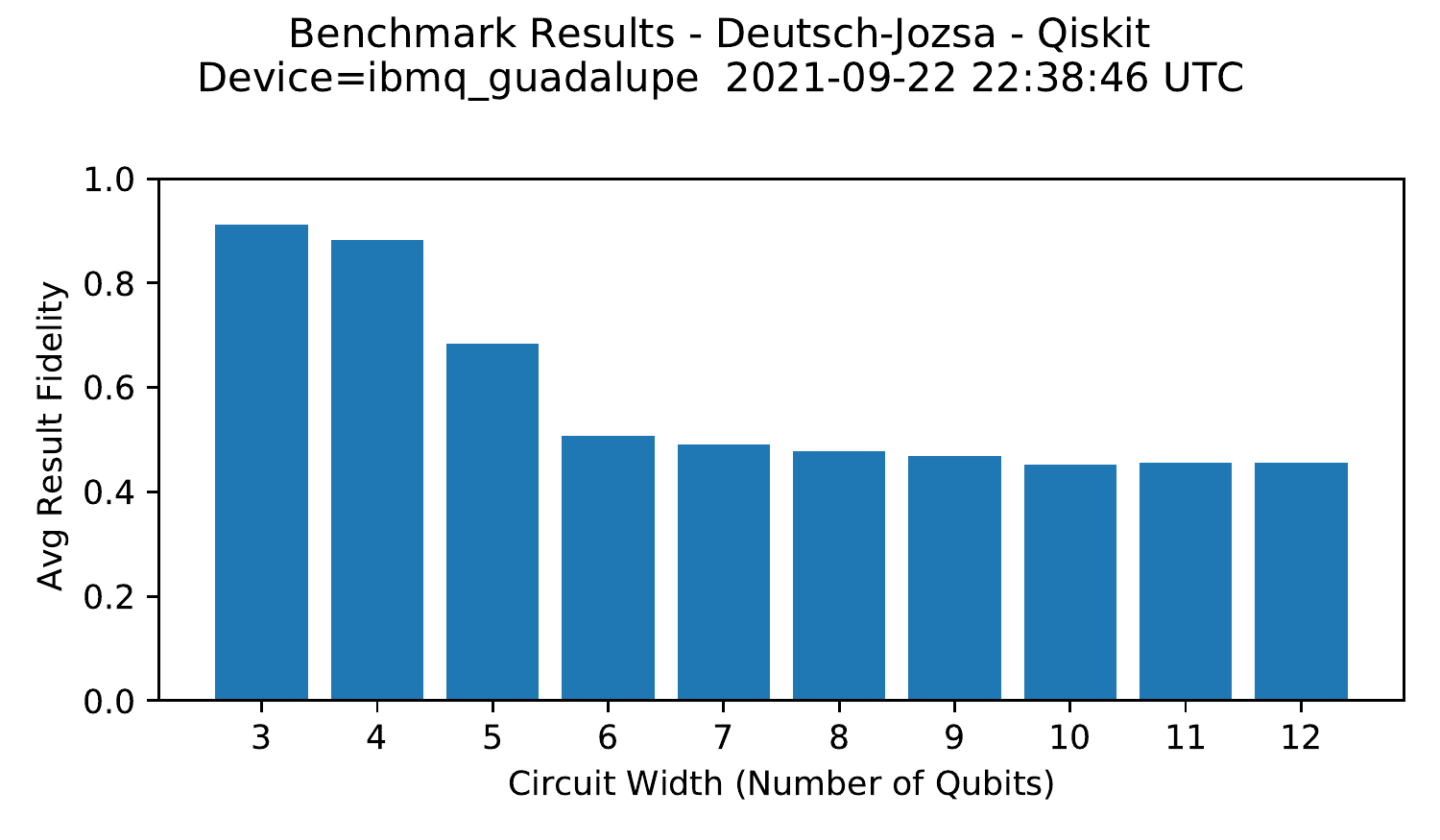}
\caption{Demonstrating the limitations of average result fidelity using an example. Results from running the  \benchmark{Deutsch-Jozsa} benchmark on a simulator (upper plot) with depolarizing errors and IBM Q Guadalupe (lower plot). Unlike in the simulation, the average result fidelity on IBM Q Guadalupe decays quickly to $\sim 0.5$ with increasing circuit width but then decays slowly. This suggests a strong asymmetry in the result fidelities for the two problem instances---the balanced and unbalanced function oracles---and, as discussed in the main text, this asymmetry is consistent with known noise mechanisms in IBM Q Guadalupe.}
\label{fig:dj_fidelity_ibmq_guadalupe_12-vplot-2}
\end{figure}

Similar effects to those discussed above were apparent in our experiments. Figure~\ref{fig:dj_fidelity_ibmq_guadalupe_12-vplot-2} shows the results of running the  \benchmark{Deutsch-Jozsa} benchmark on (a) a quantum simulator with depolarizing errors, and (b) IBM Q Guadalupe. In the simulation we observe that the average result fidelity decays towards zero as the number of qubits increases (note that our normalized result fidelity $F$ is designed so that it decays towards zero in the presence of depolarizing errors, unlike $F_{\rm s}$). In contrast, the average result fidelity for IBM Q Guadalupe decays quickly to $\sim 0.5$ with increased circuit width, and then decays slowly. The constant oracle uses a shallow $O(1)$ depth circuit containing no two-qubit gates, and its correct result is $000\dots$. In contrast, the balanced oracle uses $O(n)$ two-qubit gates and its correct result is $111\dots$. The constant oracle can therefore be expected to perform much better than the balanced oracle on IBM Q Guadalupe, which has high-fidelity one-qubit gates \cite{ibmq2021} and which (like all superconducting systems) experiences substantial T1 decay that adds a bias towards returning the $000\dots$ bit string. This strong asymmetry in the performance on the constant and balanced oracles would explain the results of Fig.~\ref{fig:dj_fidelity_ibmq_guadalupe_12-vplot-2}, and it illustrates that average result fidelities must be interpreted and compared with care.

\section{Issues with Different Quantum Computing APIs}
\label{apdx:issues_apis}
Quantum applications are being implemented using a variety of APIs, e.g., Qiskit, Cirq, Braket, and Q\#.
These APIs have many features in common, and our benchmark suite was initially designed to execute consistently across these multiple quantum programming environments.
However, certain constraints exist that make it challenging to accomplish complete consistency across all APIs.

As a consequence of these differences, all of the benchmarks have been implemented using the Qiskit API.
Some, but not all, have been implemented using the Cirq and Braket APIs. 
Throughout, all executions of our benchmarking suite on a quantum simulator are performed using the Qiskit implementation of our benchmarking suite, as it provides rich circuit analysis features.
Execution on all of the hardware was also performed using the Qiskit version of our benchmarks, with the exception of the experiments on Rigetti Computing's hardware, which used the Braket version of our benchmarks.

Our benchmarking circuits are defined in a high-level and algorithm-specific gate set, and then transpiled (1) to a standard gate set that we use to define each circuit's
`depth' [see Section~\ref{sec:circuit_depth}], and (2) to the native operations of a particular quantum computer in order to run the circuits.
There are two challenges to consistently implementing this methodology across different APIs. Firstly, the permitted gates differ between APIs. For example, in the Amazon Braket API, there is not direct support for controlled subcircuits which are used in our definitions of the  \benchmark{Monte Carlo Sampling} and \benchmark{Amplitude Estimation} benchmarks. These benchmarks are therefore not currently available in the Braket implementation of our suite. Second, we define a standarized notion of circuit depth by compiling to a standard gate set. However, the depth of the resultant circuit will depend on the performance of the compilation algorithm, which will be API dependent. The majority of our experiments to date have used the Qiskit API, limiting cross-API comparisons. An API-independent definition of circuit depth will likely be necessary to enable consistent result comparisons across different APIs.

In addition to issues related to circuit compiling, a variety of other API issues exist that make it challenging to consistently implement our benchmarking suite and to gather all of the data of interest. These issues include a requirement in Cirq to include topology information in the circuit definition, the limited availability of the transpiled circuit and detailed metrics of the execution time in Braket and Cirq, and minor issues relating to circuit visualizations.
For our work, the Qiskit platform provided the best overall set of features for implementing this suite of benchmarks and generating the analysis described in this manuscript.

\section{Advances in Quantum Computers}
\label{apdx:advances}
Developing a suite of performance benchmarks early in a technology's development establishes markers with which to measure advances in both hardware components and supporting software.
In order to provide this benefit over multiple product cycles, they must be designed to anticipate advances in technology that may emerge in the next five years. Factoring such advances into benchmarks from the beginning helps to ensure that future comparisons are meaningful.

We considered specific enhancements that have been widely discussed within the quantum computing community, and we reviewed how these would impact the code structure of the application benchmark programs.
Examples include new hardware features such as mid-circuit measurement, parameterized gate operations, and tighter integration of classical / quantum computation. Some of these advances are available now on a limited set of quantum computers. With input from the community as well as the providers of quantum computing hardware, the suite can evolve to effectively gauge the impact of performance improvements that result from these continuing advances.

\subsection{Mid-Circuit Measurements}
\label{apdx:mid_circuit_measurements}
In the first quantum computers introduced by commercial providers, classical measurement of qubit state was permitted only at the end of the circuit and qubits were not reusable.
Recently, IBM \cite{nation_johnson_2021,qiskit_mid_circuit_tutorial}, Quantinuum \cite{gaebler2021suppression,moore_2020} and others have introduced ways to perform measurement at any stage of the quantum circuit.
The measurements are recorded, qubits are reset to an initial state, and additional computation can be performed while qubit coherence is maintained by the quantum hardware.
The measurements collected during execution may be retrieved by the user with an API call.

\begin{figure}[b!]
\includegraphics[width=0.45\textwidth]{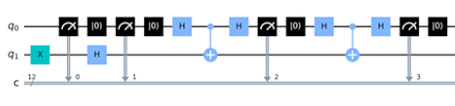}
\caption{A quantum circuit on 2 qubits that uses mid-circuit measurements to efficiently implement the Bernstein-Vazirani algorithm. This circuit implements the same computation as the standard Bernstein-Vazirani circuit on 5 qubits. This is the basis for the \benchmark{Bernstein-Vazirani(2)} benchmark.}
\label{fig:bv_5_mid_circuit}
\end{figure}

This feature provides an option for constructing algorithms with a smaller number of qubits. Ref.~\citet{nation_johnson_2021} describe an implementation of a Bernstein-Vazirani circuit on IBM hardware using only 2 qubits to encode a 12-bit secret string.
The fidelity in the measurement result is significantly greater than the equivalent 13-qubit circuit due to the use of only two qubits. Figure \ref{fig:bv_5_mid_circuit} shows a smaller version of that circuit that is equivalent to a 5-qubit implementation.

Our benchmark suite has some initial benchmarks that can test, and take advantage of, mid-circuit measurements. The current benchmarks that use mid-circuit measurement do so to implement algorithms using a smaller number of qubits. That represents the same problem as a circuit implemented with a larger number of qubits (a `qubit-equivalent' problem size).
In the \benchmark{Bernstein-Vazirani(1)} benchmark, a sweep is performed over a progressively larger number of qubits, encoding a larger secret integer each time. In the \benchmark{Bernstein-Vazirani(2)} the same sweep over problem input size is implemented using only two-qubit circuits. In this case, the depth of the circuit grows instead of the width.

Similarly, \benchmark{Shor's Order Finding(2)} benchmark implements Shor's order finding algorithm using a single qubit quantum phase estimation circuit, implemented with mid-circuit measurement. The effect of the mid-circuit measurement on Shor's algorithm can be seen in the reduction in circuit width, and in the fidelity of the results, in Figure \ref{fig:shors_vp_1}.

\subsection{Circuit Parameterization}
\begin{figure}[h!]
\includegraphics[width=0.45\textwidth]{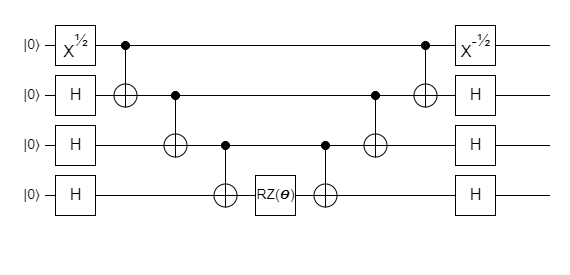}
\caption{An example of a parameterized quantum circuit, containing a single parameter ($\theta$), from a small instance of the VQE algorithm using the UCC ansatz.}
\label{fig:bm_ucc_ansatz_1}
\end{figure}

Some quantum algorithms are highly iterative, e.g., the VQE algorithm or the quantum approximate optimization algorithm (QAOA). These algorithms introduce an additional iteration layer that involves modifying the circuit before executing another round of shots. The circuit is typically changed only by modifying the rotation angles (parameters) on the gates in the circuit before executing it again. An example of this is the parameterized ansatz circuit in the VQE algorithm, an example of which is shown in Figure \ref{fig:bm_ucc_ansatz_1} taken from \cite{mccaskey2019quantum}.

Modification of rotation angles prior to execution can be done by creating the circuit from scratch in a high-level language (e.g. Qiskit), followed by compilation for the backend target.
The time taken to create, compile, and re-load the circuit for execution can be significant---on the order of a few seconds to minutes for larger circuits. For an algorithm that requires hundreds or thousands of iterations, this circuit preparation time dramatically impacts the total execution time.

The Qiskit and Cirq programming APIs both support a form of parameter definition within circuits using Python $sympy$ objects that improves performance by reducing compilation, but still requires reloading of the circuit.
It is possible that future advances in quantum computer control systems will reduce the latency between parameterized executions so that it is negligible, which would offer the optimal execution time for such applications.

The VQE-based benchmarks include the execution of a parameterized circuit. The ansatz circuit is executed repeatedly but with different angles applied to its gates. This is implemented by creating, compiling, and executing the circuit each time as a new circuit. Currently, the benchmarking suite cannot take advantage of any parameterized circuit representations and execution.
In a future work, the benchmarks could be enhanced with the addition of a method to define the ansatz as a parameterized circuit.

\subsection{Multiply-Controlled Gates}
\begin{figure}[h!]
\includegraphics[width=0.45\textwidth]{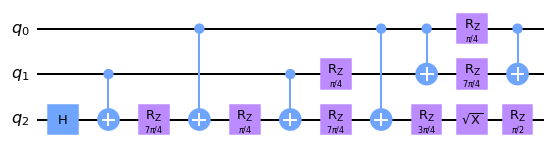}
\caption{The standard decomposition of the standard Toffoli gate into CNOT gates..}
\label{fig:ccx_decomp}
\end{figure}

Several of the benchmark circuits are defined using multiply-controlled gate operations.
This is a type of gate in which the operation applied to the target qubit is controlled by more than just a single qubit.
The many-qubit Toffoli gate is used in the definition of many of the benchmarks. None of today's hardware is able to implement these operations natively. Any multiply-controlled gate needs to be transpiled into many smaller primitive gate operations.
Figure \ref{fig:ccx_decomp} shows the decomposition of a standard Toffoli gate (an $3$-qubit Toffoli gate) which has two control qubits.

Some of our benchmarks use $n$-qubit Toffoli gates with $n$ proportional to the circuit width. The transpiled depth of an $n$-qubit Toffoli gate increases quickly with $n$. Thus the transpiled circuits grow in depth very rapidly as circuit width increases. This is readily seen in the volumetric positioning profile for the \benchmark{Grover's Search} benchmark in Figure \ref{fig:bm_grovers_vp}.

It is possible that hardware advances will lead to native implementations of some multiply-controlled gates (although note that $n$-qubit controlled gates for arbitrary $n$ cannot be implemented in constant time).
For example, \citet{article_qubit_qutrit} have described a possible native implementation of a Toffoli gate as a CCZ gate.
If multiply-controlled gates are implemented natively, benchmarks that are defined using them will quantify the performance improvement that they provide.

\subsection{Close Classical/Quantum Integration}
Increased attention is being given to the total execution time of quantum applications \cite{johnson_faro_2021,cao_hirzel_2020}.
Quantum computers are increasingly able to execute complex applications that include multiple iterations of quantum circuits as subroutines interleaved with classical computation.
The total time taken by the application to arrive at a solution, aggregated over all stages of the execution, can be hours or even days. Architectural improvements in the implementation of the hardware and firmware of the quantum computer can result in significant reduction in these execution times.

In a recent IBM blog post, \citet{johnson_faro_2021} report on enhancements to their quantum computing architecture that results in a reduction in the `time to solution' for a specific chemistry problem, from several days to nine hours.
This is significant in that it represents a focus on the evolutionary reduction in the time taken to compute a solution, and not solely on the relative asymptotic scaling time of quantum computing as compared to classical.
This trend is likely to continue, as evidenced by current research into quantum accelerators \cite{Bertels_2020} and other architectural optimizations \cite{M_ller_2017,cao_hirzel_2020}.

\citet{Karalekas_2020} described a quantum-classical-cloud architecture that supports a tight coupling between execution of the classical and quantum parts of a hybrid algorithm with adjacent hardware, to enable faster execution of such programs over the cloud. IBM's announcement of public availability of the Qiskit Runtime in IBM Quantum \cite{qiskit_runtime} is another step in this direction. In a solution of this type, a user may upload a quantum program and run it with different inputs and configurations each time each time it is executed without having to reload the program. Currently, the benchmarks have not been structured to take advantage of Qiskit Runtime. Classical and quantum computation can be integrated even further, by implementing classical computations alongside the execution of a quantum circuit, within the coherence time of the qubits \citet{wehden_gambetta_faro_2021}. For example, Ref.~\citet{takita_lekuch_corcoles_2021} discuss how such a feature would improve the implementation of a quantum phase estimation algorithm. Our benchmarking suite will require updates to fully test tightly integrated classical and quantum processing.

Precise measurement of the execution times for various stages of a combined quantum/classical application may have particular value in gauging the evolution and impact of quantum/classical computing integration. Currently, the benchmark suite collects and reports on the quantum-specific execution time reported by the target system, as well as the circuit creation, launch and queue times.
The iterative nature of the benchmarks means that the total execution time for any one benchmark program is a rough indicator for overall total time to solution for that type of application (factoring out the highly variable queue times).
An analysis of execution times aggregated over multiple executions is not reported systematically or compared in the current suite.

\clearpage

\bibliographystyle{unsrtnat}  
\bibliography{references}
\end{document}